%% file: Article.tex
\documentclass[12pt]{article}

\usepackage[utf8]{inputenc}
\usepackage[margin=1.0in]{geometry}

\usepackage{upgreek}
\usepackage{graphicx}
\usepackage{float}
\graphicspath{ {Images/} }

\usepackage{amssymb}
 \usepackage{amsmath}
\usepackage{bm}
\usepackage[T1]{fontenc}
\usepackage{longtable}
\usepackage{lineno}
\usepackage{cite}
\usepackage{braket}
\usepackage{amsmath}
\usepackage{gensymb}
\usepackage{multicol}
\usepackage{authblk}

\input{ShortCut}

\title{Theoretical and Computational Framework for the Analysis of the Relaxation Properties of Arbitrary Spin Systems. Application to High-Resolution Relaxometry}

\author[1]{Nicolas Bolik-Coulon}
\author[1]{Pavel Kade$\mathrm{\check{r}\acute{a}}$vek}
\author[1]{Philippe Pelupessy}
\author[2]{Jean-Nicolas Dumez}%
\author[1]{Fabien Ferrage}%
\author[1]{Samuel F. Cousin}%

\affil[1]{Laboratoire des Biomolécules, LBM, Département de Chimie, École Normale Supérieure, PSL University, Sorbonne Université, CNRS, 75005 Paris, France}
\affil[2]{CEISAM, CNRS, Université de Nantes, 44300 Nantes, France}

\date{}
\begin{document}
\maketitle

\begin{abstract}
A wide variety of nuclear magnetic resonance experiments rely on the prediction and analysis of relaxation processes. Recently, innovative approaches have been introduced where the sample travels through a broad range of magnetic fields in the course of the experiment, such as dissolution dynamic nuclear polarization or high-resolution relaxometry. Understanding the relaxation properties of nuclear spin systems over orders of magnitude of magnetic fields is essential to rationalize the results of these experiments. For example, during a high-resolution relaxometry experiment, the absence of control of nuclear spin relaxation pathways during the sample transfers and relaxation delays leads to systematic deviations of polarization decays from an ideal mono-exponential decay with the pure longitudinal relaxation rate. These deviations have to be taken into account to describe quantitatively the dynamics of the system. Here, we present computational tools to (1) calculate analytical expressions of relaxation rates for a broad variety of spin systems and (2) use these analytical expressions to correct the deviations arising in high-resolution relaxometry experiments. These tools lead to a better understanding of nuclear spin relaxation, which is required to improve the sensitivity of many pulse sequences, and to better characterize motions in macromolecules.
\end{abstract}

\tocless\section{Introduction}
\label{Introduction}
\input{Chapters/0_Introduction}

\tocless\section{Theory and relaxation}
\label{Theory}
\input{Chapters/1_Theory}

\tocless\section{Implementation and usage}
\label{Implementation}
\input{Chapters/2_Implementation}

\tocless\section{Application to \CDDH-methyl groups using HRR and 2F-NMR}
\label{MethylGroupRelaxometry}
\input{Chapters/3a_MethylGroup_Theory}

\label{MethylGroupTheory}
\input{Chapters/3b_MethylGroup_Results}

\label{MethylGroupResults}

\input{Chapters/Conclusion}
\label{Conclusion}

\input{Chapters/MaterialAndMethods}
\label{Methods}

\section*{Data availability}
\RedKite~can be found here: https://figshare.com/articles/RedKite/11745111 \\
The ICARUS suite (ICARUS, MCMC script and RedKite2ICARUS) can be found here: https://figshare.com/articles/ICARUS/9893912

\bibliographystyle{elsarticle-num} 
\bibliography{Bibliography_Main}

\input{Chapters/SuppInfo}

\label{SupInfo}

\end{document}

%% file: ShortCut.tex
\newcommand{\CDDH}{$\{^{13}$C$^1$H$^2$H$_2\}$}

\newcommand{\Proton}{$^1$H}
\newcommand{\Carbon}{$^{13}$C}
\newcommand{\carb}{^{13}\mathrm{C}}

\newcommand{\ubiquitin}{U-$[$$^2$H, $^{15}$N$]$, Ile-$\delta_1$[$^{13}$C$^2$H$_2\!^1$H]-Ubiquitin}

\newcommand{\omegaC}{\omega_{\mathrm{C}}}
\newcommand{\omegaH}{\omega_{\mathrm{H}}}
\newcommand{\omegaD}{\omega_{\mathrm{D}}}
\newcommand{\omegaN}{\omega_{\mathrm{N}}}
\newcommand{\Cz}{\^C$_{\mathrm{z}}$}

\newcommand{\Pup}{\hat{\hat{\mathcal{P}}}^\mathrm{up}}
\newcommand{\Pdown}{\hat{\hat{\mathcal{P}}}^\mathrm{down}}

\newcommand{\Quadr}{d_{\mathcal{Q}}} 

\newcommand{\Lorentz}[1]{\frac{#1}{1+(\omega #1)^2}}

\newcommand{\Jimpd}[2][Wikimedia]{
  \mathcal{J}_ \mathrm{#1,#2}}
\newcommand{\Jimpu}[1]{\mathcal{J}_\mathrm{#1}}

\newcommand{\diffd}{\mathrm{d}}
\newcommand{\deriv}[2]{\frac{\mathrm{d}#1}{\mathrm{d}#2}}

\newcommand\Hz{\^H$_{\mathrm{z}}$}
\newcommand\CzHz{\^C$_{\mathrm{z}}$\^H$_{\mathrm{z}}$}

\newcommand\RedKite{\textsc{RedKite}}
\newcommand{\thIJ }{\theta_{i,j}}

\newcommand{\SsD}{S_s^2}
\newcommand{\SfD}{S_f^2}
\newcommand{\SD}{S^2}

\newcommand{\TldH}[1]{\tilde{\hat{#1}}}

\newcommand{\beginsupplement}{%
        \setcounter{table}{0}
        \renewcommand{\thetable}{S\arabic{table}}%
        \setcounter{figure}{0}
        \renewcommand{\thefigure}{S\arabic{figure}}%
     }

\newcommand{\nocontentsline}[3]{}
\newcommand{\tocless}[2]{\bgroup\let\addcontentsline=\nocontentsline#1{#2}\egroup}
\newcommand{\mysection}[2]{\setcounter{section}{#1}\addtocounter{section}{-1}\section{#2}}

%% file: Chapters/0_Introduction.tex
The development of most Nuclear Magnetic Resonance (NMR) experiments requires the understanding of relaxation properties. Improvement in the sensitivity and resolution have been obtained, ranging from the use of an optimum excitation angle with respect to the longitudinal relaxation, known as the Ernst angle \cite{Ernst_1966}, to the development of Transverse Relaxation Optimized SpectroscopY (TROSY) experiments \cite{Pervushin_pnas_1997,tugarinov_jacs_2003} that exploit relaxation interferences \cite{Shimizu_JCP_1964, Goldman1984, Wimperis_MolPhys_1989, Werbelow_JMR_1973}. An in depth investigation of relaxation processes is particularly critical to design and interpret several classes of experiments which are based on moving the sample through a broad range of magnetic fields. A variety of such experiments have been designed recently: (1) The existence of Long-Lived States (LLS) \cite{Carravetta_PRL_2004,Carravetta_JCP_2005} was revealed by the combination of high-field coherent evolution and low-field relaxation; (2) In dissolution Dynamic Nuclear Polarization (dDNP) \cite{Larsen2003,Milani_JCP_2015,Hilty_2008}, the hyperpolarized sample is transferred back and forth between the polarizing magnetic center and the high-field spectrometer through magnetic fields that can be as low as the earth magnetic field; (3) Multi-scale dynamics can be characterized with Fast-Field Cycling (FFC) relaxometry \cite{Kimmich2004} where the magnetic field is switched from \textit{ca.}~1\,T down to \textit{ca.}~100\,$\mu$T; (4) A sample-shuttle apparatus can be used to combine relaxometry experiments with high-field NMR \cite{Redfield_MRC_2003,Redfield2012,chou_jmr_2012,Charlier2013} to gain atomic resolution description of molecular dynamics; (5) This kind of device can also be used to investigate relaxation properties of spin terms that are only relevant at low fields \cite{Korchak_JCP_2010}; (6) A sample shuttle may couple two magnetic centers in a two-field NMR spectrometer \cite{Cousin_PCCP_2016} to record multi-dimensional experiments where spins are manipulated at two vastly different fields \cite{Cousin_PCCP_2016, Cousin2016, KaderavekJPCL2019, Jasenakova_JBNMR_2020}. \\
Sample-shuttling experiments have been used to measure longitudinal relaxation rates over orders of magnitude of magnetic fields and characterize the dynamics of membrane vesicules \cite{Roberts_JACS_2004}, protein backbone \cite{Clarkson2009, Charlier2013} and side-chains \cite{Cousin_JACS_2018}. This type of experiments, called High-Resolution Relaxometry (HRR), consists in the measurement of relaxation rates over a broad range of magnetic field while preserving the high resolution of conventional high-field magnets (\textit{i.e.} higher than 9\,T) \cite{Redfield_MRC_2003, Redfield2012}. This approach relies on moving the NMR sample in the stray field of a commercial magnet to measure longitudinal relaxation rates over orders of magnitude of magnetic field. The sample is transfered back in the high-field magnetic center for detection, thus ensuring high sensitivity and resolution. \\
During a high-resolution relaxometry experiment, the sample is moved outside of the magnetic center where no radiofrequency pulse can be applied. Thus, relaxation decays acquired using HRR suffer from two types of systematic errors. First, the effective density operator at the beginning of the relaxation delay is usually different from the desired longitudinal operator due to cross-relaxation during the sample transfers. Second, cross-relaxation pathways during the relaxation delay may lead to multi-exponential polarization decays. Therefore, the analysis of experimental HRR rates requires to account for these systematic deviations in order to accurately determine the motional parameters of the system under study. We introduced an iterative correction procedure called Iterative Correction for the Analysis of Relaxation Under Shuttling (ICARUS) \cite{Charlier2013, Cousin2018} for the correction of HRR relaxation rates. Using symbolic expressions of magnetic-field dependent relaxation matrices, the HRR experiments are simulated and measured relaxometry relaxation rates are corrected so that a reliable analysis of the dynamic properties of the system under study can be performed. \\
Thus, the development of tools to simulate spin relaxation for a broad variety of field trajectories is of great interest, in several areas of magnetic resonance \cite{Jerschow_JMR_2005, Kuprov2006, Bengs_MagnResonChecm_2017}. Here, we present a toolbox that combines two programs. The first one, \RedKite, provides analytical expressions of relaxation rates and relaxation matrices for arbitrary spin system. The second one, ICARUS, is used to retrieve accurate estimates of longitudinal relaxometry relaxation rates that are further used to determine the parameters describing the dynamics of the system. ICARUS simulates the HRR experiments using analytical expressions obtained from \RedKite. \\
\RedKite~has been written in \textsc{Mathematica} (version 12.0) \cite{Inc2016} to perform efficiently analytical calculations using the \textsc{SpinDynamica} (version 2.15.1b10) \cite{Bengs_MagnResonChecm_2017} package and the so called "BRW engine" to simplify the computation of relaxation rates \cite{Kuprov2006}. This version of ICARUS has been written in \textsc{Python} (version 3.5). This language has the advantage of being free and easy to install, allowing for relatively fast numerical evaluations, and being easy to customize by the user. ICARUS is written as a framework so that users can define the spin systems, relaxation matrices and spectral density functions relevant for their applications. \\
In this paper, we first describe succinctly our approach to calculate relaxation rates efficiently and apply this method on an isolated $^{15}$N$^1$H spin system using \RedKite. We  illustrate the power of these tools with a detailed presentation of the recently published analysis of carbon-13 HRR in \CDDH-methyl groups in the protein Ubiquitin \cite{Cousin_JACS_2018} and test the validity of key hypotheses made during the analysis. In particular, we use two-field NMR to determine the relevant interactions to describe the relaxation properties of \CDDH-methyl groups, and verify the validy of the correction at 0.33\,T. \\

%% file: Chapters/1_Theory.tex
\tocless\subsection{Calculation of relaxation superoperators with \RedKite}
\label{BWRtheory}
The full description of the Bloch-Wangsness-Redfield (BWR) relaxation theory in liquid-state NMR is beyond the scope of this article and can be found elsewhere \cite{Kumar2000, Goldman1984, Kowalewski2006, Nicholas2010, Abragam1961}. A condensed version is presented here.\\
The evolution of the density operator $\hat{\sigma}(t)$ is described by the Liouville-von Neumann equation, in units of $\hbar$:
\begin{equation}
	\deriv{\hat{\sigma}(t)}{t}=-i[\hat{\mathcal{H}}(t),\hat{\sigma}(t)].
	\label{eq:Liouville}
\end{equation}
The Hamiltonian $\hat{\mathcal{H}}$ of the system can be expressed as the sum of a stationary part $\hat{\mathcal{H}}_0$ and a fluctuating part $\hat{\mathcal{H}}_1(t)$:
\begin{equation}
	\hat{\mathcal{H}}(t)=\hat{\mathcal{H}}_0 + \hat{\mathcal{H}_1}(t).
\end{equation}
This equation can be transformed in the interaction frame of the stationary Hamiltonian $\hat{\mathcal{H}}_0$. An operator $\hat{\mathcal{O}}$ transformed into the interaction frame is labeled with a tilde:
\begin{equation}
\tilde{\hat{\mathcal{O}}}(t)=\exp{(i\hat{\mathcal{H}}_0t)}\hat{\mathcal{O}}(t)\exp{(-i\hat{\mathcal{H}}_0t)}.
\label{eq:RepresentationChange}
 \end{equation}
The frame transformation of the full Hamiltonien $\hat{\mathcal{H}}$ requires the subtraction of the Zeeman Hamiltonien $\hat{\mathcal{H}}_0$, so that the Liouville-von Neumann equation now reads:%
\begin{equation}
	\deriv{\tilde{\hat{\sigma}}(t)}{t}=i[\tilde{\hat{\sigma}}(t),\tilde{\hat{\mathcal{H}}}_1(t)].
	\label{eq:LvN}
\end{equation}
After developing a second-order time-dependent perturbation, the Liouville-von Neumann equation in the interaction frame can be written as:
\begin{equation}
	\frac{\diffd \TldH{\sigma}(t)}{\diffd t}=
	+i \left[ \TldH{\sigma}(0),\TldH{\mathcal{H}}_1(t) \right]
      - \int \limits_0^t  \left[[\TldH{\sigma}(t'),\TldH{\mathcal{H}}_1(t')] , \TldH{\mathcal{H}}_1(t) \right]\diffd t'. %
	\label{eq:main1}
\end{equation}
In the frame of the BWR theory, the following hypotheses are made to calculate the ensemble average of the evolution of the density operator: i) for an ensemble average, denoted by the horizontal bar, $\overline{\left[ \TldH{\sigma}(0),\TldH{\mathcal{H}}_1(t) \right]}$ averages to zero, and ii) a time $t$ can be found that is short enough such that the evolution of the spin system is negligible on the interval $[0, t]$ but that is much larger than the typical correlation times for the fluctuations of $\TldH{\mathcal{H}}_1(t)$. The evolution of the density matrix $\TldH{\sigma}(t)$ over time for an ensemble average, under a perturbation Hamiltonian $\TldH{\mathcal{H}}_1(t)$, can now be expressed as:
\begin{equation}
	\frac{\overline{\mathrm{d} \TldH{\sigma}(t)}}{\mathrm{d} t}=- \int \limits_0^{\infty} \overline{ \left[ \TldH{\mathcal{H}}_1(t),[\TldH{\mathcal{H}}_1(t+\tau),\TldH{\sigma}(t)] \right]}\mathrm{d} \tau.%
    \label{eq:masterequation}
\end{equation}
This equation can be further simplified using the irreducible tensor representation in order to separate the angular and spin parts of the Hamiltonian. The perturbation Hamiltonian $\TldH{\mathcal{H}}_1(t)$ may include several interactions, identified by the label $i$. Each of them can be written as the sum of the product of time-dependent spatial variables $V_{l,-q}(t)$ and tensor spin operators $\hat{T}_{l,q}$ of rank $l$ and coherence order $q$ (which is usually simply called order):
\begin{equation}
	\hat{\mathcal{H}}_1(t)=\sum_{i}\zeta_i \sum \limits_{l} \sum^{l} \limits_{q=-l}(-1)^q V^i_{l,-q}(t) \hat{T}^i_{l,q}, 
    \label{eq:Hpertube}
\end{equation}
where $\zeta_i$ is the amplitude of the interaction $i$. The irreducible tensor $\hat{T}^i_{l,q}$ can be expressed as a linear combination of  eigenoperators $\{\hat{A}_{l,q,p}^{i}\}$ of the superoperator $[ \hat{H}_0, \cdot ]$, with eigenvalues $\omega_{l,q,p}^{(i)}$:\\
\begin{equation}
\hat{T}^i_{l,q} = \sum_{p} \hat{A}_{l,q,p}^{i}.
\end{equation}
These eigenoperators can be written in the interaction frame as:
\begin{equation}
\TldH{A}_{l,q,p}^{i} (t) = \exp{(i\hat{\mathcal{H}}_0t)}\hat{A}_{l,q,p}^{i}\exp{(-i\hat{\mathcal{H}}_0t)} = e^{i\omega_{l,q,p}^{(i)} t}\hat{A}_{l,q,p}^{i}.
\end{equation}
In the interaction frame, we now have:
\begin{equation}
\TldH{\mathcal{H}}_1(t)=\sum_{i}\zeta_i \sum \limits_{l} \sum^{l} \limits_{q=-l} \sum_p (-1)^q e^{i \omega_{l,q,p}^{(i)} t} V^i_{l,-q}(t) \hat{A}^i_{l,q, p}.
	\label{eq:H1}
\end{equation}
Since $\TldH{\mathcal{H}}_1$ is Hermitian, we can also write:
\begin{equation}
\TldH{\mathcal{H}}_1(t)=\sum_{i}\zeta_i \sum \limits_{l} \sum^{l} \limits_{q=-l} \sum_p (-1)^q e^{-i \omega_{l,q,p}^{(i)} t} V^{i,\ast}_{l,-q}(t) \hat{A}^{i,\dag}_{l,q, p},
	\label{eq:H1Cplx}
\end{equation}
where ($\dag$) denotes the hermitian conjugate of the operator, and ($\ast$) the complex conjugate. Substituting Eq.\,\ref{eq:H1} and \ref{eq:H1Cplx} into Eq.\,\ref{eq:masterequation} gives:
\begin{equation}
	\begin{aligned}
	\frac{\overline{\mathrm{d} \TldH{\sigma}(t)}}{\mathrm{d} t} = &
              - \sum \limits_{i,j} \zeta_i \zeta_j \sum_{l, l'}  \sum^{l} \limits_{q=-l} \sum^{l'} \limits_{q'=-l'} \sum_{p, p'} (-1)^{q+q'}   
	      e^{i(\omega_{l,q,p}^{(i)} - \omega_{l',q',p'}^{(j)}) t}  \times \\
		& \left[\hat{A}_{l,q,p}^{i},[\hat{A}_{l',q',p'}^{j,\dag},\tilde{\hat{\sigma}}(t)] \right]   \int \limits_0^{\infty} \langle V^i_{l,-q}(t)V_{l',-q'}^{j,\ast}(t+\tau) \rangle e^{-i\omega_{l',q',p'}^{(j)}\tau} \mathrm{d} \tau,
	\end{aligned}
	\label{eq:EvolutionFull}
\end{equation}
The correlation function $C_{i,j}$ between the interations $i$ and $j$ is defined as:
\begin{equation}
	\langle V^i_{l,-q}(t)V_{l',-q'}^{j\ast}(t+\tau) \rangle =\frac{1}{2l+1} \delta_{q,q'}\delta_{l,l'}C_{i,j}(\tau),
    \label{eq:Correlation}
\end{equation}
where $\delta$ is the Kronecker delta. Oscillating terms are neglected as they average to zero much faster than the evolution of the density operator (secular approximation) under relaxation. Thus, only secular terms for which $\omega_{l,q,p}^{(i)}=\omega_{l',q',p'}^{(j)}$ contribute to Eq.\,\ref{eq:EvolutionFull}. Only rank-2 ($l=2$) tensors are relevant to describe dipole-dipole and quadrupolar interactions. For the CSA interaction, the rank-1 tensor part (antisymmetric) is usually neglected. Note that, in the presence of highly anisotropic motions, the contribution of the antisymmetric CSA (rank-1 tensors) may account for up to 10\,\% of the contribution of the CSA rank-2 tensors to auto-relaxation \cite{Kowalewski1997, Paquin2010}. In the following, only rank-2 tensors are considered.\\
The spectral density function is defined as the Fourier tranform of the correlation function:
\begin{equation}
\mathcal{J}_{i,j}(\omega)= 2 \int \limits_0^{\infty} \frac{1}{5}C_{i,j}(\tau) e^{-i\omega \tau} \mathrm{d}\tau.
\end{equation}
Inserting the spectral density function in Eq.\,\ref{eq:EvolutionFull} and applying the above approximations leads to the following expression of the Master equation:
\begin{equation}
	\frac{\overline{\mathrm{d} \TldH{\sigma}(t)}}{\mathrm{d} t}=-\frac{1}{2} \sum \limits_{i,j} \zeta_i \zeta_j  \sum^{2} \limits_{q=-2} \sum_{p,p'} \delta_{\omega_{2,q,p}^{(i)},\omega_{2,q,p'}^{(j)}} \mathcal{J}_{i,j}\Bigl(\omega_{2,q,p}^{(i)}\Bigr)   
	 \left[ \hat{A}^i_{2,q,p},[\hat{A}_{2,q,p'}^{j,\dag}, \tilde{\hat{\sigma}}(t)] \right].
	\label{eq:FinalEvolution}
\end{equation}
The final step consists in transforming Eq.\,\ref{eq:masterequation} from the interaction representation back to the Schr\"odinger representation given in Eq.\,\ref{eq:Liouville}. For this, we invert Eq.\,\ref{eq:RepresentationChange}:
\begin{equation}
	\hat{\sigma}(t)=\exp{(-i\hat{\mathcal{H}}_0t)}\tilde{\hat{\sigma}}(t)\exp{(i\hat{\mathcal{H}}_0t)},
\end{equation}
with time-derivative:
\begin{equation}
\frac{\mathrm{d} \hat{\sigma}(t)}{\mathrm{dt}} = - i [\hat{\mathcal{H}}_0,\hat{\sigma}(t)] + \exp{(-i\hat{\mathcal{H}}_0t)} \frac{\mathrm{d} \tilde{\hat{\sigma}}(t)}{\mathrm{dt}} \exp{(i\hat{\mathcal{H}}_0t)}.
\label{eq:DensityLabFrame}
\end{equation}
Inserting Eq.\,\ref{eq:FinalEvolution} into Eq.\,\ref{eq:DensityLabFrame} leads to:
\begin{equation}
	\begin{aligned}
		\frac{\overline{\mathrm{d} \hat{\sigma}(t)}}{\mathrm{d} t}=
		& -i \overline{[\hat{\mathcal{H}}_0,\hat{\sigma}(t)]}-\\
		& \frac{1}{2} \sum \limits_{i,j} \zeta_i \zeta_j  \sum^{2} \limits_{q=-2} \sum_{p, p'}  \delta_{\omega_{2,q,p}^{(i)}, \omega_{2,q,p'}^{(j)}} \mathcal{J}_{i,j}\Bigl(\omega_{2,q,p}^{(i)}\Bigr)  
		   \left[ \hat{A}^i_{2,q,p},[\hat{A}_{2,q,p'}^{j,\dag}, \hat{\sigma}(t)] \right].
	\end{aligned}
\end{equation}
We now define the relaxation super-operator $\hat{\hat{\mathcal{R}}}$ as:
\begin{equation}
\hat{\hat{\mathcal{R}}} =\frac{1}{2} \sum \limits_{i,j} \zeta_i \zeta_j  \sum^{2} \limits_{q=-2} \sum_{p, p'}  \delta_{\omega_{2,q,p}^{(i)}, \omega_{2,q,p'}^{(j)}} \mathcal{J}_{i,j}\Bigl(\omega_{2,q,p}^{(i)}\Bigr)  
		   \left[ \hat{A}^i_{2,q,p},[\hat{A}_{2,q,p'}^{j,\dag}, \cdot ] \right].
	\label{eq:RelaxOp}
\end{equation}
The relaxation rate between operators $\hat{A}$ and $\hat{B}$ is:
\begin{equation}
		\mathcal{R}(\hat{A},\hat{B})= \frac{\langle \hat{B} | \hat{\hat{\mathcal{R}}} | \hat{A} \rangle}{\sqrt{\langle \hat{A} | \hat{A}\rangle \langle \hat{B} | \hat{B} \rangle}}.
\end{equation} 
If $\hat{A} = \hat{B}$, we speak of an auto-relaxation rate, while $\hat{A} \neq \hat{B}$ refers to a cross-relaxation rate, if $i = j$, it is an auto-correlated relaxation rate, and if $i \neq j$ a cross-correlated relaxation rate. These rates can easily be calculated analytically using the BRW engine \cite{Kuprov2006}. It consists in calculating the double commutator for each pair of spin tensors with identical eigenfrequencies and multiplying them by the spectral density function evaluated at this frequency. The implementation of this algorithm in \textsc{Mathematica} \cite{Inc2016} is detailed for an isolated $^{15}$N$^1$H spin pair (Section\,3.1) and a $^{13}$C$^1$H$^2$H$_2$ methyl group with a vicinal deuterium (Supplementary Materials).

\tocless\subsection{Expectation value of spin operators}
\label{AverageLiouvillian}
The expectation value of a specific operator after an evolution period $t$ is obtained from the calculation of the propagator:
\begin{equation}
\hat{\hat{\mathcal{P}}}(t) = e^{-\hat{\hat{\mathcal{L}}} t},
\label{eq:PropagatorGeneral}
\end{equation}
with $\hat{\hat{\mathcal{L}}}$ the Liouvillian. Eq.\,\ref{eq:PropagatorGeneral} assumes a constant Liouvillian over the interval $t$, including a constant Hamiltonian. This assumption does not hold when pulses are applied, or in field-varying experiments, such as in dDNP or relaxometry. In dDNP, the sample is polarized using a microwave source at a specific field outside the NMR spectrometer, dissolved and pushed into the spectrometer, so that the sample experiences successively: the static field of the polarizer, the fields of the trajectory between the polarizer and the spectrometer, and the static field of the NMR spectrometer \cite{Larsen2003}. In a relaxometry experiment, the fields during the polarization, relaxation and detection periods are potentially all different \cite{Kimmich2004,Redfield2012}. In these cases, the evolution time $t$ is decomposed in periods that are small enough so that the field can be considered constant, and the propagator equals:
\begin{equation}
\hat{\hat{\mathcal{P}}}(t) = \hat{\hat{d\mathcal{P}_n}}(\delta t_n, B_n) \times ... \times \hat{\hat{d\mathcal{P}_1}}(\delta t_1, B_1),
\label{eq:decomposeProp}
\end{equation}
where $\hat{\hat{d\mathcal{P}_i}}$ is the propagator during the interval $\delta t_i$ for which the magnetic field equals $B_i$. \\
When pulses are applied, which is typically the case in standard pulse sequences for the measurement of relaxation rates \cite{Ferrage_MMB_2012}, cross-relaxation pathways may no longer be active and Eq.\,\ref{eq:decomposeProp} can be simplified using averaged Liouvillian theory \cite{Levitt1994, Ghose2009}. For example, for the measurement of longitudinal relaxation rates of nitrogen-15 in a $^{15}$N-$^{1}$H spin pair, proton $\pi$-pulses are applied during the relaxation delay. In the abscence of such pulses, the Liouvillian reads:
\begin{equation}
\hat{\hat{\mathcal{L}}} =  	\begin{pmatrix}
		R_1^\mathrm{N} & \sigma_\mathrm{NH} & \delta_\mathrm{N}\\
    	\sigma_\mathrm{NH} & R_1^\mathrm{H} & \delta_\mathrm{H}\\
        \delta_\mathrm{N} & \delta_\mathrm{H} & R_\mathrm{NH}
	\end{pmatrix},
\end{equation}
where the relaxation matrix has been written in the basis formed by the spin operators $\{\hat{\mathrm{N}}_z, \hat{\mathrm{H}}_z, 2\hat{\mathrm{N}}_z\hat{\mathrm{H}}_z\}$ and $R_1^\mathrm{N}$ (respectively $R_1^\mathrm{H}$) refers to nitrogen-15 (respectively proton) longitudinal relaxation rate, $R_\mathrm{NH}$ to the two-spin order relaxation rate, $\sigma_\mathrm{NH}$ to the dipole-dipole (DD) cross-relaxation rate between the nitrogen-15 and proton, and $\delta_\mathrm{N}$ (respectively $\delta_\mathrm{H}$) to the CSA-DD cross-correlated cross-relaxation rate involving the nitrogen-15 (respectively proton) CSA. After applying a proton $\pi$-pulse, the Liouvillian is transformed according to:
\begin{equation}
\hat{\hat{\mathcal{L}'}} = \hat{\hat{P_{\pi}}} \hat{\hat{\mathcal{L}}} \hat{\hat{P_{\pi}}},
\end{equation}
where $\hat{\hat{P_{\pi}}}$ is the propagator for an ideal proton $\pi$-pulse:
\begin{equation}
\hat{\hat{P_{\pi}}} = \begin{pmatrix}
		1 & 0 & 0\\
    	0 & -1 & 0\\
        0 & 0 & -1
	\end{pmatrix}.
\end{equation}
When the evolution delay before and after the pulse are equal, the proton inversion pulse leads to the following average Liouvillian over the whole relaxation period:
\begin{equation}
\hat{\hat{\mathcal{L}}}_{av} = \begin{pmatrix}
		R_1^\mathrm{N} & 0 & 0\\
    	0 & R_1^\mathrm{H} & \delta_\mathrm{H}\\
        0 & \delta_\mathrm{H} & R_\mathrm{NH}
	\end{pmatrix}.
\end{equation}
Over this time period, the spin operator $\hat{\mathrm{N}}_z$ is an eigenvector of the relaxation matrix, and the time-evolution of its expectation value is given by:
\begin{equation}
\langle \hat{\mathrm{N}}_z \rangle (t) = e^{-R_1^\mathrm{N} t},
\end{equation}
which is the usual mono-exponential decay used for the analysis of relaxation rates measurements (note that the evolution towards an effective saturated state is obtained from the averaging of consecutive scans \cite{Levitt1994,Pelupessy_JCP_2007}). By constrast, an accurate analysis of relaxation properties in the abscence of radio-frequency pulses, or in field-varying experiments, requires the full relaxation matrix.

%% file: Chapters/2_Implementation.tex
\tocless\subsection{ \RedKite~in Mathematica}
\label{sec:RedKite}
\begin{figure*}
	\begin{center}
		\includegraphics[width=1.0\textwidth]{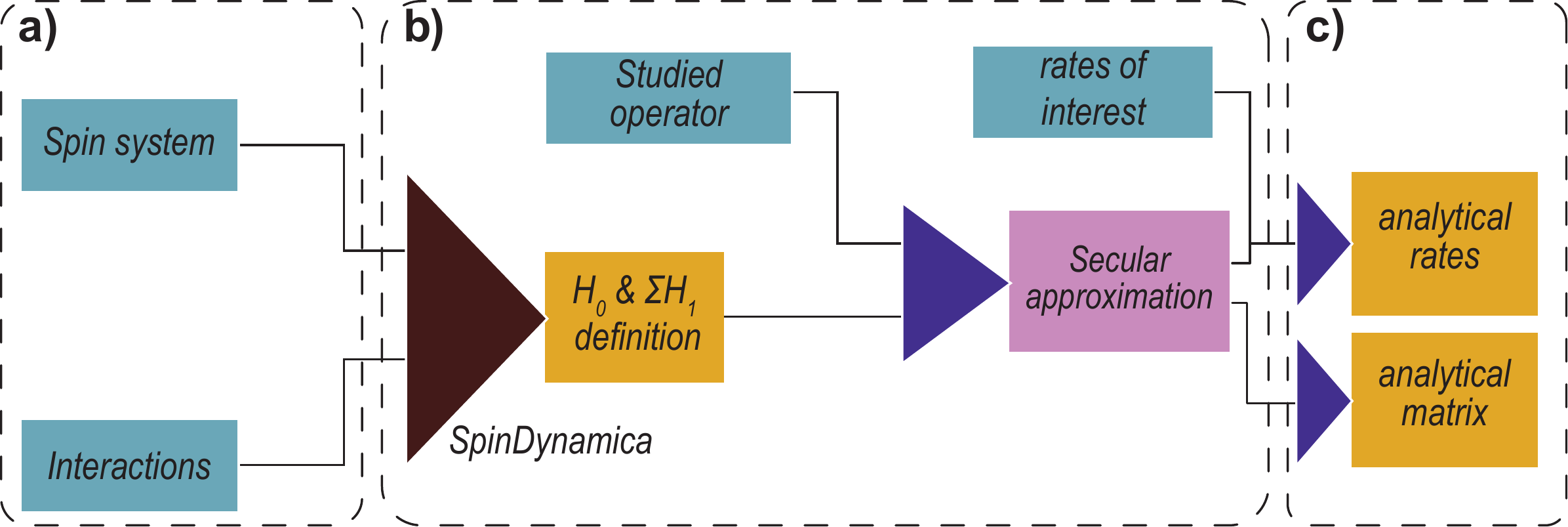}
	\end{center}
	\caption{Schematic representation of the \RedKite~calculation, describing input information and the output of the Mathematica notebook. \textbf{a)} Initial inputs from the user are the spin system (isotopes and geometry) and CSA and quadrupolar interactions. \textbf{b)} After definition of the operator basis, Hamiltonian operators are defined. After indicating the operator of the basis studied during the experiment, a reduction of the size of the basis is performed. Rates of interest are defined as well. \textbf{c)} Calculations produce analytical expressions for the relaxation rates and the relaxation matrix. Blue rectangles: user inputs. Yellow rectangles: calculated outputs. Pink rectangle and purple triangles: tasks performed by \RedKite.}  
	\label{fig:RedKitePipeline}
\end{figure*}
The computation of the relaxation rates is highly efficient with the formalism of the BRW engine \cite{Kuprov2006} which does not require an explicit expression of the Wigner matrices defining the correlation function (Eq.\,\ref{eq:Correlation}). Relaxation rates are first expressed as a function of the spectral density function $\mathcal{J}(\omega,\theta_i,\theta_j)$ where $\theta_k$ is the orientation of the interaction $k$ in the system frame (SF) of the chemical moiety. This frame corresponds to an arbitrary frame in which the orientation of the interactions are calculated. The different steps of \RedKite~are presented in the flowchart shown in Fig.\,\ref{fig:RedKitePipeline}. We will illustrate the use of \RedKite~on an isolated pair of spin-1/2 nuclei: a $^{15}$N-$^1$H pair. We have used \RedKite~to analyse HRR data recorded on $^{13}$C$^1$H$^2$H$_2$ specifically labelled isoleucine-$\delta1$ methyl groups of the protein Ubiquitin \cite{Cousin_JACS_2018}, and to study the relaxation properties of $^{13}$C$^1$H$_3$ methyl groups during a HZQC experiments \cite{BolikCoulon2019}.
\subsubsection{Definition of the spin system}
The first step is to define the spin system by specifying for each nuclear spin the nucleus type with its isotopic number, and a unique label for each spin which is used for identification. We present as an illustration the example of a simple spin system composed of an isolated $^{15}$N-$^{1}$H pair. The spin system is therefore defined as: 
\begin{center}
	Nuclei = \{\{"15N","NA"\}, \{"1H", "HA"\}\};
\end{center}
where "NA" and "HA" refer to the Nitrogen-15 and Proton respectively, before running the \textsc{SpinDynamica} \cite{Bengs_MagnResonChecm_2017} SetSpinSystem command:
\begin{center}
	SetSpinSystem[Table[\{Nuclei[[i, 2]], NuclearSpinQuantumNumber[Nuclei[[i, 1]]]\}, \{i, 1, Length[Nuclei]\}]];
\end{center}
The NuclearSpinQuantumNumber command is implemented in \textsc{SpinDynamica} \cite{Bengs_MagnResonChecm_2017} and defines the quantum spin number of the considered nucleus. \\
The geometry of the spin system is defined next. We define an array of size $n \times 3$ (where \textit{n} is the number of nuclei in the spin system, in our case 2) containing the position of each atom in a Cartesian axis system. In our example, we set the nitrogen nucleus at the origin of the axis system and the proton 1.02\,\AA~away from the nitrogen in the z-direction:
\begin{equation*}
	\mathrm{Coordinates} = \{\{0,0,0\}, \{0, 0, 1.02 \times 10^{-10} \}\};
    \label{eq:geometry}
\end{equation*}
To complete the definition of the spin system, the Chemical Shift Anisotropy (CSA) and quadrupolar properties have to be defined. The nuclei for which the CSA will be considered must be defined as such. In our example, we will only consider the nitrogen CSA:
\begin{equation*}
	\mathrm{CSAConsidered} = \{1,0\};
    \label{eq:csaconsidered}
\end{equation*}
It is possible to give a numerical value to the CSA or keep its value as an analytical parameter. We will consider this latter case here:
\begin{equation*}
	\delta_{csa} [1] = \Delta \sigma_N;
    \label{eq:csavalue}
\end{equation*}
Note that defining $\delta_{csa} [2]$ is not necessary since the proton CSA is neglected. Similarly, the strength of the quadrupolar interaction does not need to be defined (see in the Supplementary Materials for an example that includes quadrupolar interactions).\\
The orientations of the CSA tensor have to be given (either numerically or analytically). For the sake of simplicity, we choose an alignement along the N-H axis:
\begin{center}
vectorNum$^{\text{"CSA"}}_ 1$ = \{0, 0, 1\};
\end{center}
The index $1$ refers to the first spin in the spin system (\textit{i.e.} the nitrogen-15). There is also a possibility to consider asymmetric CSA tensors. In this case, the asymmetric CSA tensor is decomposed in two axially symmetric components. The longitudinal and orthogonal component of the CSA have to be defined using the variables names $\sigma \mathrm{long}[i]$ and $\sigma \mathrm{perp}[i]$ for the longitudinal and orthogonal values of the CSA tensors of isotope \textit{i}, and vectorNuml$^\text{"CSA"}_ i$ and vectorNump$^\text{"CSA"}_ i$ for the associated orientations. Table\,\ref{table:VariableExplain} contains the definitions of the different variables of \RedKite.
\subsubsection{Definition of spin tensors and Hamiltonian}
Three different types of interactions are considered in \RedKite: the dipolar couplings, the CSA (in the case where at least one spin has a CSA) and the quadrupolar couplings (in the case where spins with $m_s > 1/2$ are present in the spin system). Analytical forms of these Hamiltonian operators are calculated automatically. Other Hamiltonian operators can be defined and added if other interactions or effects are considered. \\
Calculation of Hamiltonian operators requires the definition of spin-tensor operators. \textsc{SpinDynamica} already contains their definition, but each tensor of coherence order-q is given as a linear combination of eigentensors \cite{Bengs_MagnResonChecm_2017}. Consequently, \textsc{SpinDynamica} tensors can be linear combinations of eigenvectors with different eigenfrequencies, which is an inappropriate basis to perform the secular approximation (based on the equality of eigenfrequencies of two eigenvectors). The secular approximation is better performed with complete separation of the tensor operators. The definition of each tensor has already been reported for each considered interactions (dipole-dipole, CSA and quadrupolar) \cite{Cavanagh_ProteinNMR_2007} and their definition in \textsc{Mathematica} can be found in Tables\,\ref{table:DDTensors}-\ref{table:QuadTensors}. In the case of non-equivalent homonuclear spin systems, performing the secular approximation is more complex, especially at low fields, where the oscillation frequency in Eq.\,\ref{eq:EvolutionFull} can be comparable to the relaxation rates. Numerical tools, such as \textsc{Spinach} \cite{Kuprov2006}, are available to study such systems. The Hamiltonian, as written in \textsc{RedKite}, can be found in the Supplementary Materials. \\
In the definition of the Hamiltonian, we introduce the function \textbf{M}, similarly to the BRW engine \cite{Kuprov2006}, which depends on the operator coherence order \textit{m} being considered, its associated eigenfrequency, a time \textit{t} at which the Hamiltonian is calculated, and the orientation of the interaction. The function \textbf{M} is useful when calculating the double commutators to obtain relaxation rates (as detailed in Section\,2.1). Products of the function \textbf{M} appear, which are simplified according to:
\begin{center}
M[l\_, f1\_, 0, i\_]Conjugate[M[k\_, f2\_, t\_, j\_]] := KroneckerDelta[l, k] KroneckerDelta[f1, f2] G[t, f1, i, j];
\end{center}
where $\mathrm{KroneckerDelta}[x, y] = 1$ if $x=y$ and 0 otherwise, \textit{l} and \textit{k} are associated to tensor coherence order, \textit{f1} and \textit{f2} to the tensor eigenfrequencies, \textit{t} the time at which the Hamiltonian is calculated, and \textit{i} and \textit{j} are the orientation of the interactions in the molecular frame. \textit{G[t, f1, i, j]} is the correlation function evaluated at time $t$ and is further replaced by the spectral density function evaluated at frequency $f1$. For auto-correlation, $i=j$, while cross-correlation is obtained when $i \neq j$.
\subsubsection{Operator of interest}
We define the operator of interest as the initial state where the polarization has been stored. In HRR, it is the longitudinal Zeeman term. In our case, we are interested in the nitrogen-15 longitudinal relaxation rates, which is defined by:
\begin{center}
OperatorOfInterest = opI["NA", "z"];
\end{center}
where opI is a \textsc{SpinDynamica} \cite{Bengs_MagnResonChecm_2017} command to define operators, here the N$_z$ operator.
\subsubsection{Analytical and numerical spin state restriction}
\begin{figure}
	\begin{center}
		\includegraphics[width=0.75\textwidth]{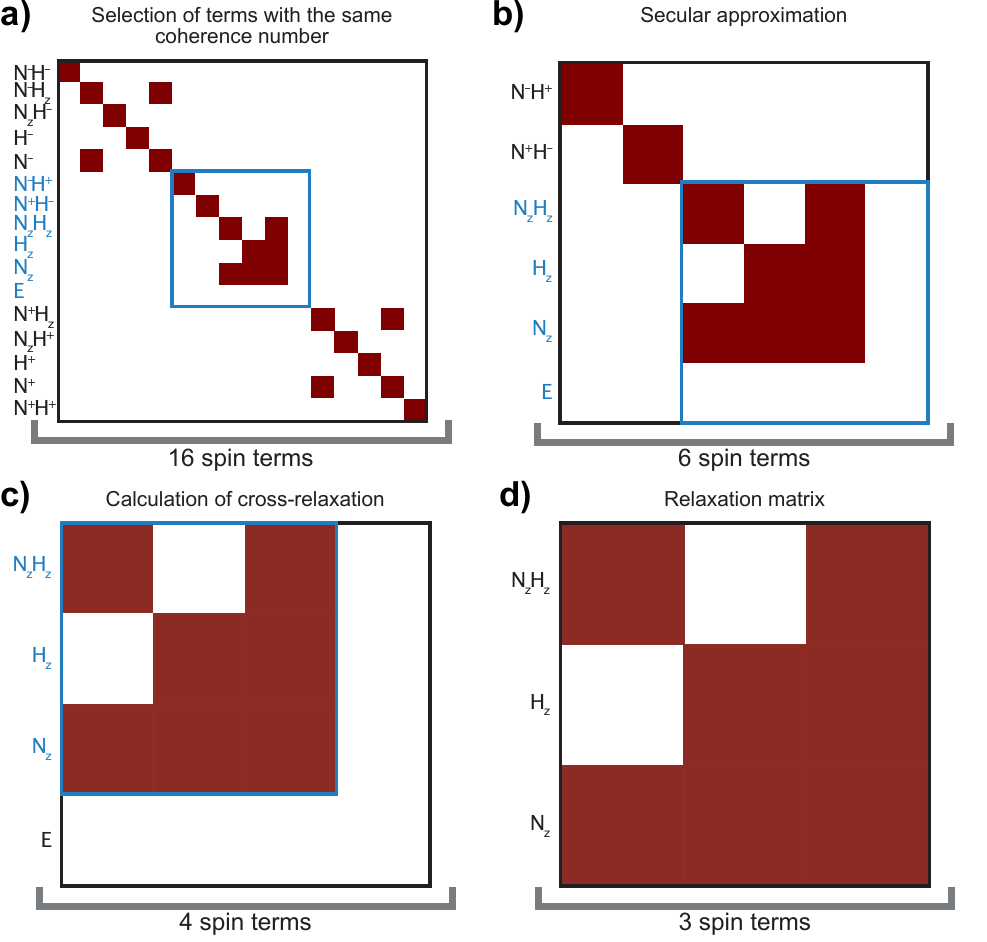}
	\end{center}
	\caption{Reduction of the matrix size for our case example of a $^{15}$N-$^1$H spin system. \textbf{a)} A $^{15}$N-$^1$H isolated spin pair has 16 operators in its basis. \textbf{b)} The first step of the matrix reduction size consists in keeping only terms that have the same coherence order as the spin-term of interest, leading to 6 terms in the basis. \textbf{c)} The secular approximation allows another level of size reduction: only terms that are secular with the Zeeman Hamiltonian are kept in the basis.  Two terms are removed at this stage. \textbf{d)} In the absence of cross-relaxation with the spin term of interest N$_z$, the identity operator is removed from the basis and the final basis contains 3 operators. In this graphical representation of the relaxation matrices, a red square indicates a non-zero value for the corresponding relaxation rate. The blue rectangles contain the selected part of the relaxation matrix after each steps of the size reduction. Normalization factors for the spin operators have been omitted for clarity.}
	\label{fig:ReducedBasis}
\end{figure}
The number of terms in the basis is equal to $4^n$ for \textit{n} spin-1/2 nuclear spins. Hence, in this two-spin system there are 16 terms, which is still a workable number. For more complex spin systems, reducing the size of the basis is essential. We only keep the terms contributing to the relaxation of the operator of interest following the scheme of Fig.\,\ref{fig:ReducedBasis}. First, only terms with the same coherence order as the operator of interest are selected (indicated in blue in Fig.\,\ref{fig:ReducedBasis}a). Then, the secular approximation removes all non-secular terms in the interaction frame (Fig.\,\ref{fig:ReducedBasis}b). Cross-relaxation rates with the operator of interest in this reduced basis are calculated (Fig.\,\ref{fig:ReducedBasis}c) and the operators with no cross-relaxation with the operator of interest are discarded from the basis (here this last step only removes the identity operator $E$, Fig.\,\ref{fig:ReducedBasis}d). This step is basis-dependent and some indirect cross-relaxation pathways affecting the operator of interest may be suppressed. An additional step can be applied for large spin systems to sort and select only major cross-relaxation pathways. In our example of an isolated $^{15}$N-$^1$H spin pair with a CSA on the nitrogen-15, only 3 terms remain in the basis:
\begin{equation*}
\mathrm{ReducedBasis} = \{\mathrm{NA_z}, \mathrm{HA_z}, \mathrm{2 NA_z HA_z}\};
\end{equation*}
\subsubsection{Calculations}
Once the basis has been defined, the relaxation matrix can be calculated:
\begin{equation*}
\mathrm{RM} = 
	\begin{pmatrix}
		R_1^\mathrm{N} & \sigma_\mathrm{NH} & \delta_\mathrm{N}\\
    	\sigma_\mathrm{NH} & R_1^\mathrm{H} & 0\\
        \delta_\mathrm{N} & 0 & R_\mathrm{NH}
	\end{pmatrix},
\label{eq:MatrixRelax}
\end{equation*}
where $R_1^\mathrm{N}$ and $R_1^\mathrm{H}$ refer to the nitrogen-15 and proton longitudinal relaxation rates respectively, $R_\mathrm{NH}$ to the auto-relaxation rate of the two-spin order, $\sigma_\mathrm{NH}$ to the dipole-dipole cross-relaxation rate between nitrogen-15 and proton and $\delta_\mathrm{N}$ to the CSA-(dipole-dipole) cross-relaxation rate due to the cross-correlation of the nitrogen-15 CSA and the dipole-dipole coupling:
\begin{eqnarray*}
R_1^\mathrm{N} & = & \frac{d_\mathrm{NH}^2 }{2} (\mathcal{J}(\omegaN-\omega_\mathrm{H}) + 6 \mathcal{J}(\omega_\mathrm{N}+\omega_\mathrm{H}) + 3 J(\omega_\mathrm{N}) ) + \frac{2\sigma_\mathrm{N}^2}{3} \Delta \sigma_\mathrm{N}^2  \omega_\mathrm{N}^2 \mathcal{J}(\omega_\mathrm{N}), \\
R_1^\mathrm{H} & = & \frac{d_\mathrm{NH}^2 }{2} (\mathcal{J}(\omega_\mathrm{N}-\omega_\mathrm{H}) + 6 \mathcal{J}(\omega_\mathrm{N}+\omega_\mathrm{H}) + 3 \mathcal{J}(\omega_\mathrm{H})), \\
R_\mathrm{NH} & = & \frac{3d_\mathrm{NH}^2}{2}  (\mathcal{J}(\omega_\mathrm{N}) + \mathcal{J}(\omega_\mathrm{H})) + \frac{2}{3} \Delta \sigma_\mathrm{N}^2 \omega_\mathrm{N}^2 \mathcal{J}(\omega_\mathrm{N}), \\
\sigma_\mathrm{NH} & = & \frac{d_\mathrm{NH}^2}{2}  (-\mathcal{J}(\omega_\mathrm{N}-\omega_\mathrm{H}) + 6 \mathcal{J}(\omega_\mathrm{N}+\omega_\mathrm{H})), \\
\delta_\mathrm{N} & = & 2 \Delta \sigma_\mathrm{N} \omega_\mathrm{N} d_\mathrm{NH} \mathcal{J}(\omega_\mathrm{N}),
\end{eqnarray*}
with $d_\mathrm{NH} =-\frac{\mu_0}{4\pi} \frac{\hbar \gamma_\mathrm{H} \gamma_\mathrm{N}}{r_\mathrm{NH}^3}$ the dipolar coefficient between the proton and the nitrogen-15, $r_\mathrm{NH}$ the distance separating the two nuclei, $\gamma_X$ the gyromagnetic ratio of nucleus $X$, $\hbar$ the Plank constant devided by $2\pi$, $\mu_0$ the permeability of free space, and $\Delta \sigma_\mathrm{N} = \sigma_\mathrm{zz} - \frac{\sigma_\mathrm{xx}+\sigma_\mathrm{yy}}{2}$ the CSA of the nitrogen-15 with $\sigma_{kk}$ the $k^{th}$ diagonal element of the chemical shift tensor. $\mathcal{J}$ is the spectral density function and is expressed as a function of the proton ($\omega_\mathrm{H}$) and nitrogen-15 ($\omega_\mathrm{N}$) Larmor frequencies.\\
All types of relaxation rates in this spin system can be calculated. In such a spin system, it is relatively easy to record longitudinal and transverse relaxation rates for the nitrogen-15 nucleus, as well as the cross-relaxation rate with the proton. These rates are calculated by:
\begin{flushleft}
		RatesOfInterest = \{ \\
		\{Rate[opI["NA", "z"], opI["NA", "z"]], "R1N"\}, \\
		\{Rate[opI["NA", "+"], opI["NA", "+"]], "R2N"\}, \\
		\{Rate[opI["NA", "z"], opI["HA", "z"]], "Sigma"\}\};
\end{flushleft}
where Rate is the implemented command to calculate relaxation rates as described in the previous section. This leads to the expression of transverse relaxation rate for nitrogen-15:
\begin{equation*}
	\begin{aligned}
R_2^\mathrm{N} =&\frac{d_\mathrm{NH}^2}{4} (\mathcal{J}(\omega_\mathrm{N}-\omega_\mathrm{H}) + 6 \mathcal{J}(\omega_\mathrm{N}+\omega_\mathrm{H}) +  3 \mathcal{J}(\omega_\mathrm{N}) + 6 \mathcal{J}(\omega_H) + 4 \mathcal{J}(0)) \\
& + \frac{\Delta \sigma_\mathrm{N}^2 \omega_\mathrm{N}^2}{9} (3 \mathcal{J}(\omega_\mathrm{N}) + 4 \mathcal{J}(0)).
	\end{aligned}
\end{equation*}
\subsubsection{Model selection and formating}
The user has to provide at least one definition of spectral density function in order to have a model for the dynamics of the system. In our case, we can use a model-free approach \cite{Lipari1982} with a correlation time for global tumbling $\tau_c$, one order parameter $S^2$ and an effective correlation time for internal motions $\tau_{int}$:
\begin{equation*}
	\mathcal{J}(\omega) = \frac{1}{5} \left( \frac{S^2 \tau_c}{1 + (\omega \tau_c)^2} + \frac{(1-S^2) \tau_{int}'}{1 + (\omega \tau_{int}')^2} \right),
    \label{eq:MFNH}
\end{equation*}
where $\tau_{int}'^{-1} = \tau_c^{-1} + \tau_{int}^{-1}$. This function is implemented in \RedKite~as:
\begin{equation*}
	\begin{aligned}
	\mathrm{JNH}[\omega\_ , i\_ , j\_] :=& \mathrm{Module}[\{\mathrm{spec}, \tau 1 \},  \\
    \tau 1 =& \tau_c  \tau_i / (\tau_c + \tau_i);  \\
    \mathrm{spec} =& \frac{1}{5} \left(S2 \frac{\tau_c}{1 + (\omega \tau_c)^2} + \right. \\
		 & \left. (1-S2) \frac{\tau 1}{1+(\omega \tau 1)^2} \right)]
	\end{aligned}
    \label{eq:MFNH_Matematica}
\end{equation*}
At this point, the relaxation rates seen above can be expressed as a function of the parameters of the dynamics of the system (order parameter and correlation times). Numerical calculations can be performed if values for the parameters of the spectral density function are provided.
\subsubsection{Preparing for ICARUS}
In order to use the results obtained in \RedKite~for the analysis of HRR, symbolic expressions have to be exported. Exporting to ICARUS requires that all variables have Latin-only characters as the interpretation of non-Latin characters is not implemented in ICARUS. During the export process, the spectral density function is provided by the user as:
\begin{equation*}
\mathrm{JofInterest} = \mathrm{JNH};
\end{equation*}
The user can export the first derivatives of the relaxation rates with respect to all the variables (magnetic field excluded as it is not useful in the following analysis). All the expressions of the relaxation matrix and the relaxation rates (and first derivatives if required) are saved in separate files named respectively RelaxationMatrix.txt for the entire relaxation matrix, \textit{Rate}.txt for the relaxation rates defined in the RatesOfInterest array, and \textit{Rate}deriv\_\textit{Variable}.txt where \textit{Rate} refers to the considered relaxation rate and \textit{Variable} to the variable name by which the rate is derivated. The first derivatives of the relaxation rates can be used in minimization procedures. An additional file named PositionOfInterest.txt is also created and contains the position of the operator of interest in the relaxation matrix ($\hat{\mathrm{N}}_z$ in our case example).

\tocless\subsection{ICARUS implementation}
\label{ICARUS}
In this paper, we show as an example how \RedKite~can be used for the analysis of HRR experiments. Other applications of \RedKite~have been published elsewhere \cite{Novakovic_jmr_2018, BolikCoulon2019}, and can be envisioned, as relaxation rates can be obtained for any spin system. We detail here the analysis of HRR relaxation rates.
\subsubsection{Accurate estimation of relaxation rates from high-resolution relaxometry measurements}
\label{SimulateShuttle}
\begin{figure}
	\begin{center}
		\includegraphics[width=0.75\textwidth]{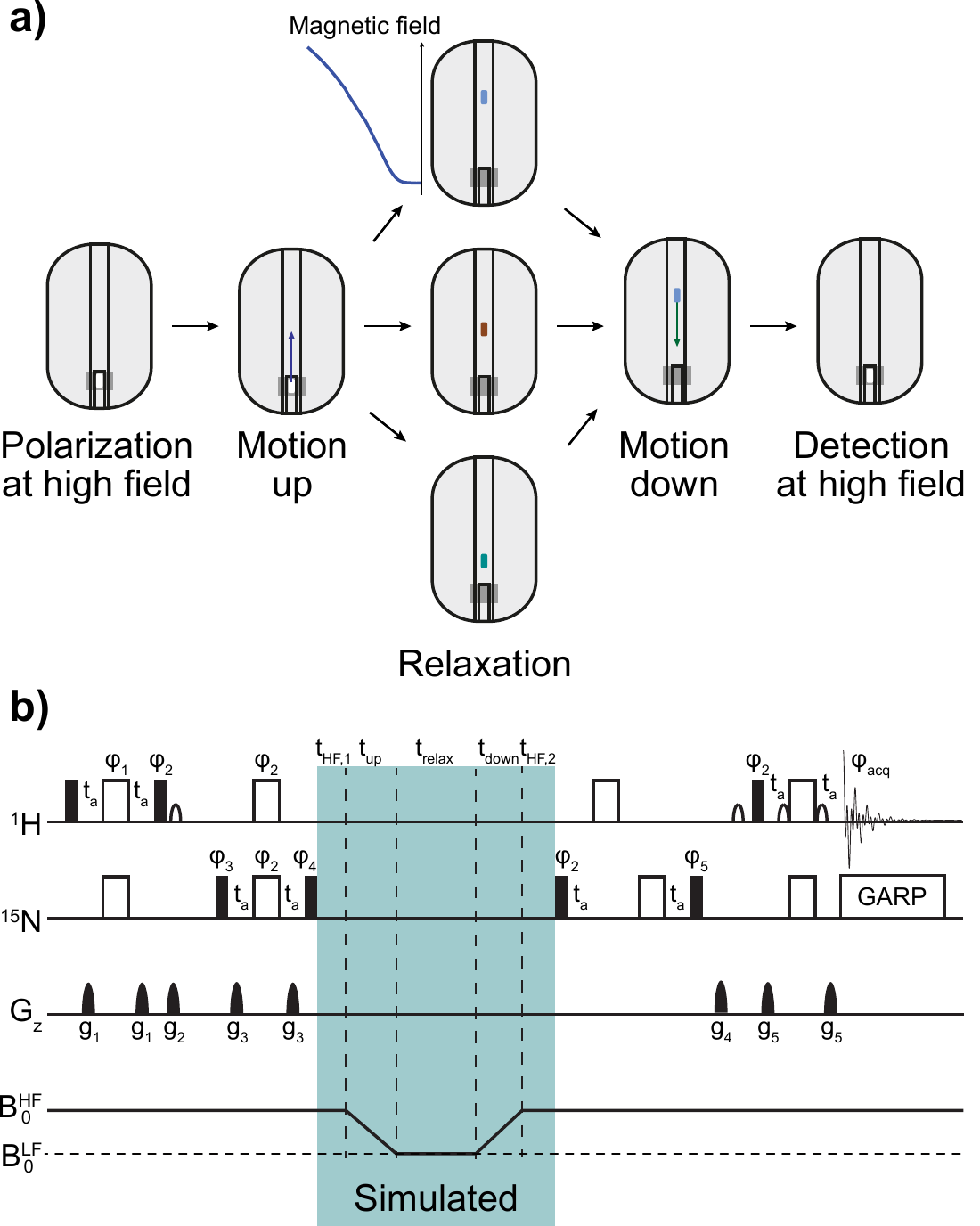}
	\end{center}
	\caption{Description of an HRR scheme. \textbf{a)} The position of the sample is changed during the course of the experiment. It is first polarized at high field, and transfered to a chosen position in the stray field of the superconducting magnet, characterized by a lower magnetic field, for relaxation. The sample is then moved back to the high-field position for detection. Pannel adapted from \cite{Cousin_JACS_2018}. \textbf{b)} A typical pulse sequence used to record HRR experiment. During the analysis of HRR rates, the highlighted part of the pulse sequence (blue) is simulated.  Black narrow (respectively wide empty) rectangles represent $\pi/2$-pulses (respectively $\pi$-pulses). Pulses are applied along the x-axis if not otherwise stated (by the $\varphi_i$). The amplitude of pulse field gradients are labeled $g_i$. Additional experimental details can be found in \cite{Cousin2018}.}
	\label{fig:PulseSequence}
\end{figure}
\begin{figure}
	\begin{center}
	\includegraphics[width=0.7\textwidth]{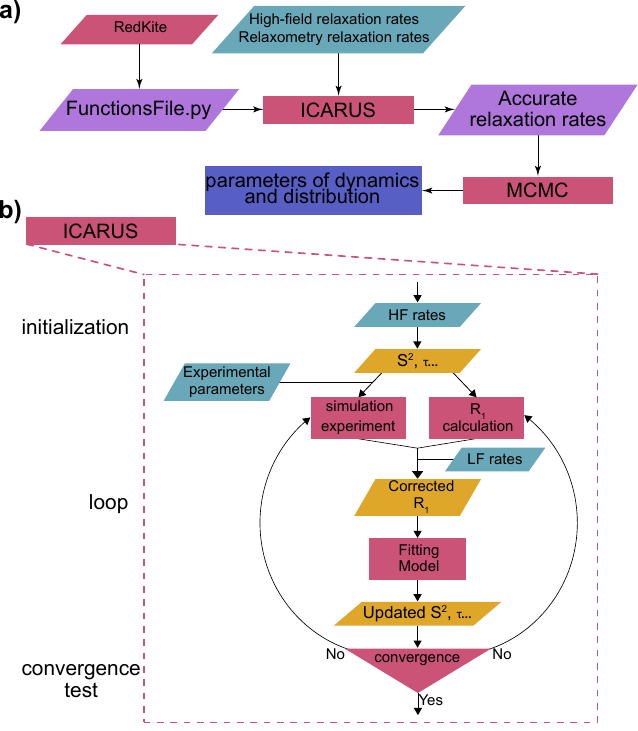}
	\end{center}
	\caption{Flow chart for the analysis of high-resolution relaxometry data with ICARUS. \textbf{a)} After a FunctionsFile has been obtained from \RedKite, ICARUS can be run, using, among other inputs, relaxation rates recorded on standard high-field spectrometers and the high-resolution relaxometry data. Accurate relaxometry relaxation rates are obtained, and a Markov-Chain Monte-Carlo (MCMC) analysis of these corrected rates and high-field relaxation rates leads to values of parameters describing the dynamics of the system and their distribution. \textbf{b)} Flow chart of the ICARUS procedure. Accurate high-field (HF) relaxation rates are used to obtain an initial set of parameters for the dynamics of the system. These parameters are used to simulate the high-resolution relaxometry experiments (using the same experimental set up, \textit{i.e.} shuttling time, delays, magnetic fields) from which biased simulated R$_1$ are extracted, and also to calculate the accurate expected R$_1$. The ratios of these two calculated rates are called correction factors. The product of experimental decay rates and correction factors are corrected experimental low field (LF) relaxometry relaxation rates. Together with the high-field relaxation data, the corrected rates are used to determine a new set of parameters of dynamics, further used in the next correction iteration. Convergence is not evaluated within ICARUS and the number of iterations remains a choice of the user. However, we recommend to verify the convergence of the correction factors, as these ones are essential in the determination of the final parameters of the dynamics. Typically three or four iterations are sufficient.
    }
	\label{fig:ICARUSPipeline2}
\end{figure}

High-resolution relaxometry can be used to obtain a precise description of the dynamics of spin systems over orders of magnitude of timescales \cite{Charlier2013, Cousin_JACS_2018, Cousin2018}. The analysis is based on the measurement of longitudinal relaxation rates over a broad range of magnetic fields (typically from a few tenths of Tesla up to about 20\,T). A reliable description of the motions requires accurate estimates of the relaxation rates. \\
During each high-resolution relaxometry experiment, the sample is transferred outside of the magnetic center to a defined position $z_{relax}$ in the stray field above the magnet (Fig.\,\ref{fig:PulseSequence}). During the two transfers (from high to low field, and back) and the relaxation delay, all relaxation pathways are active. In contrast to the example presented in Section\,2.2, measured polarization decays can be affected by cross-relaxation and therefore cannot be used as is to determine longitudinal relaxation rates accurately (this is true for any relaxation experiment where pulses can not be applied during the relaxation period). Doing so would lead to systematic deviations in the parameters used to describe the dynamics of the system. Simulating the experiment including the time when the sample is outside the superconducting coil allows one to take into account cross-relaxation pathways and to estimate accurate relaxation rates. The complete relaxation period in a high-resolution relaxometry experiment includes three delays at constant fields and two transfers through a strong gradient of magnetic field. \\
The simulation of the experiment is performed by calculating the propagator during the highlighted part of the pulse sequence in Fig.\,\ref{fig:PulseSequence}b. For convenience, it is written as a product of individual propagators:
\begin{equation}
	\begin{aligned}
    \hat{\hat{\mathcal{P}}}_{tot}(t_{\mathrm{HF,1}},t_\mathrm{up},t_\mathrm{relax},t_\mathrm{down},t_{\mathrm{HF,2}}) =&\hat{\hat{ \mathcal{P}}}^\mathrm{HF, 2}(t_{\mathrm{HF,2}}) \cdot  \Pdown(t_\mathrm{down}) \cdot \hat{\hat{\mathcal{P}}}^\mathrm{LF}(t_\mathrm{relax}) \cdot  \\
&      \Pup(t_\mathrm{up}) \cdot \hat{\hat{\mathcal{P}}}^\mathrm{HF, 1}(t_{\mathrm{HF,1}}),
	\end{aligned}
\end{equation}
where $\hat{\hat{\mathcal{P}}}^\mathrm{HF, 1}$ and $\hat{\hat{\mathcal{P}}}^\mathrm{HF, 2}$ are the propagators calculated at high field, respectively before and after shuttling, $\hat{\hat{\mathcal{P}}}^\mathrm{LF}$ is the propagator calculated at the low field position and $\Pup$ (respectively $\Pdown$) is the propagator calculated during the motion up (respectively down) from the high-field to the low-field position (respectively from the low-field to the high-field position). This decomposition allows the calculation of the segmental propagators using either Eq.\,\ref{eq:PropagatorGeneral} (for constant Liouvillian superoporators) or Eq.\,\ref{eq:decomposeProp} (for time-dependent Liouvillian superoporators). The propagators for constant-field positions (\textit{i.e.} $\hat{\hat{\mathcal{P}}}^\mathrm{HF, 1}$, $\hat{\hat{\mathcal{P}}}^\mathrm{LF}$  and $\hat{\hat{\mathcal{P}}}^\mathrm{HF, 2}$) are calculated using Eq.\,\ref{eq:PropagatorGeneral} and the relaxation matrix calculated at high field ($\hat{\hat{\mathcal{R}}}_\mathrm{HF}$) and low field ($\hat{\hat{\mathcal{R}}}_\mathrm{LF}$):
\begin{equation}
\begin{aligned}
	\hat{\hat{\mathcal{P}}}^\mathrm{HF,i} (t_\mathrm{HF,i}) &= e^{-t_\mathrm{HF,i} \hat{\hat{\mathcal{R}}}_\mathrm{HF}}, \\
    \hat{\hat{\mathcal{P}}}^\mathrm{LF} (t_\mathrm{relax}) &= e^{-t_\mathrm{relax} \hat{\hat{\mathcal{R}}}_\mathrm{LF}}.
\end{aligned}
\end{equation}
The simulation of the transfers through the magnetic field gradient is performed by subdividing the experiment into intervals of few milli-seconds $\delta t$ that still fulfill the conditions of Redfield theory. In order to stay in the Redfield hypothesis, $\delta t$ must be large compared to the correlation time of the system to extend the integration to infinity in Eq.\,\ref{eq:main1}. In addition, $\delta t$ must be sufficiently small in order to perform a discretization of the integral over the full sample trajectory. In the case of high-resolution relaxometry with a sample traveling at $\approx$10\,m.s$^{-1}$ over at most 1\,m, we considered a $\delta t$  of 1\,ms, which corresponds, at most, to a change of about 10\,\% of the magnetic field between two consecutive steps. The propagators $\mathrm{d}\hat{\hat{\mathcal{P}}}(\delta t, z(t))$ for these small steps are obtained following Eq.\,\ref{eq:PropagatorGeneral}:   
\begin{equation}
\begin{aligned}
	\mathrm{d}\hat{\hat{\mathcal{P}}}( \delta t, z(t)) &= e^{-\delta t \hat{\hat{\mathcal{R}}}(z(t))},
	\label{eq:InfiniteProp}
\end{aligned}
\end{equation}
where $\hat{\hat{\mathcal{R}}}(z(t))$ is the relaxation matrix evaluated at the position $z(t)$ along the bore of the magnet  and characterized by its magnetic field (note: the field profile can be mapped using a gaussmeter). The experimental field profile is fitted to a polynomial expansion in ICARUS. Each propagator $\mathrm{d}\hat{\hat{\mathcal{P}}}( \delta t, z(t))$ is field dependent due to the field dependence of the relaxation matrix. The propagator for the motions up to and down from the position $z_{relax}$ are defined as the products of the infinitesimal propagators $\mathrm{d}\hat{\hat{\mathcal{P}}}$:
\begin{equation}
\begin{aligned}
	\Pup &= \prod_{n=0} ^{n_\mathrm{max}^\mathrm{up}} \mathrm{d} \Pup( \delta t, (z(n \times \delta t))), \\
    \Pdown &= \prod_{n=0} ^{n_\mathrm{max}^\mathrm{down}} \mathrm{d} \Pdown ( \delta t, (z(n \times \delta t))),
\end{aligned}
\end{equation}
where $n_{max}^\mathrm{up}$ (respectively $n_{max}^\mathrm{down}$) is defined by $t_\mathrm{transfer}^\mathrm{up} = n_{max} \times \delta t$ (respectively $t_\mathrm{transfer}^\mathrm{down} = n_{max}^\mathrm{down} \times \delta t$) with $t_\mathrm{transfer}^\mathrm{up}$ (respectively $t_\mathrm{transfer}^\mathrm{down}$) the delay of transfer to the top (respectively down) position. In these calculations, the relaxation matrix is derived using the analytical expression obtained from \RedKite, a model of motions and a set of parameters of dynamics. \\
The expectation value for the operator of interest at the end of the full relaxation period (delays at high field and low field as well as the two transfers in between) can then be extracted from the calculated propagator for each relaxation delay. The simulated decay as a function of the relaxation time is fitted with a mono-exponential decay function with an effective longitudinal relaxation rate $R_{sim}$ (Table\,\ref{table:RateNomenclature} sums up our nomenclature for the different calculated and measured relaxometry relaxation rates). All relaxation pathways are active during the transfers between high and low-field positions. The initial density operator is partially projected onto the eigenvectors of the relaxation matrix (relaxation modes) of lowest eigenvalues. Thus, the simulated decay rate $R_{sim}$ is {\it a priori} lower than the pure longitudinal relaxation rate $R_{calc}$ calculated using the parameters of dynamics. We define the correction factor for each  relaxometry experiment as the ratio between these two rates for an experiment $j$ (corresponding to a specific low field $B_\mathrm{LF}^{(j)}$) and a residue $i$:
\begin{equation}
	\mathcal{C}(\mathcal{E}_j, B_\mathrm{LF}^{(j)}, \mathcal{D}_i) = \frac{R_{calc}(B_\mathrm{LF}^{(j)}, \mathcal{D}_i)}{R_{sim}(\mathcal{E}_j, B_\mathrm{LF}^{(j)}, \mathcal{D}_i)},
\end{equation}
where $\mathcal{E}_j$ are the experimental parameters (shuttling times and relaxation delays), and $\mathcal{D}_i$ are the parameters of dynamics. The correction factor is applied to each corresponding measured relaxometry data $R_{meas}(\mathcal{E}_j, B_\mathrm{LF}^{(j)})$:
\begin{equation}
R_{corr} (\mathcal{E}_j, B_\mathrm{LF}^{(j)}) = \mathcal{C}(\mathcal{E}_j, B_\mathrm{LF}^{(j)}, \mathcal{D}_i) \times R_{meas}(\mathcal{E}_j, B_\mathrm{LF}^{(j)}).
\end{equation}
The correction is performed iteratively (Fig.\,\ref{fig:ICARUSPipeline2}). The set of parameters $\mathcal{D}_i$ for the first iteration is obtained from the analysis of the accurate relaxation rates, \textit{i.e} measured with the use of pulses, typically on high-field magnets. Then corrected relaxometry relaxation rates are analyzed alongside high-field relaxation rates. A new set of parameters of dynamics is extracted from this ensemble of relaxation rates. In the next iteration, these parameters of dynamics are used to simulate the experiment and compute improved corrections of experimental rates to estimate the accurate low-field relaxation rates. This is repeated until the correction factors converge. The final set of high-field and corrected relaxometry relaxation rates can then be used to extract the distribution of the parameters of local motions in a Markov-Chain Monte-Carlo (MCMC) procedure and thus evaluate the median value and uncertainty of these parameters (see below).
\begin{table*}
	\def\arraystretch{1.5}
	\caption{Nomenclature for the relaxometry relaxation rate labels and parameters determining their values. $\{\mathcal{E}_j\}$ are the experimental parameters for experiment $j$, $B_\mathrm{LF}^{(j)}$ is the low field chosen for relaxation and $\mathcal{D}_i$ are the parameters of the spectral density function used to describe the dynamics of residue $i$.}
		\begin{center}
		\begin{tabular}{ccc}
			\hline
			Label & Parameters & Description \\
			\hline
			\hline
		 	$R_{sim}$ & $\mathcal{E}_j$, $B_\mathrm{LF}^{(j)}$ and $\mathcal{D}_i$ & Relaxation rate extrated from the fitting \\
				& & of the simulated polarization decay \\
			$R_{calc}$ & $B_\mathrm{LF}^{(j)}$ and $\mathcal{D}_i$ & Relaxation rate calculated from \\
				& & the parameters of dynamics \\
			$R_{meas}$ & $\mathcal{E}_j$ and $B_\mathrm{LF}^{(j)}$ & Measured relaxation decay rate \\
			$R_{corr}$ & $\mathcal{E}_j$, $B_\mathrm{LF}^{(j)}$ and $\mathcal{D}_i$ & Corrected relaxation decay rate \\
			\hline
		\end{tabular}
		\end{center}
	\label{table:RateNomenclature}
\end{table*}
\subsubsection{Compiling expressions in the FunctionsFile.py script}
 Information about the relaxation properties of the spin system are contained in an independent script called FunctionsFile with expressions of the relaxation rates (and their derivatives if required) and the relaxation matrix in the considered basis. The FunctionsFile can be edited and adapted to the spin system under investigation. \RedKite~outputs first need to be converted from \textsc{Mathematica} to \textsc{Python} format and compiled in this FunctionsFile.py script. \\
This task is performed by the RedKite2ICARUS.py program. Briefly, it takes as input all the output files from \RedKite~(Section\,3.1) and asks for variables names (the ones that have to be fitted, usually parameters defining the spectral density function) and the ones that characterize the system and are not fitted (e.g., CSA tensors). It is also possible to set the CSA as a fitted variable. In the case where the overall diffusion frame is asymmetric, ICARUS requires a file containing the orientations of internuclear vectors in the anisotropic diffusion frame. Creating such a file has been implemented in RedKite2ICARUS.
\subsubsection{Fitting parameters of the model of motion to relaxation rates}
The program ICARUS (Iterative Correction for the Analysis of Relaxation Under Shuttling) \cite{Charlier2013, Cousin2018} has been entirely written in \textsc{Python} (version 3.5). The detailed description on how to use ICARUS has been already published elsewhere \cite{Cousin2018}. The key parts of the code are the fitting of parameters of a user-defined model of motion using accurate (generally high field) relaxation rates and corrected relaxometry rates as experimental constraints, as well as the simulation of the experiments (as detailed in Section\,3.2.1). Fitting the parameters of the model relies on the \textit{basin-hopping} function implemented in the \textit{scipy.optimize} \textsc{Python} library with the L-BFGS-B method \cite{Wales1997} for $\chi^2$ minimization:
\begin{equation}
\chi^2 = \sum_i \frac{(R_\mathrm{model,i} - R_\mathrm{exp,i})^2}{\sigma_\mathrm{exp,i}^2},
\label{eq:chi2}
\end{equation}
where $R_\mathrm{model,i}$ are the calculated relaxation rates and $R_\mathrm{exp,i}$ are the measured relaxation rates with experimental error $\sigma_\mathrm{exp,i}$. \\
Bounds of the dynamics parameters are provided by the user in the GUI. The \textit{basin-hopping} function allows the use of first derivatives of the relaxation rates in the fitting (provided in the FunctionsFile as explained above), usually leading to faster minimization. An additional minimization based on a grid search has been implemented in order to avoid local minimum traps. This step is time-consuming and optional. \\
The core of the code does not contain information about a particular spin system nor experimental set up, such that the usage of ICARUS can be extended to any situation (spin system or model of motion). Data and experimental set up are loaded as separate text files using the GUI, and all analytical expressions of relaxation rates are contained in the independent FunctionsFile script.
\subsubsection{ICARUS output and MCMC}
\begin{figure*}
\begin{center}
	\includegraphics[width=1\textwidth]{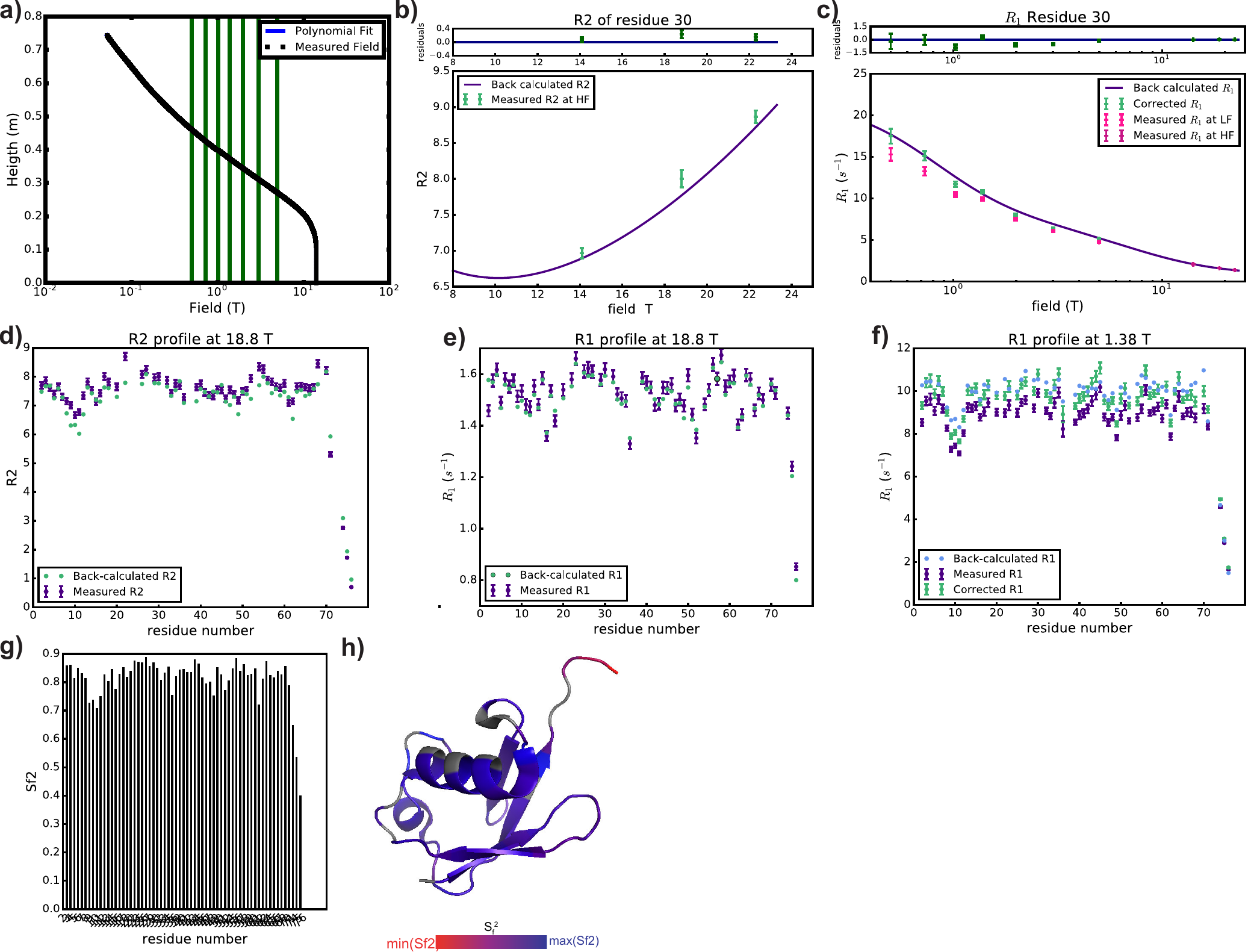}
\end{center}
	\caption{Outputs created by ICARUS for the study of motions of amide backbone $^{15}$N of Ubiquitin. \textbf{a)} Fit of the magnetic field in the spectrometer. The vertical green lines show the magnetic fields at which relaxometry measurements were performed. Checking the quality of this fit is important in order to make sure magnetic fields will be calculated correctly for each position of the sample during its trajectories. \textbf{b)} and \textbf{c)} Fit of the nitrogen-15 transverse and longitudinal relaxation rates for the residue Ile-30. \textbf{d)} Transverse and \textbf{e)} longitudinal nitrogen-15 relaxation rates measured at 18.8\,T. Measured and calculated relaxation rates using the final fitted parameters are shown in purple and green respectively. \textbf{f)} Profile of the longitudinal relaxometry nitrogen-15 relaxation rate at $\approx$1.38\,T. Measured, corrected and calculated rates using the final set of fitted parameters are shown in purple, light green and dark green respectively. \textbf{g)} Evolution of the order parameter  $S_f^2$ throughout the sequence (residues for which no data are provided are not displayed in this bar plot). \textbf{h)} Color-coding of the Ubiquitin structure (PDB ID: 1D3Z) according to the final fitted values of the order parameter $S_f^2$. Residues for which no data are provided are shown in grey.}
	\label{fig:ICARUSoutput}
\end{figure*}
The outputs of ICARUS have already been described \cite{Cousin2018}. Fig.\,\ref{fig:ICARUSoutput} shows selected figures created by ICARUS. Briefly, output figures consist of the fit of the stray field gradient (Fig.\,\ref{fig:ICARUSoutput}a), profiles of the relaxation rates (accurate, calculated, and corrected in the case of relaxometry data) at each field throughout the protein sequence (Fig.\,\ref{fig:ICARUSoutput}b,\,c,\,d), fits of all the relaxation rates for each residue (Fig.\,\ref{fig:ICARUSoutput}e,\,f), bar plots of fitted parameters (Fig.\,\ref{fig:ICARUSoutput}g). Several text files are created which contain corrected relaxometry relaxation rates, the set of fitted parameters after the fit of the accurate relaxation rates only, and of the whole data set (accurate and corrected relaxometry data) as well as the correction factors after each iteration of ICARUS. Finally, scripts are also created. One allows the user to calculate all the defined relaxation rates and the relaxation matrix using the final fitted parameters with the use of a GUI where the magnetic field and residue number of interest have to be set. The other scripts are created only if a PDB ID for the protein of interest has been provided in the GUI, and are meant to be run in \textsc{PyMOL} in order to color the structure according to the final set of fitted parameters (order parameters, correlation times, etc...) and the final $\chi^2$ (Eq.\,\ref{eq:chi2}) to facilitate the visualisation of the results over the protein structure (one file is created for each of these parameters). An example is shown in Fig.\,\ref{fig:ICARUSoutput}h.\\ 
In order to provide a better analysis of the dynamics, a Markov-Chain Monte-Carlo (MCMC) analysis of accurate and corrected relaxometry relaxation rates should be performed. We have written a script that directly reads ICARUS output folders to perform an MCMC using the \textit{emcee} \textsc{Python} library \cite{Foreman-Mackey2013}. The MCMC analysis provides a better error evaluation of the parameters of dynamics as well as potential correlations between them. A README file explaining how to use the MCMC program is provided with the script.\\
Overall, the \RedKite-ICARUS suite is intended to allow for an efficient (a complete analysis of $^{15}$N relaxometry data on Ubiquitin can be obtained within two hours on a standard laptop computer) and highly flexible (it can be extended to broad range of spin systems, with all types of model of motions and for most commonly measured relaxation rates) analysis of high-resolution relaxometry data. \RedKite~and RedKite2ICARUS create the scaffold (FunctionsFile) that is used by ICARUS. The use of ICARUS is convenient with a simple graphical user interface. After the correction of the relaxometry relaxation rate, a final Markov-Chain Monte-Carlo analysis is performed by a script that reads directly ICARUS output folders (Fig.\,\ref{fig:ICARUSPipeline2}a).

%% file: Chapters/3a_MethylGroup_Theory.tex
Motions of protein side-chains are important for their function. These motions have been investigated thanks to NMR methodological development and selective labeling strategies based on the clever use of metabolic pathways \cite{Tugarinov2006, Clore2012, Mas2013}. The averaging of the dipolar interactions arising from their fast rotation confers favourable relaxation properties to methyl groups. They make good candidates for the study of side-chain motions, in particular in the hydrophobic core of proteins where they constitute an entropy reservoir \cite{DuBay2015, Frederick2007}, or at protein-protein and protein-ligand binding interfaces where their motions can allow a re-modeling for a better complementary interaction with the binding partner. In this context, we have recently performed a detailed analysis of the motions of isoleucine-$\delta_1$ methyl-group on the selectively labeled protein \ubiquitin \,with the use of HRR and relaxation rates recorded using conventional high-field magnets \cite{Cousin_JACS_2018}. In this section, the combined \RedKite~and ICARUS analysis of HRR in \ubiquitin~is presented. 

\tocless\subsection{Theoretical framework for the dynamics of methyl group}
\subsubsection{Model of correlation function}
Different models of correlation function for a wide variety of molecular systems have been suggested in the past \cite{Lipari1982, Clore1990, Meirovitch2006, Meirovitch2007, Charlier2016, Calandrini2010, Khan2015, Hsu_Biophys_2018}. In our analysis of high field and relaxometry relaxation rates on \CDDH-methyl group of Ubiquitin, the data recorded at low fields (lower than 5\,T) allowed a better characterization of the complexity of motions that can occur in a methyl-bearing side-chain, in particular $\chi_1$/$\chi_2$ rotameric transitions in isoleucine residues on nanosecond timescales \cite{Cousin_JACS_2018}. The analysis was based on the Extended Model Free (EMF) description of the CC bond motions. Assuming (i) isotropic tumbling of the protein characterized by a correlation time $\tau_c$, (ii) EMF for CC bonds motions, (iii) perfect tetrahedral symmetry for the methyl group with a characteristic correlation time for the methyl group rotation $\tau_{met}$ associated to an order parameter $S_{met}^2 (\theta_{i,j})$ \cite{Frueh2002} and (iv) statistical independence between methyl group rotation, motions of the methyl group axis and overall rotational diffusion, the correlation function can be modeled by:
\begin{equation}
 C^{\mathrm{met}}_{i,j}(t)=C_g(t) C_\mathrm{axis}(t) C_\mathrm{rot}^{i,j}(t),
\end{equation}
where:
\begin{equation}
	\begin{aligned}
& C_g(t)= e^{-t/\tau_c},  \\
& C_\mathrm{axis}(t)  = S^2 + (1-\SfD)e^{-t/\tau_f} +\SfD(1-\SsD)e^{-t/\tau_s},  \\
& C_\mathrm{rot}^{i,j}(t) =  S_{met}^2(\theta_{i,j})+ \left(\mathcal{P}_2(\cos\theta_{i,j}) - S_{met}^2(\theta_{i,j})  \right)e^{-t/\tau_{met}},  
	\end{aligned}
\label{eq:MethylCorrFunc}
\end{equation}
with $S_{met}^2(\theta_{i,j}) =\mathcal{P}_2(\cos\theta_i) \mathcal{P}_2(\cos \theta_j) $ and $\mathcal{P}_2$ is the second order Legendre polynomial function, $\mathcal{P}_2(x)=(3x^2-1)/2 $, $\theta_k$ is the angle between the principal axis of an axially symmetric interaction $\bm{k}$ vector and the CC-axis (methyl group symmetry axis) and $\theta_{i,j}$ the angle between the principal axes of two (possibly identical) axially symmetric interactions $\bm{i}$ and $\bm{j}$. The order parameters $\SfD$ and $\SsD$ characterize motions of the system frame and are associated with the correlation times $\tau_f$ and $\tau_s$, respectively. The overall order parameter is defined as $S^2=\SfD\SsD$. The value of the angles $\theta_k$ and $\theta_{i,j}$ is constrained by the geometry of the spin system. The corresponding spectral density function is:
\begin{equation}
\begin{aligned}
	\mathcal{J}_{i,j} (\omega) = \frac{1}{5}& \left[ S_{met}^2(\theta_{i,j}) \left( \SfD \SsD \Lorentz{\tau_c} + (1-\SfD) \Lorentz{\tau'_f} + \right. \right. \\
				&  \left. \left. \SfD (1-\SsD)\Lorentz{\tau'_s} \right) + (\mathcal{P}_2\cos(\thIJ)-S_{met}^2(\theta_{i,j})) \times \right. \\
    & \left.  \left( \SfD \SsD \Lorentz{\tau'_{met}} + (1-\SfD) \Lorentz{\tau''_f}+\right.  \right. \\
	& \left. \left. \SfD(1-\SsD) \Lorentz{\tau''_s}    \right) \right],
\end{aligned}
	\label{eq:SpecMethyl}
\end{equation}
where  ${\tau'_a}^{-1} = \tau_a^{-1} + \tau_c^{-1}$ and ${\tau''_a}^{-1} = \tau_a^{-1} + \tau_c^{-1} + \tau_{met}^{-1}$. \\
In the following, $\mathcal{J}_\mathrm{AB}$ will be used to denote the dipole-dipole auto-correlation between nuclei A and B, $\mathcal{J}_{A}$ for the CSA auto-correlation of nucleus A, $\mathcal{J}_\mathrm{AB, CD}$ for the dipole-dipole/dipole-dipole cross-correlation between the spin pairs AB and CD, $\mathcal{J}_\mathrm{A, BC}$ for the cross-correlation between the CSA of nucleus A and the dipole-dipole interaction between nuclei B and C. Finally, the index $\mathcal{Q}$ will be used to denote the quadrupolar interactions. These notations follow conventions proposed by Werbelow and Grant \cite{Werbelow1975}. \\
As detailed bellow, in our treatment of the relaxometry data, the effects of the surrounding deuterium nuclei arising from the labelling of the protein have to be considered. These have been taken into account by adding a single additional deuterium nucleus in the spin system. For simplicity, while we consider the additional dipolar contributions to relaxation rates of the \CDDH~spin system, we do not include this additional nucleus in our basis. We approximated the spectral density function for the correlations involving this vicinal deuterium D$_\mathrm{vic}$ to be described by Eq.\,\ref{eq:SpecMethyl}, although it is not part of the methyl group.
\subsubsection{Relaxation rates}
In our analysis of high-field and relaxometry relaxation rates on \CDDH-methyl groups of Ubiquitin, longitudinal and transverse carbon-13 autorelaxation rates, longitudinal proton autorelaxation rates and dipolar cross-relaxation rates were used. Dipolar relaxation with an effective vicinal deuterium was considered. The set-up of \RedKite~for such a spin system is detailed in Supplementary Materials. The contribution of the proton CSA to relaxation is expected to be negligible \cite{Tugarinov_JBNMR_2004}, and is not considered in the following. The CSA tensor of the carbon-13 nucleus is assumed to be symmetric and aligned with the CC bond. Expressions of the relaxation rates are given in the following equations:
\begin{equation}
	\begin{aligned}
	R_1 (\carb) =& \frac{2}{3} \Delta \sigma_C^2\omegaC^2  \Jimpu{C} (\omegaC)  \\
    &+ \frac{1}{2} d_\mathrm{CH}^2  \left( \Jimpu{CH}(\omegaC-\omegaH) + 3 \Jimpu{CH}(\omegaC) + 6 \Jimpu{CH}(\omegaC+\omegaH)  \right) \\
    &+ \frac{8}{3} d_\mathrm{CD}^2 \left( \Jimpu{CD}(\omegaC-\omegaD)+ 3 \Jimpu{CD}(\omegaC)  + 6 \Jimpu{CD}(\omegaC+\omegaD) \right) \\
    &+ \frac{4}{3} d_\mathrm{CD_\mathrm{vic}}^2 \left( \Jimpu{CD_\mathrm{vic}}(\omegaC-\omegaD) + 3 \Jimpu{CD_\mathrm{vic}}(\omegaC) + 6 \Jimpu{CD_\mathrm{vic}}(\omegaC+\omegaD) \right),  \\
	R_2 (\carb) =& \frac{1}{9} \Delta \sigma_C^2\omegaC^2 \left( 4 \Jimpu{C}(0) + 3 \Jimpu{C}(\omegaC) \right) \\
    &+ \frac{1}{4} d_\mathrm{CH}^2 (4 \Jimpu{CH}(0) + \Jimpu{CH}(\omegaC-\omegaH) + 3 \Jimpu{CH}(\omegaC) + 6 \Jimpu{CH}(\omegaH) \\
& + 6 \Jimpu{CH}(\omegaC+\omegaH)) \\
    &+ \frac{4}{3}  d_\mathrm{CD}^2 (4 \Jimpu{CD}(0) + \Jimpu{CD}(\omegaC-\omegaD) + 3 \Jimpu{CD}(\omegaC) + 6 \Jimpu{CD}(\omegaD)  \\
& + 6 \Jimpu{CD}(\omegaC+\omegaD)) \\
    &+ \frac{2}{3} d_\mathrm{CD_\mathrm{vic}}^2 (4 \Jimpu{CD_\mathrm{vic}}(0) + \Jimpu{CD_\mathrm{vic}}(\omegaC-\omegaD) + 3 \Jimpu{CD_\mathrm{vic}}(\omegaC) \\
& + 6 \Jimpu{CD_\mathrm{vic}}(\omegaD) + 6 \Jimpu{CD_\mathrm{vic}}(\omegaC+\omegaD)),  \\
    R_1 (^{1}\mathrm{H}) =& \frac{1}{2} d_\mathrm{CH}^2 ( \Jimpu{CH}(\omegaC-\omegaH) + 3 \Jimpu{CH}(\omegaH) + 6 \Jimpu{CH}(\omegaC+\omegaH)  ) \\
    +& \frac{8}{3} d_\mathrm{HD}^2 ( \Jimpu{HD}(\omegaD-\omegaH) + 3 \Jimpu{HD}(\omegaH) + 6 \Jimpu{HD}(\omegaD+\omegaH) ) \\
    +& \frac{4}{3} d_\mathrm{HD_\mathrm{vic}}^2 ( \Jimpu{HD_\mathrm{vic}}(\omegaD-\omegaH) + 3 \Jimpu{HD_\mathrm{vic}}(\omegaH) + 6 \Jimpu{HD_\mathrm{vic}}(\omegaD+\omegaH) ), \\
	\sigma_\mathrm{CH} =& \frac{1}{2} d_\mathrm{CH}^2 (-\Jimpu{CH}(\omegaC-\omegaH) + 6 \Jimpu{CH}(\omegaC+\omegaH)),
	\end{aligned}
	\label{eq:HFrates}
\end{equation}
where $d_{AB}$ is the dipolar coefficient between atoms \textit{A} and \textit{B} and equals $-(\mu_0 \hbar \gamma_A \gamma_B)/(4 \pi r^3_{AB})$ with $\mu_0$ the permeability of free space, $\hbar$ the Planck's constant divided by $2\pi$, $\gamma_X$ the gyromagnetic ratio of nucleus \textit{X} and $r_{AB}$ the internuclear distance between nuclei \textit{A} and \textit{B}, $\Delta \sigma_C$ is the chemical shift anisotropy of the carbon-13 nucleus and $\omega_X=-\gamma_X B_0$ is the Larmor frequency for the nuclei \textit{X} at a magnetic field $B_0$. The geometry of the methyl group was assumed to be tetrahedral with $r_\mathrm{CH}=r_\mathrm{CD}=111.5$\,pm leading to $r_\mathrm{HD}=182$\,pm. The distance $r_\mathrm{CD_\mathrm{vic}}$ is determined during the ICARUS analysis as described below.
\subsubsection{Relaxation matrix}
The secularized basis for the subspace that includes $\hat{\mathrm{C}}_z$ in a \CDDH-methyl group contains 14 terms:
\begin{equation}
	\begin{aligned}
\mathcal{B}_\mathrm{secularized} =& \left\{ \frac{\hat{C}_z}{3\sqrt{3}}, \frac{\hat{H}_z}{3\sqrt{3}}, \frac{\hat{D}_{1,z}}{6\sqrt{2}}, \frac{\hat{D}_{2,z}}{6\sqrt{2}}, \frac{2\hat{C}_z\hat{H}_z}{3\sqrt{3}}, \frac{\hat{C}_z\hat{D}_{1,z}}{3\sqrt{3}}, \frac{\hat{C}_z\hat{D}_{2,z}}{3\sqrt{3}},  \frac{\sqrt{2}\hat{C}_z\hat{H}_z\hat{D}_{1,z}}{3}, \right. \\
& \left. \frac{\sqrt{2}\hat{C}_z\hat{H}_z\hat{D}_{2,z}}{3}, \frac{\hat{C}_z\hat{D}_{1,z}\hat{D}_{2,z}}{2\sqrt{3}}, \frac{\hat{C}_z\hat{D}_{1}^{+}\hat{D}_{2}^{-}}{4\sqrt{3}}, \frac{\hat{C}_z\hat{D}_{1}^{-}\hat{D}_{2}^{+}}{4\sqrt{3}}, \right. \\
& \left. \frac{3\hat{C}_z\hat{D}_{1,z} \hat{D}_{1,z} - 2\hat{C}_z}{3\sqrt{6}},   \frac{3\hat{C}_z\hat{D}_{2,z} \hat{D}_{2,z} - 2\hat{C}_z}{3\sqrt{6}}  \right\},
	\end{aligned}
\end{equation}
where $C$, $H$, $D_{1}$ and $D_{2}$ refer to the carbon, proton, deuterium 1 and deuterium 2, respectively, as defined in the spin system in \RedKite. The deuterium 1 and 2 are considered magnetically equivalent and can be exchanged by symmetry (see Fig.\,\ref{fig:EffectDeuterium}c for a visualisization of the geometry of the system).\\
As shown below, the analysis of the relaxation properties of the \CDDH-methyl groups of Ubiquitin during a relaxometry experiment can be performed with satisfactory accuracy in the subspace spanned by the three operators:
\begin{equation}
\mathcal{B}_{reduced,3}=\left\{\frac{\hat{C}_z}{3\sqrt{3}}, \frac{\hat{H}_z}{3\sqrt{3}}, \frac{2\hat{C}_z\hat{H}_z}{3\sqrt{3}}\right\},
\label{eq:FinalBasis}
\end{equation}
leading to the following relaxation matrix:
\begin{equation}
\mathcal{R}_{3}=
	\begin{pmatrix}
	R_1(^{13}\mathrm{C}) & \sigma_\mathrm{CH} & \eta_z^\mathrm{C} \\
    \sigma_\mathrm{CH} & R_1(^{1}\mathrm{H}) & 0 \\
    \eta_z^\mathrm{C} & 0 & R_\mathrm{CH}
	\end{pmatrix},
\end{equation}
where $R_1(^{13}\mathrm{C})$, $R_1(^{1}\mathrm{H})$ and $\sigma_\mathrm{CH}$ are defined above and:
\begin{equation}
	\begin{aligned}
R_\mathrm{CH} =& \frac{2}{3} \Delta \sigma_C^2 \omegaC^2 \Jimpu{C} (\omegaC) + \frac{3}{2} d_\mathrm{CH}^2 \left(\Jimpu{CH}(\omegaC) + \Jimpu{CH}(\omegaH) \right)  \\
&+ \frac{8}{3} d_\mathrm{CD}^2 \left(\Jimpu{CD}(\omegaC-\omegaD) + 3 \Jimpu{CD}(\omegaC) + 6 \Jimpu{CD}(\omegaC + \omegaD) \right)  \\
&+ \frac{8}{3} d_\mathrm{HD}^2 \left(\Jimpu{HD}(\omegaH-\omegaD) + 3 \Jimpu{HD}(\omegaH) + 6 \Jimpu{HD}(\omegaH + \omegaD) \right) \\
&+ \frac{4}{3} d_\mathrm{CD_\mathrm{vic}}^2 \left(\Jimpu{CD_{vic}}(\omegaC - \omegaD) + 3 \Jimpu{CD_{vic}}(\omegaC) + 6 \Jimpu{CD_{vic}}(\omegaC + \omegaD) \right)   \\
&+ \frac{4}{3} d_\mathrm{HD_\mathrm{vic}}^2 \left(\Jimpu{HD_{vic}}(\omegaH - \omegaD) + 3 \Jimpu{HD_{vic}}(\omegaH) + 6 \Jimpu{HD_{vic}}(\omegaH + \omegaD) \right),   \\
\eta_z^\mathrm{C} =& - 2 \Delta \sigma_C \omegaC d_\mathrm{CH} \Jimpd[C]{CH} (\omegaC).
	\end{aligned}
\end{equation}
The expression of the secularized relaxation matrix can be found in the Supplementary Materials.

%% file: Chapters/3b_MethylGroup_Results.tex
\tocless\subsection{Analysis of several aspects of the relaxation in methyl groups}
\subsubsection{Size of the relaxation matrix}
\begin{figure}
\begin{center}
	\includegraphics[width=1\textwidth]{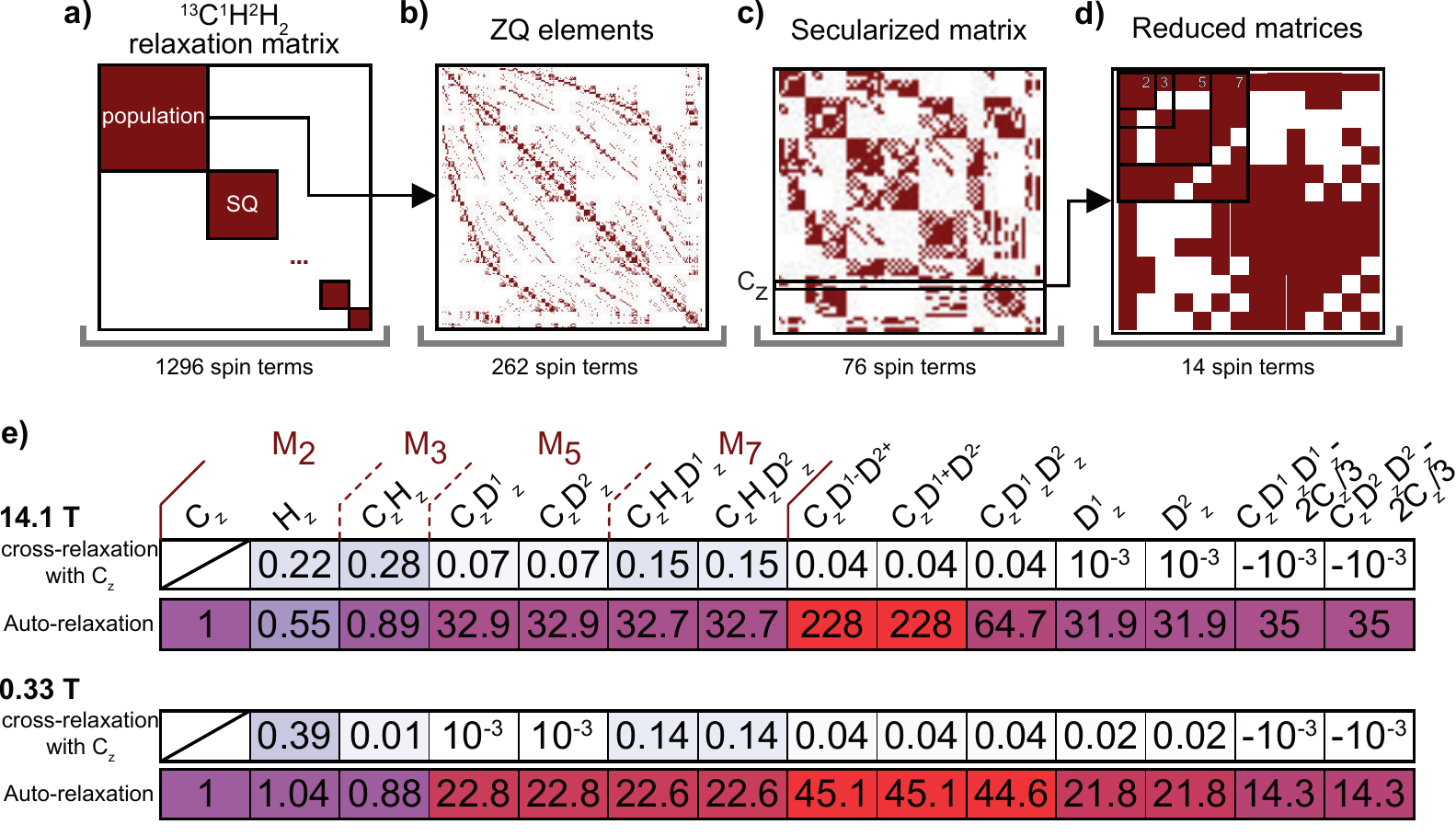}
\end{center}
\caption{Relaxation matrix size-reduction in a \CDDH-methyl group. \textbf{a)} Full relaxation matrix of a \CDDH-methyl group. \textbf{b)} Relaxation matrix of the Zero-Quantum (ZQ) coherences and populations are selected. At this stage, the matrix has a 262x262 size. \textbf{c)} Secularized relaxation matrix containing 76 secular terms in the Zeeman interaction frame. The line corresponding to the operator of interest (\Cz) is highlighted. \textbf{d)} Relaxation matrix containing only terms cross-relaxing with the operator of interest (\Cz). Evaluating the cross-relaxation rates allows another level of size reduction. \textbf{e)} Numerical values of the diagonal terms of the relaxation matrix shown in d) (auto-relaxation, bottom row) and cross-relaxation rates with \Cz~(top row) for the motional parameters of the $\delta$1 methyl group of Ile-3 in \ubiquitin~at 14.1\,T and 0.33\,T (reported in Ref.\,\cite{Cousin_JACS_2018}). Relaxation rates are normalized to the auto-relaxation rate of \Cz~at each magnetic field.}
\label{fig:relaxationMatrix}
\end{figure}

The ICARUS protocol aims at obtaining accurate estimates of low-field relaxation rates by accounting for the effects of cross-relaxation on the longitudinal relaxation decays during a high-resolution relaxometry experiment. This estimate is based on the simulation of the relaxometry experiments, where the sample travels through a broad range of magnetic fields. In order to obtain a reliable description of relaxation over orders of magnitude of magnetic fields, simulations must use appropriate relaxation matrices as well as expressions of relaxation rates, with accurate parameters for the amplitudes of interactions and the description of the spectral density function. The full Liouville space for a \CDDH~spin system is spanned by a large basis of $(2 \times \frac{1}{2} + 1 )^{2 \times n_{1/2}} \times (2 \times 1 + 1)^{2 \times n_1} = 1296$ spin terms, with $n_{1/2}$ and $n_1$ the number of spin-half and spin-one respectively (Fig.\,\ref{fig:relaxationMatrix}a). An efficient calculation requires to minimize the size of the Liouville space where the evolution of the density operator is calculated. We have reduced the size of the subspace using the steps described in Section\,3.1 for $^{15}$N-$^1$H spin systems. First, we have considered the subspace only spanned by zero-quantum coherences and population operators (Fig.\,\ref{fig:relaxationMatrix}b). We then applied the secular approximation, and calculated all cross-relaxation terms with the \Cz~operator, in order to keep only non zero terms, \textit{i.e.} terms that cross-relax with \Cz, reducing the size of the basis to 14 terms (Fig.\,\ref{fig:relaxationMatrix}d). Cross-relaxation and autorelaxation rates in this 14-element basis have been calculated at the lowest and highest magnetic fields used during our HRR experiments, \textit{i.e} 0.33\,T and 14.1\,T, using parameters obtained after a preliminary ICARUS analysis (for Ile-3) performed using $\mathcal{B}_{reduced,3}$ (Eq.\,\ref{eq:FinalBasis}, Fig.\,\ref{fig:relaxationMatrix}e). \\
The inspection of these two relaxation matrices justifies the use of a basis containing only 3 operators as cross-relaxation rates involving other operators are either negligible (cross relaxation from $\hat{\mathrm{C}}_z$ to another operator can be neglected if the ratio of this cross-relaxation rate to the auto-relaxation rate of $\hat{\mathrm{C}}_z$ is small) or involve an operator with an auto-relaxation rate much larger than the auto-relaxation rate of $\hat{\mathrm{C}}_z$ and the cross-relaxation rate with $\hat{\mathrm{C}}_z$ (see the Supplementary Materials for the proof that cross-relaxation with fast relaxing operator do not contribute to the polarization decay of slowly relaxing operators). At both magnetic fields, the largest cross-relaxation rate with the carbon-13 longitudinal polarization is the dipolar cross-relaxation with the proton longitudinal polarization. At low magnetic field (0.33\,T), even a 2-operator basis \{$\frac{1}{3}$\Cz, $\frac{1}{3}$\Hz\} would be sufficient to describe the relaxation properties of a \CDDH-methyl group as cross-relaxation towards other terms is either very small or towards fast-relaxing terms. However, the subspace should include the two-spin order 2\CzHz~at high field (14.1\,T). Thus, high-resolution relaxometry experiments in \CDDH-methyl groups have been simulated in the small subspace spanned by the three operators (\Cz, \Hz~and 2\CzHz). This subspace was used throughout our analysis of carbon-13 HRR in \CDDH~methyl groups.
\subsubsection{Proton relaxation and surrounding deuterium}
\begin{figure*}
\begin{center}
	\includegraphics[width=1.0\textwidth]{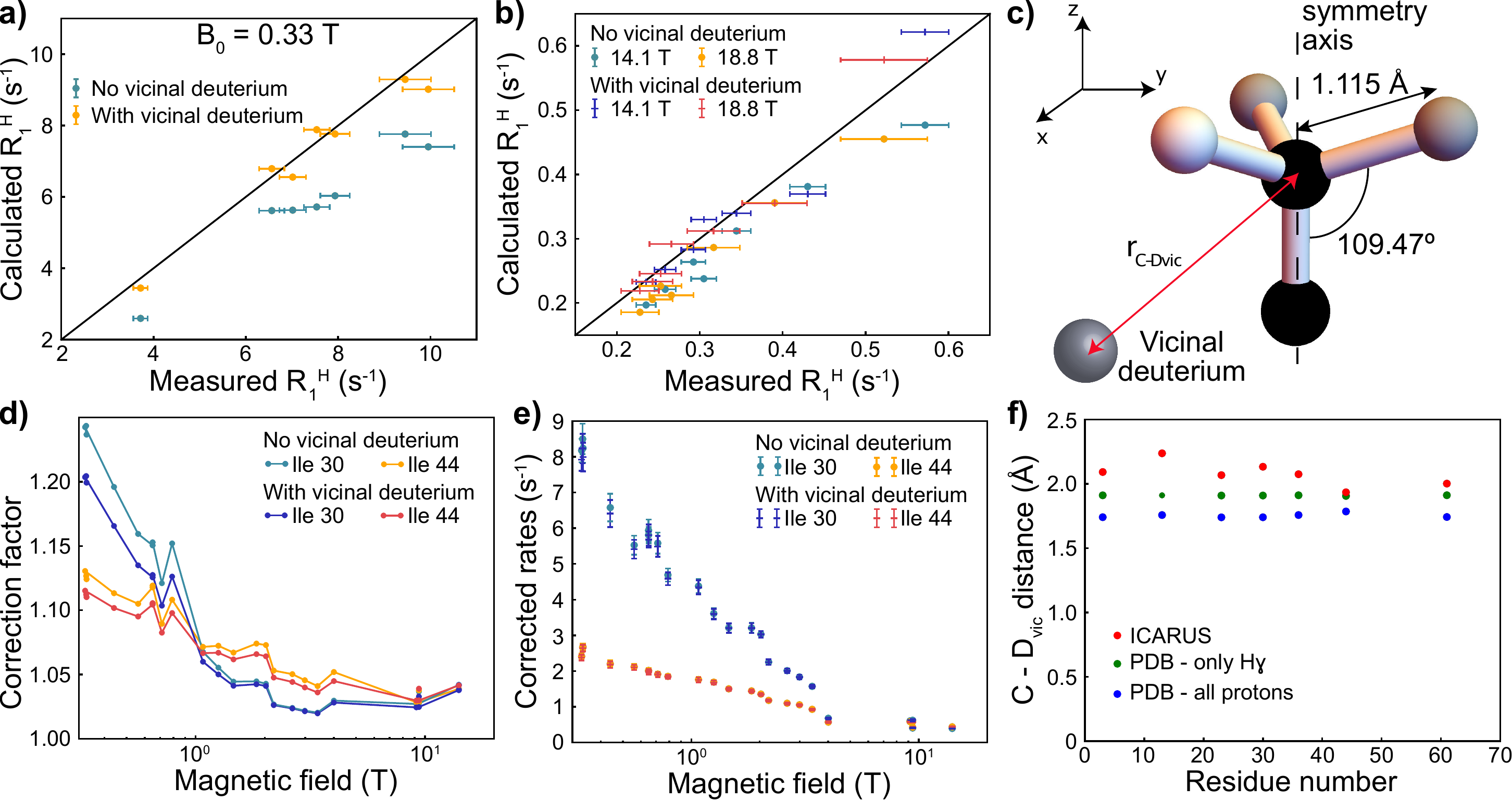}
\end{center}
\caption{Including the effect of an effective vicinal deuterium nucleus on the analysis of high-resolution relaxometry data of \ubiquitin. \textbf{a)} Correlation plot of the calculated proton longitudinal relaxation rate R$_1$ at 0.33\,T with (orange) and without (blue) including the effect of the vicinal deuterium, with the experimental R$_1$ at 0.33\,T, for the seven isoleucines of Ubiquitin. The black line is shown as a guide for perfect equality between the two rates. \textbf{b)} Correlation plots of the calculated proton longitudinal relaxation rate R$_1$ at 14.1\,T and 18.8\,T with and without including the effect of the vicinal deuterium, with the experimental R$_1$ at 14.1\,T and 18.8\,T, for the seven isoleucines of Ubiquitin. The black line is shown as a guide for perfect equality between the two rates. \textbf{c)} Geometry of the methyl group and position of the effective neighbouring deuterium. The distance $r_\mathrm{C-D_\mathrm{vic}}=\sqrt{r_\mathrm{y,D_\mathrm{vic}}^2 + r_\mathrm{z,D_\mathrm{vic}}^2}$ is determined using additional relaxation rates as explained in the main text. \textbf{d)} Correction factors as a function of the magnetic field for Ile-30 and Ile-44 with and without an effective vicinal deuterium nucleus. \textbf{e)} Corrected relaxometry relaxation rates for Ile-30 and Ile-44 with and without including an effective vicinal deuterium nucleus. \textbf{f)} Comparison of the distance of the vicinal deuterium with the carbon-13 nucleus obtained from the analysis of proton relaxation (red, ICARUS) to the calculated distance to an effective deuterium nucleus that accounts for either only the $^2$H$\gamma_1$ and $^2$H$\gamma_2$ nuclei of the isoleucine residue (green) or all the hydrogens (blue) in the structure of Ubiquitin (PDB ID: 1D3Z). In these NMR derived structures, the distances were averaged over the 10 models present in the PDB file. In each model, the distance equals $r_\mathrm{C-D_\mathrm{vic}}=\left( \sum_i \frac{1}{d_i^6}\right)^{-1/6}$ with $d_i$ the distance of the carbon-13 to proton $i$ (excluding intra-methyl group proton).}
\label{fig:EffectDeuterium}
\end{figure*}

Proton longitudinal relaxation rates $R_1(^1\mathrm{H})$ were measured at three magnetic fields (0.33, 14.1 and 18.8\,T) using standard high-field magnets (18.8\,T and 14.1\,T) and a 2F-NMR spectrometer operating at 14.1\,T and 0.33\,T \cite{KaderavekJPCL2019}. These rates were also calculated after an ICARUS analysis of high-field and HRR rates considering intra-methyl group interactions only. The predicted relaxation rates are systematically lower than those measured at 0.33\,T, 14.1\,T and 18.8\,T (Fig.\,\ref{fig:EffectDeuterium}a, b). Thus, even if relaxation rates in a \CDDH-methyl group are dominated by the contributions of internal interactions, another contribution to relaxation has to be taken into account to describe proton relaxation. The differences between the measured and calculated $R_1(^1\mathrm{H})$ rates were assigned to the effect of the neighbouring deuterium nuclei.\\
Adding the dipolar interactions with surrounding deuterium nuclei leads to non-negligible contributions to relaxation to both the proton and the carbon-13. The closest neighbouring deuterium nuclei are the $^2$H$\gamma_1$ and $^2$H$\gamma_2$ sites of the isoleucine side-chain, but other deuterium nuclei may also be in close proximity to the methyl group especially within the hydrophobic core of the protein. The correlation function for the fluctuations of the corresponding internuclear vectors are expected to vary. In particular, these interactions are expected to be affected in different ways by the fast rotation of the methyl group. We modeled the surrounding deuterium nuclei by a single deuterium at an effective distance (Fig.\,\ref{fig:EffectDeuterium}c). The interaction of the proton and carbon-13 nuclei of the methyl group with this deuterium accounts for the interaction with all the other deuterium nuclei of the protein. We used two adjustable parameters to describe its position, defining its coordinates in the Cartesian axis system: the y- and z-coordinate were fitted while the x-coordinate was fixed to 0. The position of the effective surrounding deuterium nucleus is determined independently for each residue using proton relaxation rates as well as all relaxation rates used in the ICARUS iterations (accurate and corrected) and keeping the other parameters constant (\textit{i.e.} the parameters describing the dynamics). When fitting the parameters of the model during further ICARUS analysis, the effective position of the surrounding deuterium is kept constant. Introducing the contribution of the surrounding deuterium and performing the whole ICARUS analysis again preserves the agreement between the measured and calculated proton longitudinal relaxation rates (Fig.\,\ref{fig:EffectDeuterium}a, b).\\
The surrounding deuterium has an effect on the correction factors (Fig.\,\ref{fig:EffectDeuterium}d) which leads to differences of corrected HRR rates between 0 and 4\,\% (Fig.\,\ref{fig:EffectDeuterium}e). Correction factors depend on the magnetic field and generally increase with decreasing magnetic. It must be pointed out that non-monotonous changes in the correction factors profiles in Fig.\,\ref{fig:EffectDeuterium}d are due to differences in shuttling and waiting delays at low magnetic fields (Fig.\,\ref{fig:ExpDelays}).\\
The effective distances with the surrounding deuterium nucleus are close to extracted distances from the NMR structure of Ubiquitin (Fig.\,\ref{fig:EffectDeuterium}f, PDB 1D3Z). The dipolar interaction between the methyl group and the effective deuterium is included in the following iterations of the ICARUS analysis. 
\subsubsection{Convergence of the iterative correction}
\begin{figure*}
\begin{center}
	\includegraphics[width=1\textwidth]{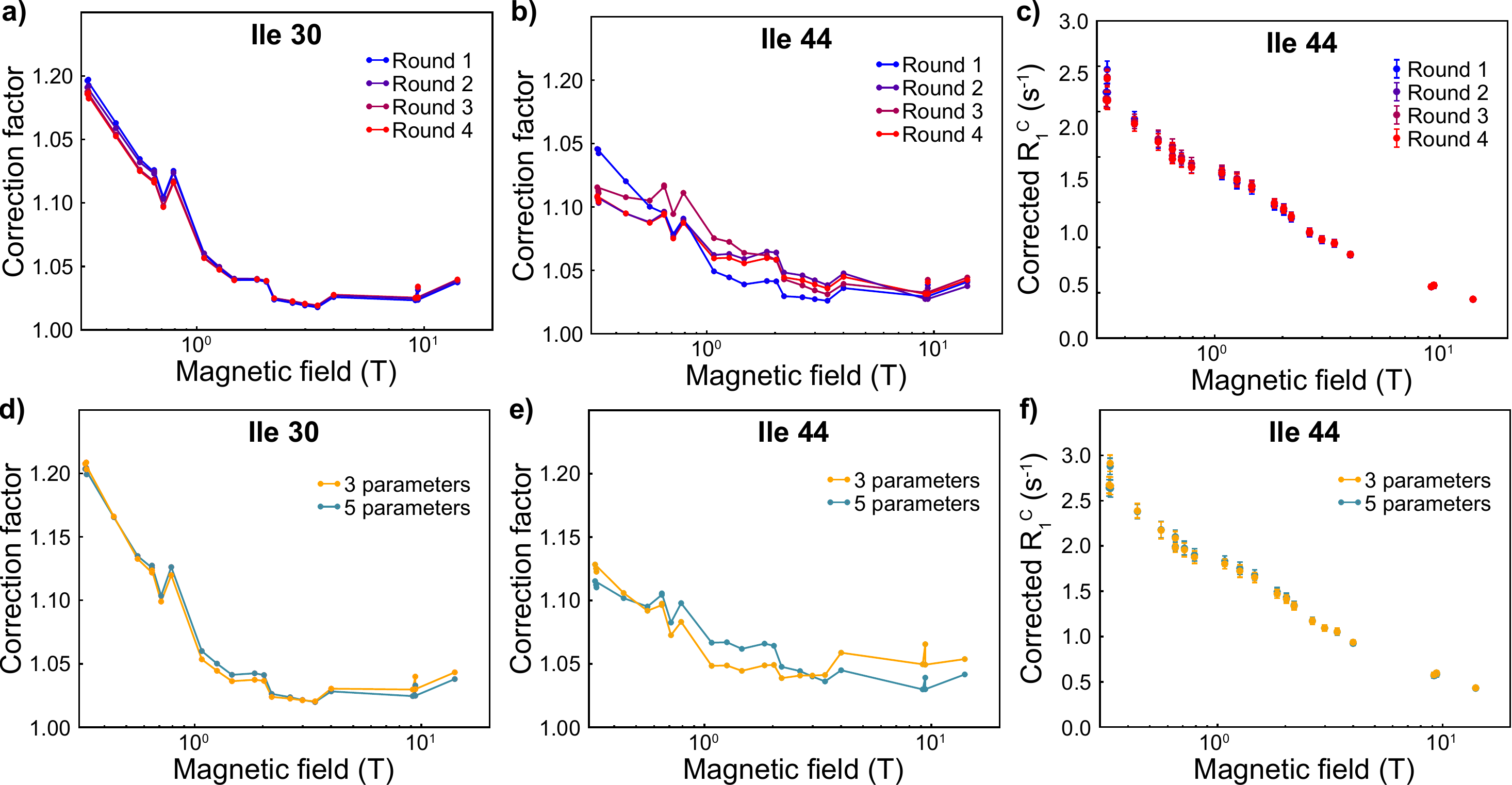}
\end{center}
\caption{Evolution of the correction with the number of iterations of ICARUS and the selected model of motions. Correction factors as a function of the magnetic field for (\textbf{a}) Ile-30 and (\textbf{b}) Ile-44 after 1 to 4 rounds of ICARUS. \textbf{c)} Evolution of the corrected relaxation rates of Ile-44 after 1 to 4 rounds of ICARUS. Correction factors as a function of the magnetic field for (\textbf{d}) Ile-30 and (\textbf{e}) Ile-44 using a model of spectral density function with 3 (Eq.\,\ref{eq:Param3} ,orange) or 5 (Eq.\,\ref{eq:SpecMethyl}, blue) parameters to describe internal dynamics. \textbf{f)} Corrected relaxation rates of Ile-44 obtained with a model with 3 (Eq.\,\ref{eq:Param3}, orange) or 5 (Eq.\,\ref{eq:SpecMethyl}, blue) parameters to describe internal dynamics.}
\label{fig:CorrectionFactors}
\end{figure*}

The number of iteration steps is expected to be dependent on the spin system under study. In the case of the \CDDH-spin system, the convergence was reached after 2 iterations (Fig.\,\ref{fig:CorrectionFactors}a) for all residues except residue 44. Some slight instability in the convergence of the correction at low field is observed for this residue (Fig.\,\ref{fig:CorrectionFactors}b) but the amplitude of change (1-2\,\% at most) has a negligible effect on the values of the corrected relaxation rates (Fig.\,\ref{fig:CorrectionFactors}c).
\subsubsection{Influence of the model of spectral density function on the correction}
Different models can be used to describe the motions  in a methyl group. Eq.\,\ref{eq:SpecMethyl} gives a rather complex description of the motion, but a simpler model can be tested by reducing the number of internal dynamics parameters to 3 by only considering the global tumbling, the methyl-group rotation with one fitted correlation time and C-C axis motions with only one fitted correlation time and one order parameter. The spectral density function for this model is:
\begin{equation}
	\begin{aligned}
	\mathcal{J}_{i,j}^{(3)} (\omega) =& \frac{1}{5} \left[ S_{met}^2(\theta_{i,j}) \left( S^2 \Lorentz{\tau_c} + (1-S^2) \Lorentz{\tau'_{int}} \right) \right. \\
	&\left. + (\mathcal{P}_2\cos(\thIJ)-S_{met}^2(\theta_{i,j})) \left( S^2 \Lorentz{\tau'_{met}}  \right. \right. \\
	& \left. \left. + (1-S^2) \Lorentz{\tau''_{int}} \right) \right],
	\end{aligned}
	\label{eq:Param3}
\end{equation}
with the same definitions as above and where $\tau_{int}$ is an internal correlation time for the motion of the C-C axis. Correction factors obtained for the two spectral density functions are shown in Fig.\,\ref{fig:CorrectionFactors}d and e. They are identical for Ile-30 where both models fit the experimental data well. In contrast, the correction is slightly different for the two models of motion for Ile-44 (Fig.\,\ref{fig:CorrectionFactors}e), where the 5-parameters model is in better agreement with the experiments. Yet, the variation on the corrected rates is small (between 1 and 2\,\%, Fig.\,\ref{fig:CorrectionFactors}f) with equally small effects on the analysis. The ICARUS analysis requires a model that accounts for the overall changes of the spectral density function on the range of frequencies probed during the experiments but it does not require that the used model reproduces all subtle details of the spectral density function: small variations of the value of the spectral density function at a specific frequency have negligible effects on the correction.
\subsubsection{Scaling of the CSA/dipole-dipole cross-correlated cross-relaxation rates}
\begin{figure*}
	\begin{center}
		\includegraphics[width=0.45\textwidth]{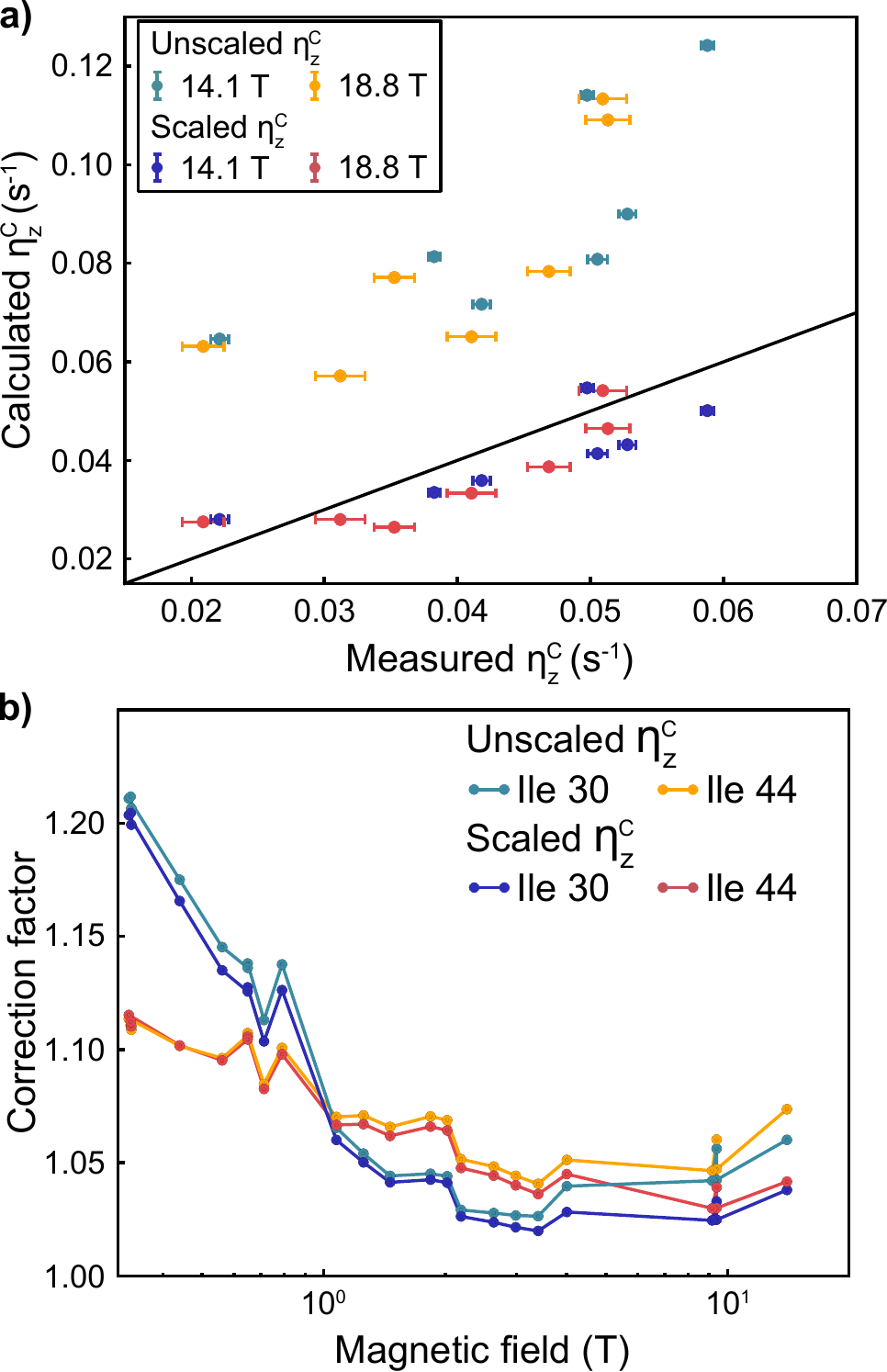}
	\end{center}
	\caption{Scaling the CSA/DD cross-correlated cross-relaxation rates. \textbf{a)} Correlation plot between the calculated unscaled and scaled longitudinal CSA/DD cross-correlated cross-relaxation rates with the measured rates at 14.1\,T and 18.8\,T. The black line is shown as a guide for perfect equality between the two rates. \textbf{b)} Correction factors as a function of the magnetic field for Ile-30 and Ile-44 with or without scaling of the longitudinal CSA/DD cross-correlated cross-relaxation rate.}
	\label{fig:CorrelationEtaZ}
\end{figure*}

Our combined analysis of low-field longitudinal and high-field transverse relaxation has allowed us to obtain the value of the CSA for each residue in addition to parameters of internal motions, except for Ile-44 for which chemical exchange prevented the analysis of the carbon-13 transverse relaxation rates \cite{Cousin_JACS_2018}. In order to validate our analysis, a series of relaxation rates were measured as detailed hereafter: accurate low field carbon longitudinal relaxation rates \cite{KaderavekJPCL2019} as well as high-field longitudinal CSA/dipole-dipole (CSA/DD) cross-correlated cross-relaxation rates (cross-relaxation between \Cz~and 2\CzHz~refered to as $\eta_z^C$). These relaxation rates were not used during the analysis of the relaxometry relaxation rates, but calculated using the set of motional parameters obtained after correction of the relaxometry data.\\
The calculated longitudinal CSA/DD cross-relaxation rates were strongly correlated to measurements at 14.1\,T and 18.8\,T but significantely overestimated (Fig.\,\ref{fig:CorrelationEtaZ}a). In order to have a better description of the CSA/DD cross-correlation, a scaling factor was applied directly to this term in the relaxation matrix. The scaling factor was calculated as the averaged inverse correlation coefficient between the unscaled and measured $\eta_z^C$ at 14.1\,T and 18.8\,T and equals 0.505. A number of hypothesis can be made to explain the origin of the scaling factor: i) the carbon-13 CSA may be overestimated since it is determined essentially from transverse relaxation rates, which may suffer from small chemical exchange contributions; ii) the carbon-13 CSA may not be perfectly alligned with the C-C bond; iii) the form of the spectral density function may not describe correctly the motions of the methyl group; iv) the amplitude of the carbon-13 CSA may be rotamer-dependent.\\
To understand the origin of this scaling factor, we also measured the carbon transverse CSA/DD cross-correlated cross-relaxation rates ($\eta_{xy}^C$). The calculated relaxation rates correlate with the measurement, with an averaged inverse correlation coefficient between the calculated and measured $\eta_{xy}^C$ at 14.1\,T and 18.8\,T of 0.629 (Supplementary Materials Fig.\,\ref{fig:EtaX}). The discrepency between the scaling factors of the longitudinal and transverse CSA/DD cross-correlated cross-relaxation rates can not be accounted for only from a miss-evaluation of the carbon-13 CSA (under our assumptions of axially symmetry and perfect alignment allong the CC bond). Thus, it is likely that the model of correlation function does not describe entirely the complexity of the motions in the methyl group, and additional work toward this direction has to be done. For example, transitions between rotamers may be better modelled with instantateous jumps. \\
The analysis of the relaxometry relaxation data was performed again after applying the scaling factor to longitudinal CSA/DD relaxation rates. As expected, the agreement between calculated and measured CSA/DD cross-relaxation rates is significantly improved by the use of a scaling factor (Fig.\,\ref{fig:CorrelationEtaZ}a). Low-field correction factors are not sensitive to the scaling of a CSA-dependent relaxation rate (Fig.\,\ref{fig:CorrelationEtaZ}b). At moderate and high field, the effect is larger with a reduction of the correction by about 2\,\% which has limited impact on the analysis.

\tocless\subsection{Validation of the correction with the suppression of cross-relaxation pathways}
\begin{figure*}
\begin{center}
	\includegraphics[width=0.45\textwidth]{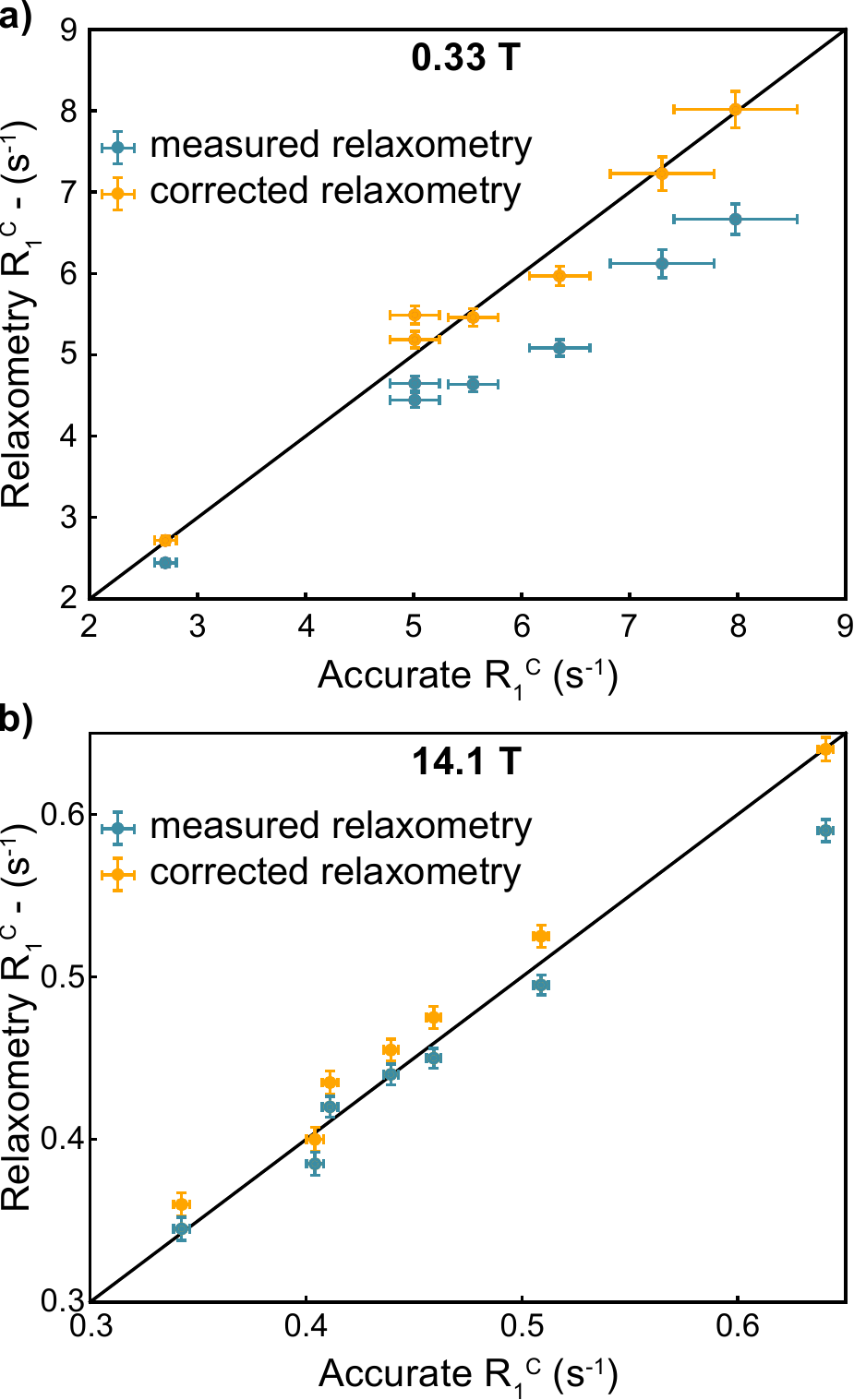}
\end{center}
\caption{Validation of the correction protocol. \textbf{a)} Correlation plot between the relaxometry uncorrected (blue) and corrected (orange) carbon R$_1$ at 0.33\,T with the measured two-field R$_1$($^{13}$C). \textbf{b)} Correlation plot between the pseudo-relaxometry uncorrected (blue) and corrected (orange) R$_1(^{13}\mathrm{C})$ with the accurate relaxation rates measured at 14.1\,T. The black line is shown as a guide for perfect equality between the two rates.}
\label{fig:CorrR1}
\end{figure*}

Using the recently developed 2F-NMR spectrometer \cite{Cousin2016, Cousin_PCCP_2016}, we measured, among other relaxation rates, the longitudinal carbon-13 relaxation rates at 0.33\,T with suppression of cross-relaxation pathways \cite{KaderavekJPCL2019}. The rates of the seven isoleucines acquired at 0.33\,T have been compared to measured and corrected relaxometry relaxation rates at the same magnetic field (Fig.\,\ref{fig:CorrR1}a). The uncorrected relaxometry rates $R_1(^{13}\mathrm{C})$ are systematically lower than the accurate relaxation rates. This stresses the fact that the relaxometry relaxation rates have to be corrected in order to reach a reliable analysis of the properties the dynamics of the system.  Corrected rates are in excellent agreement with the accurate $R_1(^{13}\mathrm{C})$ rates measured with the two-field system. This comparison validates the ICARUS approach on this spin system. In addition, experiments have been recorded at 14.1\,T with and without pulses during the relaxation delay. Corresponding relaxation rates are displayed in Fig.\,\ref{fig:CorrR1}b. The high-field experiment recorded without control of cross-relaxation pathways is similar to a shuttling experiments. Correction factors seem to be slightly overestimated  at 14.1\,T, but corrected rates are in better agreement with accurate rates than uncorrected rates (r.m.s.d of $3.8 \times 10^{-2}$\,s$^{-1}$ versus $5.7 \times 10^{-2}$\,s$^{-1}$, respectively).

%% file: Chapters/Conclusion.tex
\tocless\section{Conclusion}
In this paper, we have presented a general framework for the analysis of high-resolution relaxometry data. First, \RedKite~is a powerful \textsc{Mathematica} notebook to calculate relaxation rates and entire relaxation matrices in any nuclear spin system. We have shown how it can be used for the analysis of HRR, but it can also be applied more generally for the study of relaxation properties. Second, ICARUS is a \textsc{Python}-based program designed to analyze relaxometry datasets accounting for the effects of multiple cross-relaxation pathways. The two toolkits have been developed in order to be easily adapted to other spin systems, diffusion tensors and models of motions. Conclusions drawn here in the case of a \CDDH-methyl group with respect to the effect of the size of the relaxation matrix, the number of iteration of ICARUS or the model for the spectral density function may be different in other systems. Overall, a complete analysis by \RedKite~and ICARUS can be performed quickly, allowing one to evaluate these effects efficiently. Our approach to correct high-resolution relaxometry data has been cross-validated by the measurements of accurate low-field relaxation rates.

%% file: Chapters/MaterialAndMethods.tex
\tocless\section{Materials and methods}
Methods to obtain carbon-13 and proton longitudinal relaxation rates at 0.33\,T were previously described \cite{KaderavekJPCL2019} and are based on the use of a two-field spectrometer operating at 14.1\,T and 0.33\,T \cite{Cousin2016, Cousin_PCCP_2016}. Proton longitudinal relaxation rates at 14.1\,T and 18.8\,T were measured following methods introduced earlier \cite{Ferrage_MMB_2012}. Carbon-13 inversion pulses were applied during the relaxation period every 40\,ms and a proton inversion pulse was applied in the middle of the relaxation delay. The experiment was performed with the following relaxation delays: 0.08*, 0.24, 0.48, 0.72, 0.96, 1.28, 1.68, 2.08, 2.48, 2.88*, 3.28, 3.68, 4.08\,s (the measurements marked by a star have been performed twice). \\
The longitudinal and transverse cross-correlated cross-relaxation rates ($\eta_z^C$ and $\eta_{xy}^C$) were measured using the symmetrical reconversion principle \cite{Pelupessy_JMagnRes_2003, Pelupessy_JCP_2007}. For enhanced sensitivity, cross-relaxation experiments were accumulated with 8-times more scans than auto-relaxation experiments. The longitudinal cross-correlated cross-relaxation rate at 18.8\,T was determined with a relaxation delay of 1.5\,s, while at 14.1\,T the experiment was performed with the relaxation delays of 1.0, 1.5, and 2.0\,s. The measurement of the transverse cross-correlated cross relaxation rate was done using a spin lock irradiation with amplitudes of 2031 and 2062\,Hz at 14.1 and 18.8\,T, respectively. The alignment of the spins into the direction of the spin-lock field and back to z-direction was achieved using adiabatic half passage pulses. The calibration of the spin lock rf amplitude was done by measuring the scaling of scalar couplings under off-resonance continuous wave irradiation. The transverse cross-correlated cross relaxation rate at 18.8\,T was determined from a single experiment performed with the relaxation delay 250 ms, while the experiment was repeated twice with the relaxation delays 175 and 250\,ms at 14.1\,T. \\
The measurement of the "relaxometry-like" relaxation rate at 14.1\,T was performed with the standard pulse program to measure longitudinal relaxation rates \cite{Ferrage_MMB_2012}, but all pulses usually applied during the relaxation period were omitted. The experiment was measured twice, first with the relaxation delays 0.06*, 0.18, 0.38, 0.62, 0.94, 1.26*, 1.62, 2.02\,s, and second with relaxation delays 0.61*, 0.73, 0.93, 1.17, 1.49, 1.81*, 2.17, 2.57\,s (the star denotes measurements repeated once).

%% file: Chapters/SuppInfo.tex
\beginsupplement
\newpage
\section{\huge{Supplementary Materials}}
\label{SupInfo}

\setcounter{tocdepth}{2}
\tableofcontents

\newpage

\begin{figure*}[!ht]
\begin{center}
	\includegraphics[width=0.8\textwidth]{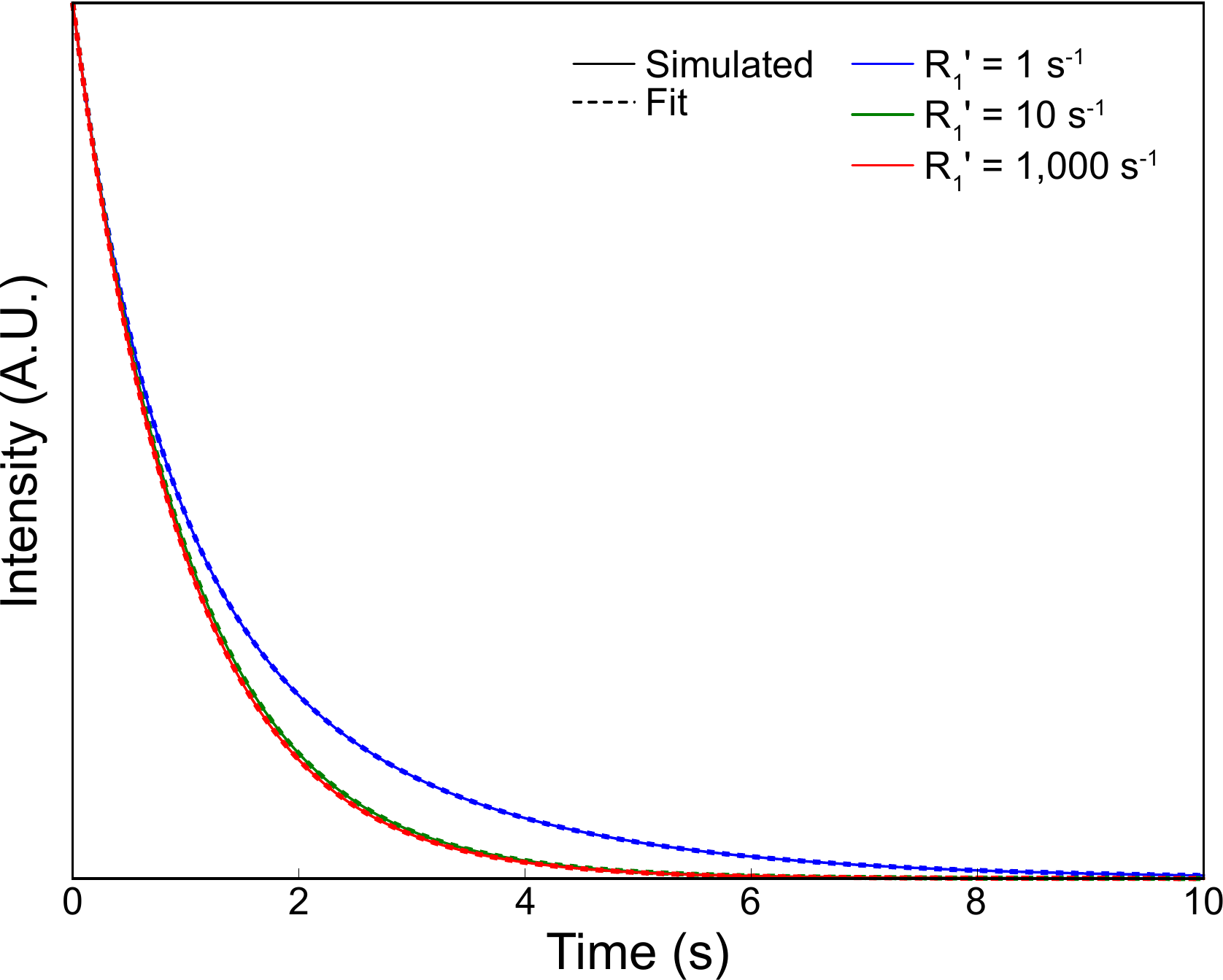}
\end{center}
\caption{Simulated polarization decay (plain) and exponential fit (dash) for different values of $R_1'$ relaxation rates.}
\label{fig:DecayPolarizationSim}
\end{figure*}

\mysection{1}{Size-reduction of relaxation matrices by removing fast-relaxing operators}
Here, we will show that fast-relaxing terms of a relaxation matrix can be discarded (as done in Section\,4.2.5 of the main text) in order to reduce the size of the relaxation matrix and save computational time. For the sake of simplicity, we consider a 2x2 Liouvillian:
\begin{equation}
	\mathcal{L} = \begin{pmatrix}
		R_1 & \sigma \\
		\sigma & R_1'
		\end{pmatrix}.
\end{equation}
The characteristic polynomial of $\mathcal{L}$ is:
\begin{equation}
\det[\mathcal{L}-\lambda \mathcal{I}] = \lambda^2 - \lambda(R_1 + R_1') - \sigma^2 + R_1 R_1',
\end{equation}
with $\mathcal{I}$ the identity matrix. The roots are given by:
\begin{equation}
\lambda_\pm = \frac{R_1 + R_1' \pm \sqrt{\Delta}}{2},
\end{equation}
with:
\begin{equation}
\Delta = R_1'^2 + R_1^2 - 2 R_1 R_1' + 4\sigma^2.
\end{equation}
Let's assume $R_1' \gg R_1, \sigma$. A first order approximation in $R_1$ and $\sigma$ of $\sqrt{\Delta}$ leads to:
\begin{equation}
\sqrt{\Delta} \approx R_1'(1 - \frac{R_1}{R_1'}) = R_1' - R_1,
\end{equation}
such that the eigenvalues of $\mathcal{L}$ are $R_1$ and $R_1'$. The associated eigenvectors approximate to $\{1,0\}$ and $\{0,1\}$ and the autorelaxation of the operator of interest can be considered mono-exponential with decay rate of $R_1$. The fast relaxing operator does not contribute to the relaxation of the slowly relaxing operator.\\
This can be verified by simulating the polarization decay. We will set $R_1 = 1\,s^{-1}$, $\sigma=0.5\,s^{-1}$ and vary $R_1'$. We can compute the polarization decay (associated with the operator of interest with autorelaxation rate $R_1$) following Section\,2.2 of the main text (Fig.\,\ref{fig:DecayPolarizationSim}).
The polarization decay can be fitted to a mono-exponential decay, and fitted relaxation rates are reported in Table\,\ref{table:fitR1sim}. It is clear that the fast relaxing operator has negligeable effects on the polarization decay when $R_1'\gg R_1$.
\begin{table}[!ht]
	\def\arraystretch{1.5}
	\begin{center}
	\caption{Fitted relaxation rates from the simulated polarization decay for different values of $R_1'$} \label{table:fitR1sim}
	\begin{tabular}{cc}
		$R_1'$ ($s^{-1}$) & fitted relaxation rate ($s^{-1}$) \\
		\hline
		1 & 0.73 \\
		10 & 0.97 \\
		1,000 & 1.00
	\end{tabular}
	\end{center}
\end{table}

\section{Correlation functions and spectral density functions}
The choice of the model of motions is a key step in the analysis of relaxation rates to characterize quantitatively protein dynamics. The description of models of correlation functions can be found elsewhere \cite{Lipari1982, Clore1990, Meirovitch2006, Meirovitch2007, Charlier2016, Calandrini2010, Khan2015}. Any analytical form of the spectral density function can be used in \RedKite~and ICARUS. Assuming that different types of motions are statistically independent, the overall correlation function $\mathcal{C}_{i,j}$ associated to auto- or cross-correlation of interaction(s) (\textit{i}, \textit{j}) can be written as the product of the correlation function of overall rotation $\mathcal{C}_g$, assumed here to be isotropic, and of the individual motions $\mathcal{C}^n_{i,j}$, all supposed to be independent and isotropic: 
\begin{equation}
	\mathcal{C}_{i,j}(t)=\mathcal{C}_g(t) \prod_n \mathcal{C}^n_{i,j}(t).
\end{equation}
In model-free approaches, the overall rotation correlation function $\mathcal{C}_{i,j}$ is described by a single exponential decay for isotropic diffusion, or a sum of exponentials for axially symmetric or fully anisotropic rotational diffusion \cite{Tjandra1995}. The correlation function used for the model-free $\mathcal{C}_{i,j}^{\mathrm{MF}}$ and extended model-free $\mathcal{C}_{i,j}^{\mathrm{EMF}}$ approaches are:
\begin{equation}
	\begin{aligned}
\mathcal{C}_{i,j}^\mathrm{MF}(t)=&e^{-t/\tau_g}\left( S^2 + \left(\mathcal{P}_2(\cos\theta_{i,j})-\SD\right)e^{-t/\tau_\mathrm{int}} \right), \\
\mathcal{C}_{i,j}^\mathrm{EMF}(t)=&e^{-t/\tau_g}\left(\SfD\SsD + \left(\mathcal{P}_2(\cos\theta_{i,j})-\SfD\right)e^{-t/\tau_f} \right. \\
	&\left. +\SfD(\mathcal{P}_2(\cos\theta_{i,j})-\SsD)e^{-t/\tau_s} \right),
	\end{aligned}
\end{equation}
where $\theta_{i,j}$ is the angle between the principal axes of the two interactions, $\mathcal{P}_2(x)$ is the second order Legendre polynomial $\mathcal{P}_2(x) = (3x^2-1)/2$, $\tau_g$ the correlation time for the global tumbling. The correlation function for the model-free approach is defined by the effective correlation time $\tau_\mathrm{int}$ and the order parameter $\SD$. In the extended model-free correlation function, $\tau_s$ (respectively $\tau_f$) is the correlation time associated with the order parameter $\SsD$ (respectively $\SfD$) for the slower (respectively faster) motion. The corresponding spectral density functions $\mathcal{J}^{\mathrm{MF}}_{i,j}(\omega)$ and $\mathcal{J}^{\mathrm{EMF}}_{i,j}(\omega)$ can be used for both auto- and cross-correlation of interactions:
\begin{equation}
	\begin{aligned}
\mathcal{J}^{\mathrm{MF}}_{i,j}(\omega) =& \frac{1}{5}\left( \frac{\SD \tau_g}{1+(\omega \tau_g)^2} + \frac{\left(\mathcal{P}_2(\cos\theta_{i,j}) - \SD \right) \tau'}{1+(\omega \tau')^2} \right), \\
\mathcal{J}^{\mathrm{EMF}}_{i,j}(\omega) =& \frac{1}{5}\left( \frac{\SfD\SsD\tau_g}{1+(\omega \tau_g)^2} + \frac{\left(\mathcal{P}_2(\cos\theta_{i,j}) - \SfD \right) \tau'_f}{1+(\omega \tau'_f)^2} + \frac{\SfD(1-\SsD)\tau'_s}{1+(\omega \tau'_s)^2} \right),
	\end{aligned}
\end{equation}
where $\tau'_a$ is the effective correlation time defined as ${\tau'_a}^{-1}=\tau_a^{-1}+\tau_g^{-1}$.  \\
Other correlation functions can be used depending on the system under study. For example, the correlation function can be written as a sum of exponential functions:
\begin{equation}
\mathcal{C}_{\sum \exp}(t)=\sum_{i=1}^n A_i e^{-t/\tau_i} .
\end{equation}
The corresponding spectral density is:
\begin{equation}
\mathcal{J}_{\sum \exp}(t)=\frac{1}{5}\sum_{i=1}^n A_i \frac{\tau_i}{1+(\omega\tau_i)^2}.
\end{equation}
In the case of relaxation in a methyl group, assuming the statistical independence of the methyl group rotation, the motions of the methyl group axis and the overall rotational diffusion, the correlation function $C^{\mathrm{met}}_{i,j}$ can be expressed as the product of the three corresponding correlation functions: $C_g$ for the global tumbling, $C_\mathrm{rot}^{i,j}$ for the methyl group rotation, $C_\mathrm{axis}$ for the complex motions of the methyl group. The correlation function was given in the main text (Eq.\,\ref{eq:MethylCorrFunc}). The rotation of the methyl group is an anisotropic motion characterized by the correlation time $\tau_{met}$ and the order parameter $S_{met}^2 (\theta_{i,j})$ imposed by the geometry of the methyl group (supposed to be a tetrahedron, three corners of which are occupied by the proton and the two deuterium nuclei and the center by the carbon-13) and the relative orientations of the principal axes of interactions $i$ and $j$ with respect to the methyl axis. Motions of the methyl group axis are described by an extended model-free correlation function, with the parameters $\SfD$, $\tau_f$, $\SsD$, and $\tau_s$, as is detailed in the main text.
\section{Set up of \RedKite~for the $\{^{13}\mathrm{C}^1\mathrm{H}^2\mathrm{H}_2\}$-methyl groups of Ubiquitin with a vicinal deuterium}
Here, we show the most important command lines used to calculate relaxation rates and relaxation matrix of a \CDDH-methyl group with a vicinal deuterium nucleus.
\subsection{Definition of the spin system}
\begin{center}
	Nuclei = \{\{"13C","CA"\}, \{"1H", "HA"\},  \{"2H", "DA"\}, \{"2H", "DB"\}, \{"2H", "DC"\}\};
\end{center}
The deuterium $\mathrm{DC}$ is associated with the vicinal deuterium here. The \textit{SetSpinSystem} command is then run as explained in the main text without any changes. We define the intermediate constants:
\begin{equation*}
	\begin{aligned}
		&\alpha = 109.47\pi/180; \\
		&\mathrm{aCH} = \pi - \alpha; \\
		&\mathrm{rCH} = 1.115 \times 10^{-10}; \\
		&\mathrm{rCD} = 1.115 \times 10^{-10}; \\
		&\mathrm{hCH} = \mathrm{rCH}\times\mathrm{Cos}[\mathrm{aCH}];\\
		&\mathrm{hCD} = \mathrm{rCD}\times\mathrm{Cos}[\mathrm{aCH}];\\
		&\mathrm{OH} = \mathrm{Sqrt}[\mathrm{rCD}^2 - \mathrm{hCH}^2];\\
		&\mathrm{OD} = \mathrm{Sqrt}[\mathrm{rCD}^2 - \mathrm{hCD}^2];\\
		&\mathrm{ryCD} := \mathrm{rxyCDvic};\\
		&\mathrm{rzCD} := \mathrm{rzCDvic};\\
	\end{aligned}
\end{equation*}
before definition of the atoms coordinates:
\begin{equation*}
	\begin{aligned}
	\mathrm{Coordinates} =& \{\{0,0,0\},\\
	  & \{0, -\mathrm{OH}, \mathrm{hCH} \}, \\
	 &\{ (\mathrm{Sqrt}[3]/2) \mathrm{OD}, 2 \mathrm{OD}/2, \mathrm{hCD} \}, \\
	& \{ -(\mathrm{Sqrt}[3]/2) \mathrm{OD}, 2 \mathrm{OD}/2, \mathrm{hCD} \}, \\
	& \{0,\mathrm{ryCD}, \mathrm{rzCD} \};
	\end{aligned}
\end{equation*}
The carbon-13 is set at the origin of the Cartesian axis system, the \Proton~is in the \textit{Oyz} plan, as is the vicinal deuterium, which position is determined by two unknown (later optimized) variables describing its position along axes \textit{Oy} and \textit{Oz} ($\mathrm{ryCD}$ and $\mathrm{rzCD}$, respectively). The two deuterium nuclei of the methyl group are mirror image of one another with respect to the \textit{Oyz} plane.\\
We define a System Frame with z-axis along the symmetry axis of the methyl group, \textit{i.e.} the \textit{Oz} axis:
\begin{equation*}
	\mathrm{SF} = \{0,0,1\};
\end{equation*}
The orientation of the interactions relative to the System Frame is important when studyng the dynamics of the methyl groups, in particular their rotation around the symmetry axis, and are used in the definition of the spectral density function (see main text). \\
We only consider the CSA for the carbon-13 nucleus, assumed to be axially symmetric:
\begin{equation*}
	\mathrm{CSAConsidered} = \{1,0, 0, 0, 0\};
\end{equation*}
with value $\mathrm{CSAValue}$ which will be a variable optimized during the analysis of relaxation data:
\begin{equation*}
	\delta_{csa} [1] = \mathrm{CSAValue};
\end{equation*}
and oriented along the CC bond (\textit{i.e.} the symmetry axis):
\begin{center}
vectorNum$^{\text{"CSA"}}_ 1$ = \{0, 0, 1\};
\end{center}
Finally, we consider the quadrupolar interaction of the methyl deuterium nuclei, but not for the vicinal deuterium \cite{Mittermaier_JACS_1999}:
\begin{equation*}
	\begin{aligned}
		\Quadr[1] &=0; \\
		\Quadr[2] &=0; \\
		\Quadr[3] &=167000*2*\pi; \\
		\Quadr[4] &=167000*2*\pi; \\
		\Quadr[5] &=0; \\
	\end{aligned}
\end{equation*}
and we define the orientations of the considered quadrupolar interactions:
\begin{center}
	\begin{tabular}{l}
		vectorNum$^{\text{"Quad"}}_3$= Vec["CA", "DA"]; \\
		vectorNum$^{\text{"Quad"}}_4$= Vec["CA", "DB"]; \\
		vectorNum$^{\text{"Quad"}}_5$= \{0, 0, 0\}; 
	\end{tabular}
\end{center}
where the command \textit{Vec} extracts the vector between the two entries (the two nuclei). In the following analytical expressions of relaxation rates, the intensity of the quadrupolar interaction will be labelled $\zeta_{\mathcal{Q}}$.
\subsection{Spectral density function}
We used the same spectral density function written in Eq.\,\ref{eq:SpecMethyl} of the main text. We assumed the vicinal deuterium nucleus follows the same model of spectral density function, even if it is not sensitive to the rotation of the methyl group as the \Carbon, \Proton~and deuterium nuclei are. Note that the two parameters used to position the effective vicinal deuterium nucleus change the effect of the methyl group rotation on relative correlation functions.
\subsection{Relaxation matrix}
The longitudinal relaxation rates measured during the relaxometry experiment correspond to the operator $\hat{\mathrm{C}}_z$. Thus:
\begin{center}
OperatorOfInterest = opI["CA", "z"];
\end{center}
The basis contains 11,664 terms, and is first reduced to 24 terms, as detailed in the main text. Calculations shows that the decays of the $\hat{\mathrm{C}}_z$ longitudinal polarization is well described using the subspace $\left\{\frac{\hat{C}_z}{3\sqrt{3}}, \frac{\hat{H}_z}{3\sqrt{3}}, \frac{2\hat{C}_z\hat{H}_z}{3\sqrt{3}}\right\}$, as detailed in the main text. The relaxation matrix is computed using this basis.
\subsection{Relaxation rates}
During the course of the analysis of \ubiquitin~dynamics, \Carbon~and \Proton~longitudinal relaxation rates, \Carbon~transverse relaxation rate and \Carbon-\Proton~cross-relaxation rates were measured. This leads to:
\begin{center}
	\begin{tabular}{l}
		RatesOfInterest = \{ \\
		\{Rate[opI["HA", "z"], opI["HA", "z"]], "R1H"\}, \\
		\{Rate[opI["CA", "z"], opI["HA", "z"]], "R1C"\}, \\
		\{Rate[opI["CA", +], opI["CA", +]], "R2C"\}, \\
		\{Rate[opI["CA", "z"], opI["HA", "z"]], "Sigma"\}\}
	\end{tabular}
\end{center}
\subsection{Export}
Export has to be done carefully as the introduction of numerically unknown positions for the vicinal deuterium introduces complications when automatically detecting the variables of the system (important in order to calculate the derivatives). This has to be corrected manually within \RedKite.
\subsection{Conversion to a FunctionsFile}
When defining the \Carbon-CSA, it was chosen to keep it as a variable that would be further optimized during the analysis of the relaxometry relaxation rates.
\section{Expression of the relaxation matrix in the reduced basis}
\subsection{Relaxation matrix}
Operators in the secularized basis are:
\begin{equation}
	\begin{aligned}
\mathcal{B}_\mathrm{secularized} =& \left\{\frac{\hat{C}_z}{3\sqrt{3}}, \frac{\hat{H}_z}{3\sqrt{3}}, \frac{2\hat{C}_z\hat{H}_z}{3\sqrt{3}}, \frac{\sqrt{2}\hat{C}_z\hat{H}_z\hat{D}_{1,z}}{3}, \frac{\sqrt{2}\hat{C}_z\hat{H}_z\hat{D}_{2,z}}{3},  \frac{\hat{C}_z\hat{D}_{1,z}}{3\sqrt{3}}, \frac{\hat{C}_z\hat{D}_{2,z}}{3\sqrt{3}}, \right. \\
& \left.  \frac{\hat{D}_{1,z}}{6\sqrt{2}}, \frac{\hat{D}_{2,z}}{6\sqrt{2}}, \frac{\hat{C}_z\hat{D}_{1}^{-}\hat{D}_{2}^{+}}{4\sqrt{3}}, \frac{\hat{C}_z\hat{D}_{1}^{+}\hat{D}_{2}^{-}}{4\sqrt{3}},  \frac{\hat{C}_z\hat{D}_{1,z}\hat{D}_{2,z}}{2\sqrt{3}}, \frac{3\hat{C}_z\hat{D}_{1,z} \hat{D}_{1,z} - 2\hat{C}_z}{3\sqrt{6}}, \right. \\
& \left. \frac{3\hat{C}_z\hat{D}_{2,z} \hat{D}_{2,z} - 2\hat{C}_z}{3\sqrt{6}}\right\}.
	\end{aligned}
\end{equation}
Note that numerical simulations were carried out in a reduced basis formed with elements $\frac{\hat{C}_z}{3 \sqrt{3}}$, $\frac{\hat{H}_z}{3 \sqrt{3}}$ and $\frac{2\hat{C}_z \hat{H}_z}{3 \sqrt{3}}$ of the secularized basis. The relaxation matrix is:
\begin{equation*}
\rotatebox{90}{$\mathcal{R}=\begin{bmatrix}
    \begin{array}{@{}*{14}{c}@{}}
	R_1(^{13}\mathrm{C}) & \sigma_\mathrm{CH} & \eta_z^\mathrm{C} & \kappa^\mathrm{C} & \kappa^\mathrm{C} & \eta_z^\mathrm{CD} & \eta_z^\mathrm{CD} & \sigma_\mathrm{CD} & \sigma_\mathrm{CD} & \lambda & \lambda & \nu_z & \mu & \mu\\
    \sigma_\mathrm{CH} & R_1(^{1}\mathrm{H}) & 0 & \kappa^\mathrm{H} & \kappa^\mathrm{H} & 0 & 0 & \sigma_\mathrm{HD} & \sigma_\mathrm{HD} & 0 & 0 & 0 & 0 & 0\\
    \eta_z^\mathrm{C} & 0 & R_\mathrm{CH} & \kappa^\mathrm{CH} & \kappa^\mathrm{CH} & \delta & \delta & 0 & 0 & 0 & 0 & 0 & 0 & 0\\
    \kappa^\mathrm{C} & \kappa^\mathrm{H} & \kappa^\mathrm{CH} & R_\mathrm{CHD} & \kappa^\mathrm{CHD} & \eta_z^\mathrm{CHD} & 0 & \sigma_\mathrm{CHD} & 0 & \lambda^{(1)} & \lambda^{(1)} & \nu_z^{(1)} & \mu^{(1)} & 0\\
    \kappa^\mathrm{C} & \kappa^\mathrm{H} & \kappa^\mathrm{CH} & \kappa^\mathrm{CHD} & R_\mathrm{CHD} & 0 & \eta_z^\mathrm{CHD} & 0 & \sigma_\mathrm{CHD} & \lambda^{(1)} & \lambda^{(1)} & \nu_z^{(1)} & 0 & \mu^{(1)}\\
    \eta_z^\mathrm{CD} & 0 & \delta & \eta_z^\mathrm{CHD} & 0 & R_\mathrm{CD} & \kappa^\mathrm{CD} & 0 & 0 & 0 & 0 & \nu_z^{(2)} & \mu^{(2)}& 0\\
    \eta_z^\mathrm{CD} & 0 & \delta & 0 & \eta_z^\mathrm{CHD} & \kappa^\mathrm{CD} & R_\mathrm{CD} & 0 & 0 & 0 & 0 & \nu_z^{(2)} & 0 & \mu^{(2)}\\
    \sigma_\mathrm{CD} & \sigma_\mathrm{HD} & 0 & \sigma_\mathrm{CHD} & 0 & 0 & 0 & R_\mathrm{D} & \sigma_\mathrm{DD} & \lambda^{(2)} & \lambda^{(2)} & \nu_z^{(3)} & \mu^{(3)} & 0\\
    \sigma_\mathrm{CD} & \sigma_\mathrm{HD} & 0 & 0 & \sigma_\mathrm{CHD} & 0 & 0 & \sigma_\mathrm{DD} & R_\mathrm{D} & \lambda^{(2)} & \lambda^{(2)} & \nu_z^{(3)} & 0 & \mu^{(3)}\\
    \lambda & 0 & 0 & \lambda^{(1)} & \lambda^{(1)} & 0 & 0 & \lambda^{(2)} & \lambda^{(2)} & R_\mathrm{CDD}^{(1)} & \kappa^\mathrm{CDD} & \nu_z^{(4)} & \mu^{(4)} & \mu^{(4)}\\
    \lambda & 0 & 0 & \lambda^{(1)} & \lambda^{(1)} & 0 & 0 & \lambda^{(2)} & \lambda^{(2)} & \kappa^\mathrm{CDD} & R_\mathrm{CDD}^{(1)} & \nu_z^{(4)} & \mu^{(4)} & \mu^{(4)} \\
    \nu_z & 0 & 0 & \nu_z^{(1)} & \nu_z^{(1)} & \nu_z^{(2)} & \nu_z^{(2)} & \nu_z^{(3)} & \nu_z^{(3)} & \nu_z^{(4)} & \nu_z^{(4)} & R_\mathrm{CDD}^{(2)} & \mu^{(5)} & \mu^{(5)} \\
    \mu & 0 & 0 & \mu^{(1)} & 0 & \mu^{(2)}& 0 & \mu^{(3)} & 0 & \mu^{(4)} & \mu^{(4)} & \mu^{(5)} & R & 0\\
    \mu& 0 & 0& 0 & \mu^{(1)}& 0 & \mu^{(2)} & 0 & \mu^{(3)} & \mu^{(4)} & \mu^{(4)} & \mu^{(5)} & 0 & R
   \end{array}
	\end{bmatrix}$}
\end{equation*}

\subsection{Auto-relaxation rates}
\begin{eqnarray*}
	R_1 (\carb) &=& \frac{2}{3} \Delta \sigma_C^2\omegaC^2  \Jimpu{C} (\omegaC)  \\
    &+& \frac{1}{2} d_\mathrm{CH}^2  \left( \Jimpu{CH}(\omegaC-\omegaH) + 3 \Jimpu{CH}(\omegaC) + 6 \Jimpu{CH}(\omegaC+\omegaH)  \right) \\
    &+& \frac{8}{3} d_\mathrm{CD}^2 \left( \Jimpu{CD}(\omegaC-\omegaD)+ 3 \Jimpu{CD}(\omegaC)  + 6 \Jimpu{CD}(\omegaC+\omegaD) \right) \\
    &+& \frac{4}{3} d_\mathrm{CD_\mathrm{vic}}^2 \left( \Jimpu{CD_\mathrm{vic}}(\omegaC-\omegaD) + 3 \Jimpu{CD_\mathrm{vic}}(\omegaC) + 6 \Jimpu{CD_\mathrm{vic}}(\omegaC+\omegaD) \right), \\
   R_1 (^{1}\mathrm{H}) &=& \frac{1}{2} d_\mathrm{CH}^2 ( \Jimpu{CH}(\omegaC-\omegaH) + 3 \Jimpu{CH}(\omegaH) + 6 \Jimpu{CH}(\omegaC+\omegaH)  ) \\
    &+& \frac{8}{3} d_\mathrm{HD}^2 ( \Jimpu{HD}(\omegaD-\omegaH) + 3 \Jimpu{HD}(\omegaH) + 6 \Jimpu{HD}(\omegaD+\omegaH) ) \\
    &+& \frac{4}{3} d_\mathrm{HD_\mathrm{vic}}^2 ( \Jimpu{HD_\mathrm{vic}}(\omegaD-\omegaH) + 3 \Jimpu{HD_\mathrm{vic}}(\omegaH) + 6 \Jimpu{HD_\mathrm{vic}}(\omegaD+\omegaH) ), \\
R_\mathrm{CH} &=& \frac{2}{3} \Delta \sigma_C^2 \omegaC^2 \Jimpu{C} (\omegaC) + \frac{3}{2} d_\mathrm{CH}^2 \left(\Jimpu{CH}(\omegaC) + \Jimpu{CH}(\omegaH) \right)  \\
&+& \frac{8}{3} d_\mathrm{CD}^2 \left(\Jimpu{CD}(\omegaC-\omegaD) + 3 \Jimpu{CD}(\omegaC) + 6 \Jimpu{CD}(\omegaC + \omegaD) \right)  \\
&+& \frac{8}{3} d_\mathrm{HD}^2 \left(\Jimpu{HD}(\omegaH-\omegaD) + 3 \Jimpu{HD}(\omegaH) + 6 \Jimpu{HD}(\omegaH + \omegaD) \right) \\
&+& \frac{4}{3} d_\mathrm{CD_\mathrm{vic}}^2 \left(\Jimpu{CD_{vic}}(\omegaC - \omegaD) + 3 \Jimpu{CD_{vic}}(\omegaC) + 6 \Jimpu{CD_{vic}}(\omegaC + \omegaD) \right)   \\
&+& \frac{4}{3} d_\mathrm{HD_\mathrm{vic}}^2 \left(\Jimpu{HD_{vic}}(\omegaH - \omegaD) + 3 \Jimpu{HD_{vic}}(\omegaH) + 6 \Jimpu{HD_{vic}}(\omegaH + \omegaD) \right),  \\
	R_\mathrm{CHD} &=& \frac{3}{8} \zeta_{\mathcal{Q}}^2 \left(\Jimpu{\mathcal{Q}}(\omegaD) + 8 \Jimpu{\mathcal{Q}} (2 \omegaD) \right) + \frac{2}{3}\Delta \sigma_C^2 \omegaC^2 \Jimpu{C}(\omegaC) \\
	&+& \frac{4}{3} d_\mathrm{DD}^2 ( \Jimpu{DD}(0) + 3 \Jimpu{DD}(\omegaD) + 6 \Jimpu{DD} (2 \omegaD)) + 3 d_\mathrm{CH}^2 (\Jimpu{CH}(\omegaC)+\Jimpu{CH}(\omegaH)) \\
    &+& \frac{1}{6} d_\mathrm{CD}^2 ( 11 \Jimpu{CD}(\omegaC-\omegaD) + 9 \Jimpu{CD}(\omegaD) + 60 \Jimpu{CD}(\omegaC) + 66 \Jimpu{CD}(\omegaC+\omegaD)) \\
    &+& \frac{1}{6} d_\mathrm{HD}^2 (11 \Jimpu{HD}(\omegaH-\omegaD) + 9 \Jimpu{HD}(\omegaD) + 60 \Jimpu{HD}(\omegaH) + 66 \Jimpu{HD}(\omegaH+\omegaD)) \\
    &+& \frac{4}{3} d_\mathrm{DD_\mathrm{vic}}^2 (\Jimpu{DD_{vic}}(0) + 3 \Jimpu{DD_{vic}}(\omegaD) + 6 \Jimpu{DD_{vic}} (2 \omegaD)) \\
    &+& \frac{4}{3} d_\mathrm{CD_\mathrm{vic}}^2 (3 \Jimpu{CD_{vic}}(\omegaC) + \Jimpu{CD_{vic}}(\omegaC-\omegaD) + 6 \Jimpu{CD_{vic}}(\omegaC+\omegaD)) \\
    &+& \frac{4}{3} d_\mathrm{HD_\mathrm{vic}}^2 (3 \Jimpu{HD_{vic}}(\omegaH) + \Jimpu{HD_{vic}}(\omegaH-\omegaD) + 6 \Jimpu{HD_{vic}}(\omegaH+\omegaD)),
\end{eqnarray*}
\begin{eqnarray*}
    R_\mathrm{CD} &=& \frac{3}{8} \zeta_{\mathcal{Q}}^2 \left(\Jimpu{\mathcal{Q}}(\omegaD) + 4 \Jimpu{\mathcal{Q}} (2 \omegaD) \right) + \frac{2}{3}\Delta \sigma_C^2 \omegaC^2 \Jimpu{C}(\omegaC) \\
&+& \frac{1}{2} d_\mathrm{CH}^2 (\Jimpu{CH}(\omegaC-\omegaH) + 3 \Jimpu{CH}(\omegaC) + 6 \Jimpu{CH}(\omegaC+\omegaH))\\
    &+&  \frac{1}{2} d_\mathrm{HD}^2 (\Jimpu{HD}(\omegaH-\omegaD) + 3 \Jimpu{HD}(\omegaD) + 6 \Jimpu{HD}(\omegaH+\omegaD)) \\
&+& \frac{4}{3} d_\mathrm{DD}^2 (\Jimpu{DD}(0) + 3 \Jimpu{DD}(\omegaD) + 6 \Jimpu{DD}(2 \omegaD)) \\
    &+& \frac{1}{6} d_\mathrm{CD}^2 (11 \Jimpu{CD}(\omegaC-\omegaD) + 9 \Jimpu{CD}(\omegaD) + 60 \Jimpu{CD}(\omegaC) + 66 \Jimpu{CD}(\omegaC+\omegaD)) \\
    &+&  \frac{4}{3} d_\mathrm{DD_\mathrm{vic}}^2 (\Jimpu{DD_{vic}}(0) + 3 \Jimpu{DD_\mathrm{vic}}(\omegaD) + 6 \Jimpu{DD_\mathrm{vic}}(2 \omegaD)) \\
    &+& \frac{4}{3} d_\mathrm{CD_\mathrm{vic}} (\Jimpu{CD_\mathrm{vic}}(\omegaC-\omegaD) + 3 \Jimpu{CD_\mathrm{vic}}(\omegaC) + 6 \Jimpu{CD_\mathrm{vic}}(\omegaC+\omegaD)), \\
    R_\mathrm{D} &=& \frac{3}{8} \zeta_{\mathcal{Q}}^2 (\Jimpu{\mathcal{Q}}(\omegaD) + 4 \Jimpu{\mathcal{Q}}(2 \omegaD)) + \frac{4}{3} d_\mathrm{DD}^2 (\Jimpu{DD}(0) + 3 \Jimpu{DD}(\omegaD) + 6 \Jimpu{DD}(2 \omegaD)) \\
    &+& \frac{1}{2} d_\mathrm{CD}^2 (\Jimpu{CD}(\omegaC-\omegaD) + 3 \Jimpu{CD}(\omegaD) + 6 \Jimpu{CD}(\omegaC+\omegaD)) \\
    &+& \frac{1}{2} d_\mathrm{HD}^2 (\Jimpu{HD}(\omegaH-\omegaD) + 3 \Jimpu{HD}(\omegaD) + 6 \Jimpu{HD}(\omegaH+\omegaD)) \\
    &+& \frac{4}{3} d_\mathrm{DD_\mathrm{vic}}^2 (\Jimpu{DD_\mathrm{vic}}(0) + 3 \Jimpu{DD_\mathrm{vic}}(\omegaD) + 6 \Jimpu{DD_\mathrm{vic}}(2 \omegaD)), \\
    R_\mathrm{CDD}^{(1)} &=& \frac{1}{24} \zeta_{\mathcal{Q}}^2 (3 \Jimpu{\mathcal{Q}}(0) + 5 \Jimpu{\mathcal{Q}}(\omegaD) + 2 \Jimpu{\mathcal{Q}}(2 \omegaD)) + \frac{3}{2} d_\mathrm{DD} \zeta_{\mathcal{Q}} (2 \Jimpd[DD]{\mathcal{Q}}(0) + 3 \Jimpd[DD]{\mathcal{Q}}(\omegaD)) \\
&+& \frac{2}{3}\Delta \sigma_C^2 \omegaC^2 \Jimpu{C}(\omegaC) + \frac{1}{2} d_\mathrm{CH}^2 (\Jimpu{CH}(\omegaC-\omegaH) + 3 \Jimpu{CH}(\omegaC) + 6 \Jimpu{CH}(\omegaC + \omegaH)) \\
    &+& \frac{1}{2} d_\mathrm{DD}^2 (7 \Jimpu{DD}(0) + 18 \Jimpu{DD}(\omegaD) + 12 \Jimpu{DD}(2 \omegaD))\\
    &+& \frac{1}{2} d_\mathrm{HD}^2 (4 \Jimpu{HD}(0) + \Jimpu{HD}(\omegaH-\omegaD) + 3 \Jimpu{HD}(\omegaD)  + 6 \Jimpu{HD}(\omegaH)+ 6 \Jimpu{HD}(\omegaH+\omegaD) ) \\
    &+& \frac{1}{2} d_\mathrm{CD}^2 (4 \Jimpu{CD}(0) + 5 \Jimpu{CD}(\omegaC-\omegaD) + 3 \Jimpu{CD}(\omegaD) + 6 \Jimpu{CD}(\omegaC) + 30 \Jimpu{CD}(\omegaC+\omegaD) ) \\
    &+& \frac{4}{3} d_\mathrm{CD_\mathrm{vic}} (\Jimpu{CD_\mathrm{vic}}(\omegaC-\omegaD) + 3 \Jimpu{CD_\mathrm{vic}}(\omegaC) + 6 \Jimpu{CD_\mathrm{vic}}(\omegaC+\omegaD)) - 2 d_\mathrm{CD}^2 \Jimpd[CD_1]{CD_2}(0)\\
    &+& \frac{4}{3} d_\mathrm{DD_\mathrm{vic}}^2 (5 \Jimpu{DD_\mathrm{vic}}(0) + 9 \Jimpu{DD_\mathrm{vic}}(\omegaD) + 6 \Jimpu{DD_\mathrm{vic}}(2 \omegaD))  \\
&-& d_\mathrm{HD}^2 (2 \Jimpd[HD_1]{HD_2}(0) + 3 \Jimpd[HD_1]{HD_2}(\omegaH)) -  \frac{8}{3} d_\mathrm{DD}^2 (2 \Jimpd[D_1D_\mathrm{vic}]{D_2D_\mathrm{vic}}(0) + 3 \Jimpd[D_1D_\mathrm{vic}]{D_2D_\mathrm{vic}}(\omegaD)),
\end{eqnarray*}
\begin{eqnarray*}
    R_{CDD}^{(2)} &=& \frac{3}{4} \zeta_{\mathcal{Q}} (\Jimpu{\mathcal{Q}}(\omegaD) + 4 \Jimpu{\mathcal{Q}}(2 \omegaD)) + \frac{2}{3} \Delta \sigma_C^2 \omegaC^2 \Jimpu{C}(\omegaC) \\
&+& \frac{1}{2} d_\mathrm{CH}^2 (\Jimpu{CH}(\omegaC-\omegaH) + 3 \Jimpu{CH}(\omegaC) + 6 \Jimpu{CH}(\omegaC+\omegaH)) \\
    &+& d_\mathrm{CD}^2 (\Jimpu{CD}(\omegaC-\omegaD) + 3 \Jimpu{CD}(\omegaD) + 6 \Jimpu{CD}(\omegaC+\omegaD) + 12 \Jimpu{CD}(\omegaC)) \\
    &+& d_\mathrm{HD}^2 (\Jimpu{HD}(\omegaH-\omegaD) + 3 \Jimpu{HD}(\omegaD) + 6 \Jimpu{HD}(\omegaH+\omegaD)) \\
&+& d_\mathrm{DD}^2 (\Jimpu{DD}(0) + 12 \Jimpu{DD}(\omegaD) + 6 \Jimpu{DD}(2\omegaD) ) \\
    &+& \frac{4}{3} d_\mathrm{CD_\mathrm{vic}}^2 (\Jimpu{CD_\mathrm{vic}}(\omegaC-\omegaD) + 3 \Jimpu{CD_\mathrm{vic}}(\omegaC) + 6 \Jimpu{CD_\mathrm{vic}}(\omegaC+\omegaD)) \\
    &+& \frac{8}{3} d_\mathrm{DD_\mathrm{vic}} (\Jimpu{DD_\mathrm{vic}}(0) + 3 \Jimpu{DD_\mathrm{vic}}(\omegaD) + 6 \Jimpu{DD_\mathrm{vic}}(2\omegaD)), \\
    R &=& \frac{9}{8} \zeta_{\mathcal{Q}}^2 \Jimpu{\mathcal{Q}}(\omegaD) + \frac{2}{3}\Delta \sigma_C^2 \omegaC^2 \Jimpu{C}(\omegaC) \\
&+& \frac{1}{2} d_\mathrm{CH}^2 (\Jimpu{CH}(\omegaC-\omegaH) + 3 \Jimpu{CH}(\omegaC) + 6 \Jimpu{CH}(\omegaC+\omegaH)) \\
    &+& \frac{3}{2} d_\mathrm{CD}^2 (\Jimpu{CD}(\omegaC-\omegaD) + 3 \Jimpu{CD}(\omegaD) + 4 \Jimpu{CD}(\omegaC) + 6 \Jimpu{CD}(\omegaC+\omegaD)) \\
    &+& \frac{3}{2} d_\mathrm{HD}^2 (\Jimpu{HD}(\omegaH-\omegaD) + 3 \Jimpu{HD}(\omegaD) + 6 \Jimpu{HD}(\omegaH+\omegaD)) \\
&+& 4 d_\mathrm{DD}^2 (\Jimpu{DD}(0) + 3 \Jimpu{DD}(\omegaD) + 6 \Jimpu{DD}(2 \omegaD) \\
    &+& \frac{4}{3} d_\mathrm{CD_\mathrm{vic}}^2 (\Jimpu{CD_\mathrm{vic}}(\omegaC-\omegaD) + 3 \Jimpu{CD_\mathrm{vic}}(\omegaC) + 6 \Jimpu{CD_\mathrm{vic}}(\omegaC+\omegaD)) \\
    &+& 4 d_\mathrm{DD_\mathrm{vic}}^2(\Jimpu{DD_\mathrm{vic}}(0) + 3 \Jimpu{DD_\mathrm{vic}}(\omegaD) + 6 \Jimpu{DD_\mathrm{vic}}(2 \omegaD).
\end{eqnarray*}

\subsection{Cross-relaxation rates}
Cross-relaxation rates with the operator $\hat{C}_z$ are:
\begin{eqnarray*}
\begin{aligned}
	\sigma_\mathrm{CH} =& \frac{1}{2} d_\mathrm{CH}^2 (-\Jimpu{CH}(\omegaC-\omegaH) + 6 \Jimpu{CH}(\omegaC+\omegaH)), \\
	\eta_z^\mathrm{C} =& - 2 \Delta \sigma_C \omegaC d_\mathrm{CH} \Jimpd[C]{CH} (\omegaC), \\
	\kappa^\mathrm{C} =& 2 \sqrt{6} d_\mathrm{CH} d_\mathrm{CD} \Jimpd[CH]{CD} (\omegaC), \\
	\eta_z^\mathrm{CD} =& - 4 \sqrt{\frac{2}{3}} d_\mathrm{CD}\Delta \sigma_C \omegaC \Jimpd[C]{CD}(\omegaC),\\
	\sigma_\mathrm{CD} =& \sqrt{\frac{2}{3}} d_\mathrm{CD}^2 (- \Jimpu{CD}(\omegaC-\omegaD) + 6 \Jimpu{CD}(\omegaC+\omegaD)),\\
	\lambda =& \frac{4}{3} d_\mathrm{CD}^2 (\Jimpd[CD]{CD}(\omegaC-\omegaD) + 6 \Jimpd[CD]{CD}(\omegaC+\omegaD)),\\
	\nu_z =& 8 d_\mathrm{CD}^2 \Jimpd[CD]{CD}(\omegaC),\\
	\mu =& \frac{\sqrt{2}}{3} d_\mathrm{CD}^2 \left(- \Jimpu{CD}(\omegaC-\omegaD) + 6 \Jimpu{CD}(\omegaC)  - 6 \Jimpu{CD}(\omegaC+\omegaD) \right).
\end{aligned}
\end{eqnarray*}

Finally, other cross-relaxation rates are:
\begin{equation*}
	\begin{aligned}
	\kappa^H =& 2 \sqrt{6} d_\mathrm{CH} d_\mathrm{HD} \Jimpd[CH]{HD} (\omegaH),\\
    \kappa^\mathrm{CH} =& - 4 \sqrt{\frac{2}{3}} d_\mathrm{CD} \Delta \sigma_C \omegaC  \Jimpd[CD]{CC} (\omegaC),\\
    \kappa^\mathrm{CD} =& 8 d_\mathrm{CD}^2 \Jimpd[CD_1]{CD_2}(\omegaC) - \frac{4}{3} d_{DD}^2(\Jimpu{DD}(0) - 6 \Jimpu{DD}(2\omegaD)),\\
    \kappa^\mathrm{CHD} =& 8 d_\mathrm{CD}^2 \Jimpd[CD_1]{CD_2}(\omegaC) + 8 d_\mathrm{HD}^2 \Jimpd[HD_1]{HD_2}(\omegaH) - \frac{4}{3} d_\mathrm{DD}^2 (\Jimpu{DD}(0) - 6 \Jimpu{DD}(2 \omegaD)),\\
    \kappa^\mathrm{CDD} =& -\frac{3}{2} d_\mathrm{DD}^2 \Jimpu{DD}(0), \\
    \sigma_\mathrm{HD} =& \sqrt{\frac{2}{3}} d_\mathrm{HD}^2 (- \Jimpu{HD}(\omegaH-\omegaD) + 6 \Jimpu{HD}(\omegaH+\omegaD)),\\
    \sigma_\mathrm{DD} =& \frac{4}{3} d_\mathrm{DD}^2 (- \Jimpu{DD}(0) + 6 \Jimpu{DD}(2 \omegaD)),\\
    \sigma_\mathrm{CHD} =& 3 d_\mathrm{CD} d_\mathrm{HD} \Jimpd[CD]{HD}(\omegaD),\\
    \eta_z^\mathrm{CHD} =& - 2 d_{CH} \Delta \sigma_C \omegaC \Jimpd[C]{CH}(\omegaC),\\
    \delta =& \frac{12}{\sqrt{6}} d_\mathrm{CH} d_\mathrm{CD} \Jimpd[CH]{CD}(\omegaC) - \frac{2}{\sqrt{6}} d_\mathrm{HD}^2 (\Jimpu{HD}(\omegaH-\omegaD) - 6 \Jimpu{HD}(\omegaH+\omegaD)),
	\end{aligned}
\end{equation*}
\begin{equation*}
	\begin{aligned}
    \lambda^{(1)} =& \sqrt{\frac{1}{6}} d_\mathrm{HD}^2 (\Jimpd[HD_1]{HD_2}(\omegaH-\omegaD) - 6 \Jimpd[HD_1]{HD_2}(\omegaH+\omegaD)) \\
	&+ \frac{2}{\sqrt{6}} d_\mathrm{HD} d_\mathrm{DD} ( 2 \Jimpd[HD]{DD}(0) - 3 \Jimpd[HD]{DD}(\omegaD) ),\\
    \lambda^{(2)} =&  - \frac{2}{\sqrt{6}} d_\mathrm{CD} d_\mathrm{DD} (2 \Jimpd[CD]{DD}(0) + 3 \Jimpd[CD]{DD}(\omegaD)),\\
    \nu_z^{(1)} =& -\sqrt{\frac{2}{3}} d_\mathrm{HD}^2 (\Jimpu{HD}(\omegaH-\omegaD) - 6 \Jimpu{HD}(\omegaH+\omegaD)) \\
	&+ \sqrt{3} \left( d_\mathrm{CH} d_\mathrm{CD} \Jimpd[CH]{CD}(\omegaC) + d_\mathrm{HD} d_\mathrm{DD} \Jimpd[HD]{DD}(\omegaD) \right), \\
    \nu_z^{(2)} =& - 4 \sqrt{\frac{2}{3}} d_\mathrm{CD} \Delta \sigma_C \omegaC \Jimpd[C]{CD}(\omegaC), \\
    \nu_z^{(3)} =& 2 \sqrt{6} d_\mathrm{CD} d_\mathrm{DD} \Jimpd[CD]{DD}(\omegaD), \\
    \nu_z^{(4)} =& d_\mathrm{DD}^2 (\Jimpu{DD}(0) - 3 \Jimpu{DD}(\omegaD)) - \frac{3}{4} d_\mathrm{DD} \zeta_{\mathcal{Q}} (\Jimpd[DD]{\mathcal{Q}}(0) - 3 \Jimpd[DD]{\mathcal{Q}}(\omegaD) + 6 \Jimpd[DD]{\mathcal{Q}}(2 \omegaD)) \\
    &- \frac{3}{2} d_\mathrm{CD}^2 \Jimpd[CD]{CD}(\omegaD) - \frac{1}{2} d_\mathrm{HD}^2 (\Jimpd[HD]{HD}(\omegaH-\omegaD) + 3 \Jimpd[HD]{HD}(\omegaD) + 6 \Jimpd[HD]{HD}(\omegaH+\omegaD)) \\
    &- \frac{4}{3} d_\mathrm{DD_\mathrm{vic}}^2 (\Jimpd[D_1D_\mathrm{vic}]{D_2D_\mathrm{vic}}(0) + 3 \Jimpd[D_1D_\mathrm{vic}]{D_2D_\mathrm{vic}}(\omegaD) + 6 \Jimpd[D_1D_\mathrm{vic}]{D_2D_\mathrm{vic}}(2 \omegaD)),\\
    \mu^{(1)} =& -\frac{\sqrt{3}}{2} d_\mathrm{HD}^2 (\Jimpu{HD}(\omegaH-\omegaD) - 6 \Jimpu{HD}(\omegaH+\omegaD)) + 2\sqrt{3} d_\mathrm{CH} d_\mathrm{CD} \Jimpd[CH]{CD}(\omegaC) \\
&- \frac{3 \sqrt{3}}{2} d_\mathrm{HD} \zeta_{\mathcal{Q}} \Jimpd[\mathcal{Q}]{HD}(\omegaD),\\
    \mu^{(2)} =& - \frac{4}{\sqrt{3}} d_\mathrm{CD} \Delta \sigma_C \omegaC \Jimpd[C]{CD}(\omegaC), \\
    \mu^{(3)} =& \frac{\sqrt{3}}{6} d_\mathrm{CD}^2 (\Jimpu{CD}(\omegaC-\omegaD) - 6 \Jimpu{CD}(\omegaC+\omegaD)) - \frac{3 \sqrt{3}}{2} d_\mathrm{CD} \zeta_{\mathcal{Q}} \Jimpd[\mathcal{Q}]{CD}(\omegaD), \\
    \mu^{(4)} =& -\frac{\sqrt{2}}{2} d_\mathrm{DD}^2 (2 \Jimpu{DD}(0) +3 \Jimpu{DD}(\omegaD)) - \frac{\sqrt{2}}{3} d_\mathrm{CD}^2 (\Jimpd[CD]{CD}(\omegaC-\omegaD) + 6 \Jimpd[CD]{CD}(\omegaC+\omegaD)) \\
    +& \frac{3}{4 \sqrt{2}} d_\mathrm{DD} \zeta_{\mathcal{Q}} (\Jimpd[\mathcal{Q}]{DD}(0) + \Jimpd[\mathcal{Q}]{DD}(\omegaD) - 2 \Jimpd[\mathcal{Q}]{DD}(2 \omegaD)),\\
    \mu^{(5)} =& -\sqrt{2} d_\mathrm{DD}^2 (\Jimpu{DD}(0) - 6 \Jimpu{DD}(2 \omegaD)) + 4 \sqrt{2} d_\mathrm{CD}^2 \Jimpd[CD_1]{CD_1}(\omegaC) - 3\sqrt{2} d_\mathrm{DD} \zeta_{\mathcal{Q}} \Jimpd[\mathcal{Q}]{DD} (\omegaD).
	\end{aligned}
\end{equation*}

\newpage
\section{Figures}
\begin{figure*}[!ht]
\begin{center}
	\includegraphics[width=0.8\textwidth]{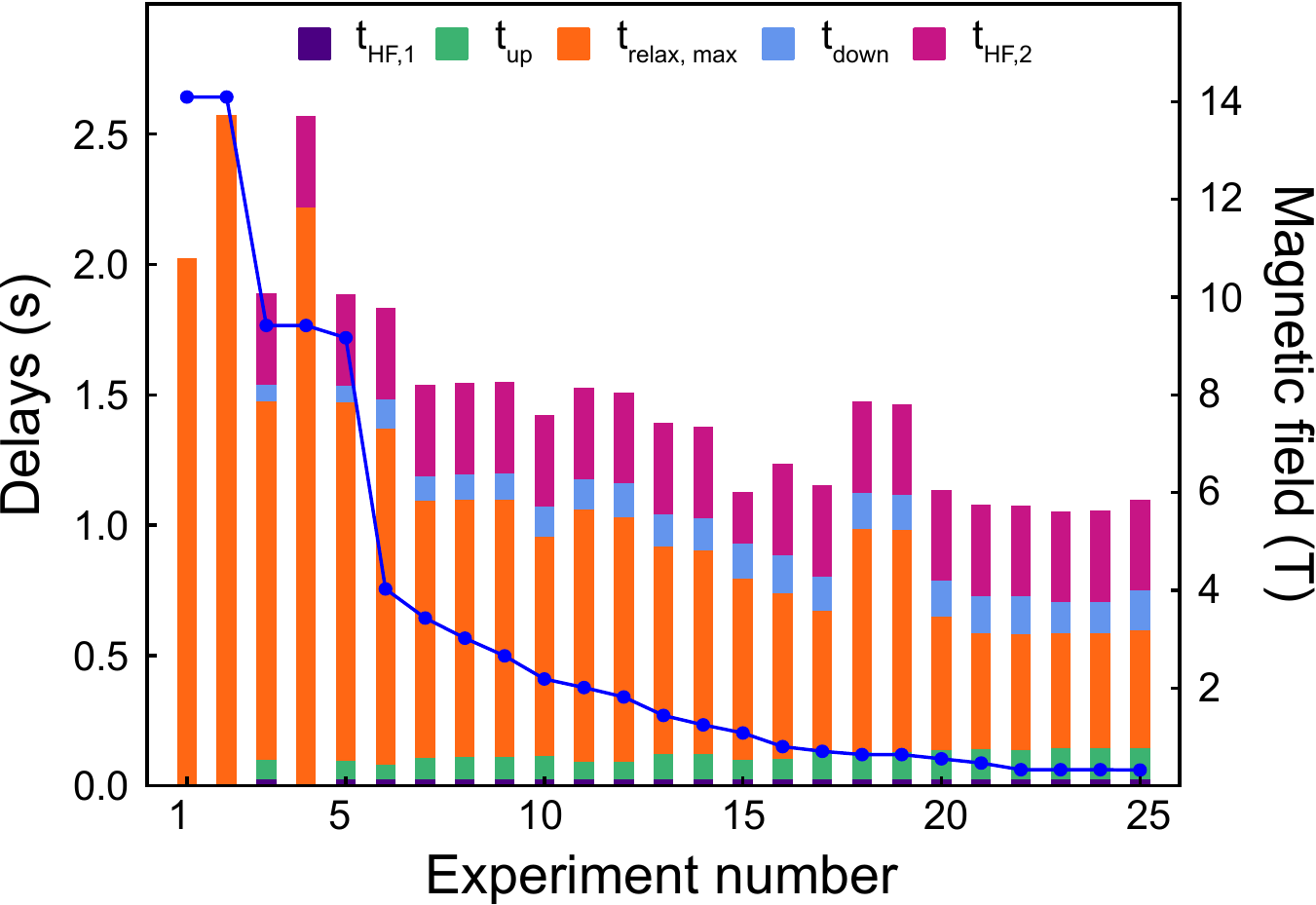}
\end{center}
\caption{Experimental delays for the 25 experiments used in the analysis of the dynamics of isoleucine-$\delta$1-methyl groups of Ubiquitin, and ordered from the highest magnetic field at which relaxation takes place to the lowest. The time labels refer to the decomposition of the free-relaxation part of the pulse-sequence, as shown in Fig.\,\ref{fig:PulseSequence} of the main text. The blue curve (right y-axis) shows the variation of the magnetic field for each experiment (associated with an increase of shuttling height). Experiments 1, 2 and 4 were performed on high-field spectrometers, with no shuttle.}
\label{fig:ExpDelays}
\end{figure*}

\begin{figure*}[!ht]
\begin{center}
	\includegraphics[width=0.8\textwidth]{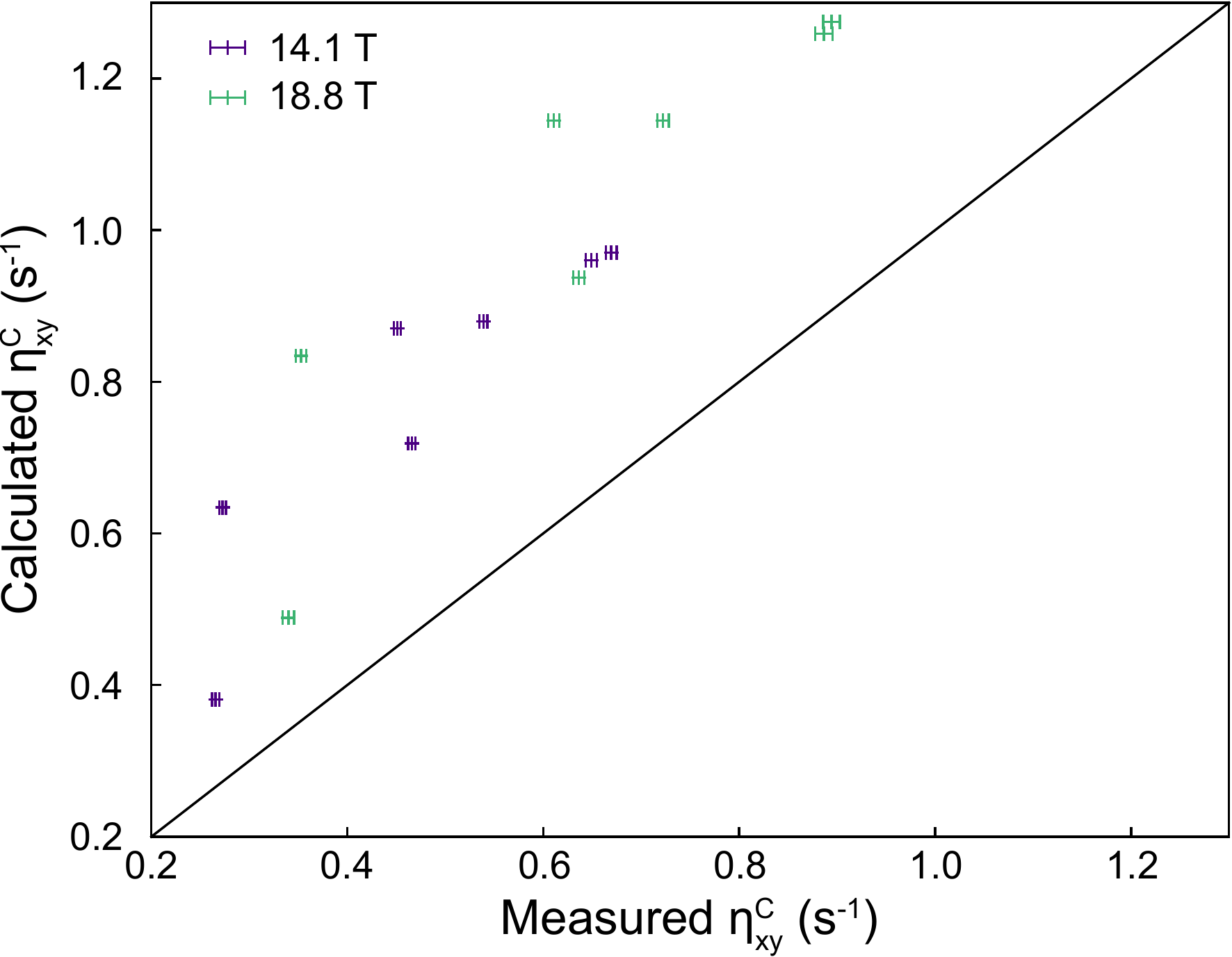}
\end{center}
\caption{Correlation plot between the calculated and measured transverse CSA/DD cross-correlated cross-relaxation rates at 14.1\,T and 18.8\,T, with no scaling of the CSA.}
\label{fig:EtaX}
\end{figure*}

\newpage
\section{Hamiltonian in \textsc{RedKite}}
We report here the definition of the Hamiltonian as written in \textsc{RedKite}. Constants are defined in Table\,\ref{table:VariableExplain}.\\
For the dipolar interaction:
\begin{center}
HDD[i\_, j\_, t\_] := $\sqrt{6}$ dDD[Nuclei[[i, 2]],Nuclei[[j, 2]]] $\times$ Sum[
 $(-1)^m$ M[m, opTDipFreq[\{Nuclei[[i,1]],Nuclei[[j,1]]\},\{-m,k\}], t,$\Phi$[Nuclei[[i,2]], Nuclei[[j,2]]]] 
opTDip[\{Nuclei[[i,1]],Nuclei[[j,1]]\},\{-m,k\}], 
 \{m, -2, 2\}], \{k, Min[0, Abs[m]-1], Min[1,  2 - Abs[m]]\}]; \\
HDDtot[t\_] := Sum[HDD[i,j,t], \{i, 1,NumberofAtoms-1\}, \{j, i+1,NumberofAtoms\}];
\end{center}
For the CSA interaction, in the case of an axially symmetric tensor:
\begin{center}
HCSA[t\_] := Sum[CSAConsidered[[n]] Sum[
 $(-1)^m \Delta_{\mathrm{Nuclei}[[n, 2]]}$ M[m, opTCSAFreq[Nuclei[[n,1]],\{-m,0\}], t,AngleCSA[[n, 1]]] 
opTCSA[Nuclei[[n,1]],\{-m,0\}], \{m, -2, 2\}], \{n, 1, NumberofAtoms\}];
\end{center}
and for an asymmetric tensor:
\begin{center}
HCSA[t\_] := Sum[CSAConsidered[[n]] Sum[
 $(-1)^m \bigg(\sigma \mathrm{ln}_{\mathrm{Nuclei}[[n, 2]]}$ M[m, opTCSAFreq[Nuclei[[n,1]],\{-m,0\}], t,AngleCSA[[n, 1]]] +  
 $\sigma \mathrm{pn}_{\mathrm{Nuclei}[[n, 2]]}$ M[m, opTCSAFreq[Nuclei[[n,1]],\{-m,0\}], t,AngleCSA[[n, 2]]] \bigg)
 opTCSA[Nuclei[[n,1]],\{-m,0\}],  \{m, -2, 2\}], \{n, 1, NumberofAtoms\}];
\end{center}
and for the quadrupolar interaction:
\begin{center}
HQuad[i\_, t\_] := $\frac{d_\mathcal{Q}[\mathrm{AtomsQuadConsidered}[[i, 2]]}{4 \mathrm{QuantumNumberConsidered}[[i]] (2 \mathrm{QuantumNumberConsidered}[[i]] - 1)}$ 
 Sum[$(-1)^k$ M[m, opTQuadFreq[Atoms[[i, 1]], \{-m, 0\}], t, AngleQ[[i]]] $V_k$ opTQuad[AtomsQuadConsidered[[n,2]],\{-m,0\}],  \{m, -2, 2\}], \{k, -2, 2\}];
\end{center}

\newpage
\section{Tables}
\begin{center}
	\begin{longtable}{l l l}
	\caption[Variable names used in \RedKite.]{Variable names used in \RedKite.} \label{table:VariableExplain} \\

	\hline \multicolumn{1}{c}{Name} & \multicolumn{1}{c}{definition} & \multicolumn{1}{c}{User-defined?} \\ \hline 
	\endfirsthead

	\multicolumn{3}{c}%
	{{\tablename\ \thetable{} -- continued from previous page}} \\
	\hline \multicolumn{1}{c}{Name} &
	\multicolumn{1}{c}{definition} &
	\multicolumn{1}{c}{User-defined?} \\ \hline 
	\endhead

	\hline \multicolumn{3}{r}{{Continued on next page}} \\ \hline
	\endfoot

	\hline

	\endlastfoot

	Atoms & Table containing the spins present in the system & Yes\\
		& and their associated labels  \\
	NumberofAtoms & number of spins considered & No \\
	LF & vector orienting the System Frame in the & Yes \\
		& Cartesian axis system \\
	Coordinates & Table containing the position of the spins in  & Yes \\
		& the Cartesian axis system \\
	CSAConsidered & Table filled with 1 (CSA is considered)  & Yes\\
		& or 0 (CSA is neglected) \\
	$\delta_{csa}[i]$ & value of the axially symmetric CSA  & Yes\\
		& associated with nucleus \textit{i} \\
	$\sigma \mathrm{long}[i]$ & value of the longitudinal component of & Yes\\
		& an asymmetric CSA associated with nucleus \textit{i}  \\
	$\sigma \mathrm{perp}[i]$ & value of the orthogonal component & Yes  \\
		& of an asymmetric CSA associated with nucleus \textit{i}\\
	vectorNum${^\mathrm{"CSA"}}_i$ & orientation of the principal axis of  & Yes\\
		& a symmetric CSA tensor for spin \textit{i} \\
	vectorNuml${^\mathrm{"CSA"}}_i$ & orientation of the longitudinal component & Yes\\
		& of a symmetric CSA tensor for spin \textit{i}  \\
	vectorNump${^\mathrm{"CSA"}}_i$ & orientation of the longitudinal component & Yes  \\
		& of a symmetric CSA tensor for spin \textit{i}\\
	$d_\mathcal{Q}[i]$ & strength of the quadrupolar interaction for spin \textit{i} & Yes \\
	vectorNum${^\mathrm{"Quad"}}_i$ & orientation of the quadrupolar interaction for spin \textit{i} & Yes \\
	opTDip & tensors associated with dipolar interactions & No \\
	opTCSA & tensors associated with CSA interactions& No \\
	opTQuad & tensors associated with quadrupolar interactions & No \\
	opTDipFreq & frequencies associated to tensors OpTDip & No \\
	opTCSAFreq & frequencies associated to tensors OpTCSA  & No \\
	opTQuadFreq & frequencies associated to tensors OpTQuad  & No \\
	dDD[i, j] & dipolar coefficient for the interaction of spins \textit{i} and \textit{j} & No \\
	$\Phi[i, j]$ & vector linking spins \textit{i} and \textit{j} & No \\
	$\Delta_i$ & symmetric CSA value in Hz: $\sqrt{2/3} \delta_{csa}[i] \omega[i]$ & No \\
	$\sigma \mathrm{ln}_i$ & longitudinal component of an asymmetric  & No\\
		& CSA value in Hz: $\sqrt{2/3} \sigma \mathrm{long}[i] \omega[i]$ \\
	$\sigma \mathrm{pn}_i$ & orthogonal component of an asymmetric  & No\\
		& CSA value in Hz: $\sqrt{2/3} \sigma \mathrm{perp}[i] \omega[i]$ \\
	$\omega[i]$ & Larmor frequency associated with spin \textit{i} & No \\
	AngleCSA[n, 1] & orientation of the longitudinal component  & No\\
		& of the CSA of spin \textit{i} \\
	AngleCSA[n, 2] & orientation of the orthogonal component  & No \\
		& of the CSA  of spin \textit{i}& No \\
	AngleQ[n, 2] & orientation of the quadrupolar interaction of spin \textit{i} \\
	\textit{M} & function depending on variables detailed in main text  & No\\
		& to perform the calculations \\
	SpinTermOfInterest & Studied operator during the relaxation experiments & Yes
	\end{longtable}
\end{center}

\begin{table}
	\def\arraystretch{1.5}
	\begin{center}
	\caption{Tensor operators for the dipole-dipole interaction and associated frequency as written in \RedKite. Tensors are of rank 2 and with coherence order \textit{q}. The letter \textit{p} refers to the decomposition of the tensors in the irreducible tensor operator basis. Tensors are written $\mathrm{opTDip}[\{i\_,j\_\},\{q, p\}]$ for the interaction between nuclei \textit{i} and \textit{j}. The associated frequencies are $\mathrm{opTDipFreq}[\{i\_,j\_\},\{q, p\}]$. We define $\omega[i] = -\gamma_i B_0$ in \RedKite. $B_0$ is the magnetic field. } \label{table:DDTensors}
		\begin{tabular}{cccc}
			coherence order & p & Tensor & Frequency \\
			\hline
		 	2  & 0 &$ \frac{1}{2} \mathrm{opI}[i, "+"].\mathrm{opI}[j, "+"]$ & $\omega[i] + \omega[j]$ \\
		 	1  & 0 &$ -\frac{1}{2} \mathrm{opI}[i, "z"].\mathrm{opI}[j, "+"]$ & $\omega[j]$ \\
		 	1  & 1 & $ -\frac{1}{2} \mathrm{opI}[i, "+"].\mathrm{opI}[j, "z"]$ & $\omega[i]$ \\
		 	0  & -1 &$ -\frac{1}{2\sqrt{6}} \mathrm{opI}[i, "-"].\mathrm{opI}[j, "+"]$ & $\omega[j] - \omega[i]$ \\
		 	0  & 0 &$ \frac{2}{\sqrt{6}} \mathrm{opI}[i, "z"].\mathrm{opI}[j, "z"]$ & 0 \\
		 	0  & 1 &$ -\frac{1}{2\sqrt{6}} \mathrm{opI}[i, "+"].\mathrm{opI}[j, "-"]$ & $\omega[i] - \omega[j]$ \\
		 	-1  & 0 & $ \frac{1}{2} \mathrm{opI}[i, "z"].\mathrm{opI}[j, "-"]$ & $\omega[j]$ \\
		 	-1  & 1 & $ \frac{1}{2} \mathrm{opI}[i, "-"].\mathrm{opI}[j, "z"]$ & $\omega[i]$ \\
		 	-2  & 0 &$ \frac{1}{2} \mathrm{opI}[i, "-"].\mathrm{opI}[j, "-"]$ & $-\omega[i] - \omega[j]$ \\
		\end{tabular}
	\end{center}
\end{table}

\begin{table}
	\def\arraystretch{1.5}
	\begin{center}
	\caption{Tensor operators for the Chemical Shift Anisotropy (CSA) interaction and associated frequency as written in \RedKite. Tensors are of rank 2 and with coherence order \textit{q}. The letter \textit{p} refers to the decomposition of the tensors in the irreducible tensor operator basis. Tensors are written $\mathrm{opTDip}[\{i\_,j\_\},\{q, p\}]$ for the interaction between nuclei \textit{i} and \textit{j}. The associated frequencies are $\mathrm{opTDipFreq}[\{i\_,j\_\},\{q, p\}]$. We define $\omega[i] = -\gamma_i B_0$ in \RedKite. $B_0$ is the magnetic field. } \label{table:CSATensors}
		\begin{tabular}{cccc}
			coherence order & p & Tensor & Frequency \\
			\hline
		 	2  & 0 & 0 & $2 \omega[i] $ \\
		 	1  & 0 &$ -\frac{1}{2} \mathrm{opI}[i, "+"]$ & $\omega[i]$ \\
		 	0  & 0 &$ \frac{2}{\sqrt{6}} \mathrm{opI}[i, "z"] $ & 0 \\
		 	-1  & 0 & $ \frac{1}{2} \mathrm{opI}[i, "-"]$ & $-\omega[i]$ \\
		 	-2  & 0 &  0  & $-2 \omega[i] $ \\
		\end{tabular}
	\end{center}
\end{table}

\begin{table}
	\def\arraystretch{1.5}
	\begin{center}
	\caption{Tensor operators for the quadrupolar interaction and associated frequency as written in \RedKite. Tensors are of rank 2 and with coherence order \textit{q}. The letter \textit{p} refers to the decomposition of the tensors in the irreducible tensor operator basis. Tensors are written $\mathrm{opTDip}[\{i\_,j\_\},\{q, p\}]$ for the interaction between nuclei \textit{i} and \textit{j}. The associated frequencies are $\mathrm{opTDipFreq}[\{i\_,j\_\},\{q, p\}]$. We define $\omega[i] = -\gamma_i B_0$ in \RedKite. $B_0$ is the magnetic field. } \label{table:QuadTensors}
		\begin{tabular}{cccc}
			coherence order & p & Tensor & Frequency \\
			\hline
		 	2  & 0 & $\frac{1}{2} \mathrm{opI}[i, "+"].\mathrm{opI}[i, "+"]$ & $2 \omega[i] $ \\
		 	1  & 0 & $ -\frac{1}{2} ( \mathrm{opI}[i, "z"].\mathrm{opI}[i, "+"]$ & $\omega[i]$\\
				& & $+  \mathrm{opI}[i, "+"].\mathrm{opI}[i, "z"] )$  \\
		 	0  & 0 & $ \frac{1}{\sqrt{6}} (2 \mathrm{opI}[i, "z"]. \mathrm{opI}[i, "z"]$ \\
				& & $-  \mathrm{opI}[i, "x"]. \mathrm{opI}[i, "x"] $  & 0 \\
				& & $-\mathrm{opI}[i, "y"]. \mathrm{opI}[i, "y"] )$  \\
		 	-1  & 0 & $\frac{1}{2} ( \mathrm{opI}[i, "z"].\mathrm{opI}[i, "-"]$ & $-\omega[i]$ \\
				& & $+ \mathrm{opI}[i, "-"].\mathrm{opI}[i, "z"] )$  \\
		 	-2  & 0 &  $\frac{1}{2} \mathrm{opI}[i, "-"].\mathrm{opI}[i, "-"]$  & $-2 \omega[i] $ \\
		\end{tabular}
	\end{center}
\end{table}

\begin{table}[!ht] 
	\def\arraystretch{1.5}
	\begin{center}
	\caption{Values of the parameters describing the position of the effective surrounding deuterium nucleus for each isoleucine residue in the Cartesian axis system which origin is occupied by the \Carbon.}%
		\begin{tabular}{cccccccc}
			Residue & 3 & 13 & 23 & 30 & 36 & 44 & 61 \\
			\hline
		 	r$_\mathrm{y,CD_{vic}}$ (\AA)    & -1.96 & -1.97 & -1.88 & -2.00 & -1.97 & -1.17 & -1.39     \\%
			r$_\mathrm{z,CD_{vic}}$ (\AA) & -0.73 & -1.06 & -0.86 & -0.74 & -0.65 & -1.54 & -1.44 \\%
		\end{tabular}
	\end{center}
	\label{table:ParamDvic}
\end{table}

\begin{table}[!ht]
	\def\arraystretch{1.5}
	\begin{center}
\caption{Longitudinal and transverse cross-correlated cross-relaxation rates between $^{13}$C and $^{13}$C, $^1$H two spin order for the 7 isoleucine residues of Ubiquitin measured at 14.1 and 18.8\,T.}
	\begin{tabular}{ccccc}
		residue & $\eta_z^C / s^{-1}$ (14.1\,T) &  $\eta_z^C / s^{-1}$ (18.8\,T) & $\eta_{xy}^C / s^{-1}$ (14.1\,T) &  $\eta_{xy}^C / s^{-1}$ (18.8\,T) \\
		\hline
		 3 & 0.0413 $ \pm $ 0.0006 &  0.0312 $ \pm $ 0.0019 & 0.669 $ \pm $ 0.006 & 0.894 $ \pm $ 0.008\\
		13 & 0.0524 $ \pm $ 0.0005 &  0.0469 $ \pm $ 0.0016 & 0.466 $ \pm $ 0.004 & 0.636 $ \pm $ 0.006\\
		23 & 0.0208 $ \pm $ 0.0007 &  0.0209 $ \pm $ 0.0016 & 0.273 $ \pm $ 0.003 & 0.353 $ \pm $ 0.005\\
		30 & 0.0505 $ \pm $ 0.0007 &  0.0411 $ \pm $ 0.0018 & 0.649 $ \pm $ 0.006 & 0.886 $ \pm $ 0.009\\
		36 & 0.0585 $ \pm $ 0.0007 &  0.0513 $ \pm $ 0.0017 & 0.539 $ \pm $ 0.004 & 0.722 $ \pm $ 0.006\\
		44 & 0.0492 $ \pm $ 0.0003 &  0.0509 $ \pm $ 0.0018 & 0.266 $ \pm $ 0.004 & 0.340 $ \pm $ 0.006\\
		61 & 0.0376 $ \pm $ 0.0006 &  0.0353 $ \pm $ 0.0015 & 0.451 $ \pm $ 0.004 & 0.611 $ \pm $ 0.006\\
	\end{tabular}
	\end{center}
	\label{table:etaZ}
\end{table}

\begin{table}[!ht]
	\def\arraystretch{1.5}
	\begin{center}
	\caption{Proton longitudinal relaxation rates of the 7 isoleucine residues of Ubiquitin measured at 14.1 and 18.8\,T.}
	\begin{tabular}{ccc}
		residue & $R_1 (^{1}H) / s^{-1}$ (14.1\,T) &  $ R_1 (^{1}H) / s^{-1}$ (18.8\,T) \\
		\hline
		3 & 0.235 $ \pm $ 0.003 &  0.228 $ \pm $ 0.001 \\
		13 & 0.344 $ \pm $ 0.003 &  0.317 $ \pm $ 0.001 \\
		23 & 0.572 $ \pm $ 0.005 &  0.522 $ \pm $ 0.002 \\
		30 & 0.258 $ \pm $ 0.003 &  0.243 $ \pm $ 0.001 \\
		36 & 0.305 $ \pm $ 0.003 &  0.266 $ \pm $ 0.001 \\
		44 & 0.292 $ \pm $ 0.003 &  0.253 $ \pm $ 0.001 \\
		61 & 0.430 $ \pm $ 0.004 &  0.390 $ \pm $ 0.001 \\
	\end{tabular}
	\end{center}
	\label{table:1HR1}
\end{table}

\begin{table}[!ht]
	\def\arraystretch{1.5}
	\begin{center}
	\caption{$^{13}$C relaxation rate measured at 14.1\,T following a relaxometry scheme (\textit{i.e.} without control of the cross-relaxation pathways). The rate $R_1^\mathrm{app}$ was measured with the same delays as used in the standard relaxation experiment. The rate $R_1^{\prime \mathrm{app}}$ was measured by adding an extra relaxation delay of 550\,ms in all experiments.}
	\begin{tabular}{ccc}
		residue & $ R_1^\mathrm{app} (^{13}C)/ s^{-1} $ & $ R_1^{\prime \mathrm{app}} (^{13}C)/ s^{-1}$ \\
		\hline
		3  & 0.349 $ \pm $ 0.009 &  0.344 $ \pm $ 0.011 \\
		13 & 0.455 $ \pm $ 0.007 &  0.452 $ \pm $ 0.010 \\
		23 & 0.603 $ \pm $ 0.008 &  0.576 $ \pm $ 0.011 \\
		30 & 0.385 $ \pm $ 0.009 &  0.391 $ \pm $ 0.011 \\
		36 & 0.445 $ \pm $ 0.008 &  0.431 $ \pm $ 0.010 \\
		44 & 0.429 $ \pm $ 0.008 &  0.412 $ \pm $ 0.010 \\
		61 & 0.497 $ \pm $ 0.007 &  0.493 $ \pm $ 0.010 \\
	\end{tabular}
	\end{center}
	\label{table:13CR1_600MHz}
\end{table}

%% file: Article.bbl
\providecommand{\noopsort}[1]{}\providecommand{\singleletter}[1]{#1}%
\begin{thebibliography}{10}
\expandafter\ifx\csname url\endcsname\relax
  \def\url#1{\texttt{#1}}\fi
\expandafter\ifx\csname urlprefix\endcsname\relax\def\urlprefix{URL }\fi
\expandafter\ifx\csname href\endcsname\relax
  \def\href#1#2{#2} \def\path#1{#1}\fi

\bibitem{Ernst_1966}
R.~R. Ernst, W.~A. Anderson, Rev.\,Sci.\,Instrumental. 37 (1966) 93.

\bibitem{Pervushin_pnas_1997}
K.~Pervushin, R.~Riek, G.~Wider, K.~Wuthrich, Proc.\,Natl.\,Acad.\,Sci.\,USA 94
  (1997) 12366.

\bibitem{tugarinov_jacs_2003}
V.~Tugarinov, P.~M. Hwang, J.~E. Ollerenshaw, L.~E. Kay, J.\,Am.\,Chem.\,Soc.
  125 (2003) 10420.

\bibitem{Shimizu_JCP_1964}
H.~Shimizu, J.\,Chem.\,Phys. 40 (1964) 3357.

\bibitem{Goldman1984}
M.~Goldman, J.\,Magn.\,Reson. 60 (1984) 437.

\bibitem{Wimperis_MolPhys_1989}
S.~Wimperis, G.~Bodenhausen, Molec.\,Phys. 66 (1989) 897.

\bibitem{Werbelow_JMR_1973}
L.~G. Werbelow, A.~G. Marshall, J.\,Magn.\,Reson. 11 (1973) 299.

\bibitem{Carravetta_PRL_2004}
M.~Carravetta, O.~G. Johannessen, M.~H. Levitt, Phys.\,Rev.\,Lett. 92 (2004)
  153003.

\bibitem{Carravetta_JCP_2005}
M.~Carravetta, M.~H. Levitt, J.\,Chem.\,Phys 122 (2005) 214505.

\bibitem{Larsen2003}
J.~H. Ardenkjaer-Larsen, B.~Fridlund, A.~Gram, G.~Hansson, L.~Hansson, M.~H.
  Lerche, R.~Servin, M.~Thaning, K.~Golman, Proc.\,Natl.\,Accad.\,Sci.\,USA 100
  (2003) 10158.

\bibitem{Milani_JCP_2015}
J.~Milani, B.~Vuichoud, A.~Bornet, P.~Mi\'{e}ville, R.~Mottier, S.~Jannin,
  G.~Bodenhausen, Rev.\,Sci.\,Instrum. 86 (2015) 024101.

\bibitem{Hilty_2008}
S.~Bowen, C.~Hilty, Angew.\,Chem.\,Int.\,Ed. 47 (2008) 5235.

\bibitem{Kimmich2004}
R.~Kimmich, E.~Anoardo, Prog.\,Nucl.\,Magn.\,Reson.\,Spectrosc. 44 (2004) 257.

\bibitem{Redfield_MRC_2003}
A.~G. Redfield, Magn.\,Reson.\,Chem. 41 (2003) 753.

\bibitem{Redfield2012}
A.~G. Redfield, J.\,Biomol.\,NMR. 52 (2012) 159.

\bibitem{chou_jmr_2012}
C.-Y. Chou, M.~Chu, C.-F. Chang, T.-h. Huang, J.\,Magn.\,Reson. 214 (2012) 302.

\bibitem{Charlier2013}
C.~Charlier, S.~N. Khan, T.~Marquardsen, P.~Pelupessy, V.~Reiss,
  D.~Sakellariou, G.~Bodenhausen, F.~Engelke, F.~Ferrage, J.\,Am.\,Chem.\,Soc.
  135 (2013) 18665.

\bibitem{Korchak_JCP_2010}
S.~Korchak, K.~Ivanov, A.~Yurkovskaya, H.-M. Vieth, J.\,Chem.\,Phys. 133 (2010)
  194502.

\bibitem{Cousin_PCCP_2016}
S.~F. Cousin, C.~Charlier, P.~Kade\v{r}\'{a}vek, T.~Marquardsen, J.-M. Tyburn,
  P.-A. Bovier, S.~Ulzega, T.~Speck, D.~Wilhelm, F.~Engelke, W.~Maas,
  D.~Sakellariou, G.~Bodenhausen, P.~Pelupessy, F.~Ferrage,
  Phys.\,Chem.\,Chem.\,Phys. 18 (2016) 33187.

\bibitem{Cousin2016}
S.~S.~F. Cousin, P.~Kade\v{r}\'{a}vek, B.~Haddou, C.~Charlier, T.~Marquardsen,
  J.-M. J.-M. Tyburn, P.-A. P.-A. Bovier, F.~Engelke, W.~Maas, G.~Bodenhausen,
  P.~Pelupessy, F.~Ferrage, Angew.\,Chem.\,Int.\,Ed. 55 (2016) 9886.

\bibitem{KaderavekJPCL2019}
P.~Kade\v{r}\'{a}vek, N.~Bolik-Coulon, S.~F. Cousin, T.~Marquardsen, J.-M.
  Tyburn, J.-N. Dumez, F.~Ferrage, J.\,Phys.\,Chem.\,Lett. 10 (2019) 5917.

\bibitem{Jasenakova_JBNMR_2020}
Z.~Jase\v{n}\'akov\'{a}, V.~Zapletal, P.~Padrta, M.~Zachrdla, N.~Bolik-Coulon,
  T.~Marquardsen, J.-M. Tyburn, L.~\v{Z}\'{i}dek, F.~Ferrage,
  P.~Kade\v{r}\'{a}vek, J.\,Biomol.\,NMR,\href
  {http://dx.doi.org/10.1007/s10858-019-00298-6}
  {\path{doi:10.1007/s10858-019-00298-6}}.

\bibitem{Roberts_JACS_2004}
M.~F. Roberts, A.~G. Redfield, J.\,Am.\,Chem.\,Soc 126 (2004) 13765.

\bibitem{Clarkson2009}
M.~W. Clarkson, M.~Lei, E.~Z. Eisenmesser, W.~Labeikovsky, A.~Redfield,
  D.~Kern, J.\,Biomol.\,NMR. 45 (2009) 217.

\bibitem{Cousin_JACS_2018}
S.~F. Cousin, P.~Kade\v{r}\'{a}vek, N.~Bolik-Coulon, Y.~Gu, C.~Charlier,
  L.~Carlier, L.~Bruschweiler-Li, T.~Marquardsen, J.-M. Tyburn,
  R.~Br{\"{u}}schweiler, F.~Ferrage, J.\,Am.\,Chem.\,Soc. 140 (2018) 13456.

\bibitem{Cousin2018}
S.~F. Cousin, P.~Kade\v{r}\'{a}vek, N.~Bolik-Coulon, F.~Ferrage, Determination
  of protein ps-ns motions by high-resolution relaxometry, in: Methods in
  Molecular Biology, Vol. 1688, Spinger, 2018, p. 169.

\bibitem{Jerschow_JMR_2005}
A.~Jerschow, J.\,Magn.\,Reson. 176 (2005) 7.

\bibitem{Kuprov2006}
I.~Kuprov, N.~Wagner-Rundell, P.~Hore, J.\,Magn.\,Reson. 184 (2007) 196.

\bibitem{Bengs_MagnResonChecm_2017}
C.~Bengs, M.~H. Levitt, Magn.\,Reson.\,Chem. 56 (2018) 374.

\bibitem{Inc2016}
I.~Wolfram~Research, Mathematica (2016).

\bibitem{Kumar2000}
A.~Kumar, C.~R.~R. Grace, P.~K. Madhu, Prog.\,Nucl.\,Magn.\,Reson.\,Spectrosc.
  37 (2000) 191.

\bibitem{Kowalewski2006}
J.~Kowalewski, L.~M{\"a}ler, Nuclear Spin Relaxation in Liquids: Theory,
  Experiments, and Applications, Taylor \& Francis, 2006.

\bibitem{Nicholas2010}
M.~P. Nicholas, E.~Eryilmaz, F.~Ferrage, D.~Cowburn, R.~Ghose,
  Prog.\,Nucl.\,Magn.\,Reson.\,Spectrosc. 57 (2010) 111.

\bibitem{Abragam1961}
A.~Abragam, The Principles of Nuclear Magnetism, Oxford University Press,
  London, 1961.

\bibitem{Kowalewski1997}
J.~Kowalewski, L.~Werbelow, J.\,Magn.\,Reson. 128 (1997) 144.

\bibitem{Paquin2010}
R.~Paquin, P.~Pelupessy, L.~Duma, C.~Gervais, G.~Bodenhausen, J.\,Chem.\,Phys.
  133 (2010) 034506.

\bibitem{Ferrage_MMB_2012}
F.~Ferrage, Methods\,Mol.\,Biol. 831 (2012) 141.

\bibitem{Levitt1994}
M.~H. Levitt, L.~D. Bari, Bull.\,Magn\,Reson. 16 (1994) 94.

\bibitem{Ghose2009}
R.~Ghose, Concepts\,Magn.\,Reson. 12 (2000) 152.

\bibitem{Pelupessy_JCP_2007}
P.~Pelupessy, F.~Ferrage, G.~Bodenhausen, J.\,Chem.\,Phys. 126 (2007) 134.

\bibitem{BolikCoulon2019}
N.~Bolik-Coulon, S.~F. Cousin, P.~Kade\v{r}\'{a}vek, J.-N. Dumez, F.~Ferrage,
  J.\,Chem.\,Phys. 150 (2019) 224202.

\bibitem{Cavanagh_ProteinNMR_2007}
J.~Cavanagh, W.~J. Fairbrother, A.~G. Palmer, M.~Rance, N.~J. Skelton, Protein
  NMR Spectroscopy: Principles and Practice, Elsevier Academic Press, 2007.

\bibitem{Lipari1982}
G.~Lipari, A.~Szabo, J.\,Am.\,Chem.\,Soc. 104 (1982) 4546.

\bibitem{Novakovic_jmr_2018}
M.~Novakovic, S.~F. Cousin, M.~J. Jaroszewicz, R.~Rosenzweig, L.~Frydman,
  J.\,Magn.\,Reson. 294 (2018) 169.

\bibitem{Wales1997}
D.~J. Wales, J.~P.~K. Doye, J.\,Phys.\,Chem.\,A 101 (1997) 5111.

\bibitem{Foreman-Mackey2013}
D.~Foreman-Mackey, D.~W. Hogg, D.~Lang, J.~Goodman, Publ.\,Astron.\,Soc.\,Pac.
  125 (2013) 306.

\bibitem{Tugarinov2006}
V.~Tugarinov, V.~Kanelis, L.~E. Kay, Nat.\,Protoc. 1 (2006) 749.

\bibitem{Clore2012}
M.~J. Plevin, J.~Boisbouvier, Isotope-Labelling of Methyl Groups for NMR
  Studies of Large Proteins, Royal Society of Chemistry, 2012, Ch.~1, pp.
  1--24.

\bibitem{Mas2013}
G.~Mas, E.~Crublet, O.~Hamelin, P.~Gans, J.~Boisbouvier, J.\,Biomol.\,NMR 57
  (2013) 251.

\bibitem{DuBay2015}
K.~H. DuBay, G.~R. Bowman, P.~L. Geissler, Acc.\,Chem.\,Res. 48 (2015) 1098.

\bibitem{Frederick2007}
K.~K. Frederick, M.~S. Marlow, K.~G. Valentine, J.~Wand, Nature 448 (2007) 325.

\bibitem{Clore1990}
G.~M. Clore, A.~Szabo, A.~Bax, L.~E. Kay, P.~C. Driscoll, A.~M. Gronenborn,
  J.\,Am.\,Chem.\,Soc. 112 (1990) 4989.

\bibitem{Meirovitch2006}
E.~Meirovitch, Y.~E. Shapiro, A.~Polimeno, J.~H. Freed, J.\,Phys.\,Chem.\,A 110
  (2006) 8366.

\bibitem{Meirovitch2007}
E.~Meirovitch, Y.~E. Shapiro, A.~Polimeno, J.~H. Freed, J.\,Phys.\,Chem.\,B 111
  (2007) 12865.

\bibitem{Charlier2016}
C.~Charlier, S.~F. Cousin, F.~Ferrage, Chem.\,Soc.\,Rev. 45 (2016) 2410.

\bibitem{Calandrini2010}
V.~Calandrini, D.~Abergel, G.~R. Kneller, J.\,Chem.\,Phys. 133 (2010) 145101.

\bibitem{Khan2015}
S.~N. Khan, C.~Charlier, R.~Augustyniak, N.~Salvi, V.~D{\'{e}}jean,
  G.~Bodenhausen, O.~Lequin, P.~Pelupessy, F.~Ferrage, Biophys.\,J. 109 (2015)
  988.

\bibitem{Hsu_Biophys_2018}
A.~Hsu, F.~Ferrage, A.~G. Palmer, Biophys.\,J. 115 (2018) 2301.

\bibitem{Frueh2002}
D.~Frueh, Prog.\,Nucl.\,Magn.\,Reson.\,Spectrosc. 41 (2002) 305.

\bibitem{Werbelow1975}
L.~G. Werbelow, D.~M. Grant, J.\,Chem.\,Phys. 63 (1975) 544.

\bibitem{Tugarinov_JBNMR_2004}
V.~Tugarinov, C.~Scheurer, R.~Br{\"{u}}schweiler, L.~E. Kay, J.\,Biomol.\,NMR
  30 (2004) 397.

\bibitem{Pelupessy_JMagnRes_2003}
P.~Pelupessy, G.~M. Espallargas, G.~Bodenhausen, J.\,Magn.\,Res. 161 (2003)
  258.

\bibitem{Tjandra1995}
N.~Tjandra, S.~E. Feller, R.~W. Pastor, A.~Bax, J.\,Am.\,Chem.\,Soc. 117 (1995)
  12562.

\bibitem{Mittermaier_JACS_1999}
A.~Mittermaier, L.~E. Kay, J.\,Am.\,Chem.\,Soc 121 (1999) 10608.

\end{thebibliography}
